\documentclass[twocolumn,showpacs,preprintnumbers,amsmath,amssymb]{revtex4}
\usepackage{graphicx}% Include figure files
\usepackage{dcolumn}% Align table columns on decimal point
\usepackage{bm}% bold math

\begin{document}

\preprint{APS/123-QED}

\title{Anomalous scaling of a passive scalar advected by the turbulent velocity
field with finite correlation time and uniaxial small-scale
anisotropy}

\author{E.\,Jur\v{c}i\v{s}inov\'a$^{1,2}$, and
        M.\,Jur\v{c}i\v{s}in$^{1,3}$}
\address{$^{1}$ Institute of Experimental Physics, Slovak Academy of Sciences,
         Watsonova 47, 040 01 Ko\v{s}ice, Slovakia \\
         $^{2}$ Laboratory of Information Technologies, Joint Institute for Nuclear Research,
         141 980 Dubna, Moscow Region, Russia \\
         $^{3}$ N.N. Bogoliubov Laboratory of Theoretical Physics, Joint Institute for
         Nuclear Research, 141 980 Dubna, Moscow Region, Russia
        } \draft

%\author{M.~Hnatich, M. Jurcisin, M. Repasan Ann  Author}
%\affiliation{%

%\affiliation{

\date{\today}% It is always \today, today,

\begin{abstract}
The influence of uniaxial small-scale anisotropy on the stability of
the scaling regimes and on the anomalous scaling of the structure
functions of a passive scalar advected by a Gaussian solenoidal
velocity field with finite correlation time is investigated by the
field theoretic renormalization group and operator product expansion
within one-loop approximation. Possible scaling regimes are found
and classified in the plane of exponents $\varepsilon-\eta$, where
$\varepsilon$ characterizes the energy spectrum of the velocity
field in the inertial range $E\propto k^{1-2\varepsilon}$, and
$\eta$ is related to the correlation time of the velocity field at
the wave number $k$ which is scaled as $k^{-2+\eta}$. It is shown
that the presence of anisotropy does not disturb the stability of
the infrared fixed points of the renormalization group equations
which are directly related to the corresponding scaling regimes. The
influence of anisotropy on the anomalous scaling of the structure
functions of the passive scalar field is studied as a function of
the fixed point value of the parameter $u$ which represents the
ratio of turnover time of scalar field and velocity correlation
time. It is shown that the corresponding one-loop anomalous
dimensions, which are the same (universal) for all particular models
with concrete value of $u$ in the isotropic case, are different
(nonuniversal) in the case with the presence of small-scale
anisotropy and they are continuous functions of the anisotropy
parameters, as well as the parameter $u$. The dependence of the
anomalous dimensions on the anisotropy parameters of two special
limits of the general model, namely, the rapid-change model and the
frozen velocity field model, are found when $u\rightarrow \infty$
and $u\rightarrow 0$, respectively.
\end{abstract}

\maketitle

\section{\label{sec:level1}Introduction}

One of the last unsolved problem in the framework of classical
physics still remains the theoretical understanding of turbulence.
Within one part of the comprehensive concept of turbulence, namely,
fully developed turbulence, one of the most interesting and still
open question is the theoretical explanation and understanding of
the possible deviations from the classical phenomenological
Kolmogorov-Obukhov theory which are suggested by both natural, as
well as numerical experiments
\cite{MonYagBook,McComb,Frisch,SreAnt97}. Such a behavior is
contained in concepts intermittency and anomalous scaling. During
the last two decades this problem was intensively studied within the
scope of the models of passively advected scalar field
(concentration of an admixture, or temperature are examples) by a
velocity field with prescribed Gaussian statistics. The reason is
twofold. First, it is well known that the deviation from the
classical Kolmogorov-Obukhov theory is even more strongly noticeable
for passively advected scalar field then for the velocity field
itself, see, e.g.,
Ref.\,\cite{SreAnt97,AnHoGaAn84,Sre91,HolSig94,Pumir94,TonWar94,ElKlRo90te,Warhaft00,ShrSig00,MoWiAnTa01},
and second, the problem of passive advection of a scalar or vector
field is considerably easier for theoretical investigation than the
original problem of anomalous scaling in the framework of
Navier-Stokes velocity field. On the other hand,  these simplified
models reproduce many of the anomalous features of genuine turbulent
heat or mass transport observed in experiments. Thus, on one hand,
the theoretical study of the models of passive scalar (or also
vector) advection can be treated as the first step on the long way
of the investigation of intermittency and anomalous scaling in fully
developed turbulence but, on the other hand, the problem of
advection has its own practical importance (see, e.g.,
Ref.\,\cite{Warhaft00} and references cited therein).

The central role in the studies of passive advection was played by a
simple model of passive scalar quantity advected by a random
Gaussian velocity field, white in time and self-similar in space,
the so-called Kraichnan rapid-change model \cite{Kraichnan68}.
Namely, in the framework of the rapid-change model, for the first
time, the anomalous scaling was established on the basis of a
microscopic model \cite{Kraichnan94} and corresponding anomalous
exponents were calculated within controlled approximations
\cite{all,Pumir} (see also survey paper \cite{FaGaVe01} and
references cited therein).

An effective method for investigation of self-similar scaling
behavior is the renormalization group (RG) technique
\cite{Collins,ZinnJustin,Vasiliev}. It plays crucial role in the
explanation of the origin of critical scaling in the theory of
critical phenomena, as well as it allows to calculate some universal
quantities (e.g., critical dimensions). The RG technique can be also
used in the theory of fully developed turbulence and related
problems \cite{Vasiliev,deDoMa79,AdAnVa96,AdAnVa99} (passive
advection is an example). It is important to note that there are
many different RG methods, with the same idea but with technical
differences, but perhaps the most formalized one is the so-called
"quantum field theory" RG which is also known as "field theoretic"
RG \cite{Vasiliev}. It is based on the standard renormalization
procedure, i.e., on the elimination of ultraviolet (UV) divergences.

In Ref.\,\cite{AdAnVa98} the field theoretic RG and operator-product
expansion (OPE) was used in the systematic investigation of the
Kraichnan's rapid-change model, where it was shown that within the
field theoretic RG approach the anomalous scaling is related to the
existence in the model of the composite operators with negative
critical dimensions in the OPE which are usually termed as {\it
dangerous} operators (see, e.g., \cite{Vasiliev,AdAnVa96,AdAnVa99}
for details). In Ref.\,\cite{AdAnVa98} the anomalous exponents were
calculated to order $\varepsilon^{2}$ (two-loop approximation)
within the $\varepsilon$ expansion where parameter $\varepsilon$
describes a given equal-time pair correlation function of velocity
field (see subsequent section) but quite early after this important
work papers \cite{AdAnBaKaVa01} have appeared where the power of the
field theoretic RG was fully demonstrated, namely, the anomalous
exponents of the Kraichnan model were calculated  to order
$\varepsilon^{3}$ (three-loop approximation) within the
$\varepsilon$ expansion. This result was not achieved by any other
method yet and as far as we know this is the only known three-loop
result in fully developed turbulence and related problems at all.

Afterwards, various descendants of the Kraichnan model, namely,
models with inclusion of small scale anisotropy \cite{AdAnHnNo00},
compressibility \cite{AdAn98,AnHo01}, finite correlation time of
velocity field \cite{Antonov99,Antonov00,AdAnHo02,HnJuJuRe06}, and
helicity \cite{ChHnJuJuMaRe06ab} were studied by field theoretic
approach. Moreover, advection of passive vector field by Gaussian
self-similar velocity field (with and without large and small scale
anisotropy, pressure, compressibility, and finite correlation time)
has been also investigated and all possible asymptotic scaling
regimes and cross-over among them have been classified and anomalous
scaling was investigated \cite{all1}. General conclusion of all
these investigations is that the anomalous scaling, which is the
most intriguing and important feature of the Kraichnan rapid change
model, remains valid for all generalized models.

The Kraichnan model works with white in time ($\delta$ correlated in
time) and self-similar in space Gaussian statistics of the velocity
field. In Ref.\,\cite{Antonov99} the field theoretic RG technique
and OPE method was applied in the analysis of more general model of
passively advected scalar field by a self-similar Gaussian velocity
field with finite correlation time first proposed in
Ref.\,\cite{HolSig94}. This model contains the Kraichnan model as a
special limit case (see next section). Maybe the most interesting
conclusion from the view of anomalous scaling analysis obtained in
Ref.\,\cite{Antonov99} is that within the one-loop approximation the
anomalous behavior of all particular models of the general one (the
Kraichnan model is an example) is the same, i.e., the corresponding
critical dimensions associated with needed composite operators
within the OPE are the same. This conclusion is held in isotropic
model, as well as in the  model with large-scale anisotropy with
incompressible (solenoidal) velocity field. This universality of the
anomalous behavior is destroyed, e.g., by the assumption that
velocity field is nonsolenoidal as was shown in
Ref.\,\cite{Antonov00} or by the assumption of the presence of
small-scale anisotropy of the velocity field what will be
demonstrated explicitly in present work. But first let us motivate
the importance of such investigations.

In Ref.\,\cite{AdAnHnNo00} the field theoretic RG and OPE were
applied to the rapid change model of passive scalar advected by
Gaussian strongly anisotropic velocity field where the anomalous
exponents of the structure functions were calculated to the first
order in $\varepsilon$ expansion. It was shown that in the presence
of small-scale anisotropy the corresponding exponents are
nonuniversal, i.e., they are functions of the anisotropy parameters,
and they form the hierarchy with the leading exponent related to the
most "isotropic" operator. The importance of these investigations is
dictated by the question of the influence of anisotropy on
inertial-range behavior of passively advected fields
\cite{Pumir,Antonov99,Antonov00,LaMa99,CeLaMaVe00,AnLaMa00,ArBiPr00,ArLvPoPr00,alll},
as well as the velocity field itself \cite{SaVe94BoOr96,all5,YoKa01}
(see also the survey paper \cite{BiPr05} and references cited
therein, as well as recent astrophysical investigations, e.g, in
Refs.\,\cite{BiBiGaVe06,SoCaBrVe06}). On one hand, it was shown that
for the even structure (or correlation) functions the exponents
which describe the inertial-range scaling exhibit universality and
they are ordered hierarchically in respect to degree of anisotropy
with leading contribution given by the exponent from the isotropic
shell but, on the other hand, the survival of the anisotropy in the
inertial-range is demonstrated by the behavior of the odd structure
functions, namely, the so-called skewness factor decreases down the
scales slower than expected earlier in accordance with the classical
Kolmogorov-Obukhov theory.

Let us describe briefly the solution of the problem in the framework
of the field theoretic approach \cite{Vasiliev,AdAnVa96,AdAnVa99}.
It can be divided into two main stages. On the first stage the
multiplicative renormalizability of the corresponding field
theoretic model is demonstrated and the differential RG equations
for its correlation functions are obtained. The asymptotic behavior
of the latter on their ultraviolet argument $(r/\ell)$ for
$r\gg\ell$ and any fixed $(r/L)$ is given by infrared stable fixed
points of those equations. Here $\ell$ and $L$ are inner
(ultraviolet) and outer (infrared) scales (lengths). It involves
some {}``scaling functions'' of the infrared argument $(r/L)$, whose
form is not determined by the RG equations. On the second stage,
their behavior at $r\ll L$ is found from the OPE within the
framework of the general solution of the RG equations. There, the
crucial role is played by the critical dimensions of various
composite operators, which give rise to an infinite family of
independent aforementioned scaling exponents (and hence to
multiscaling).

In Ref.\,\cite{Antonov99} the problem of a passive scalar advected
by Gaussian self-similar velocity field with finite correlation time
\cite{all2} was studied by field theoretic RG method. There, the
systematic study of the possible scaling regimes and anomalous
behavior was present at one-loop level. The two-loop corrections to
the anomalous exponents were obtained in Ref.\,\cite{AdAnHo02}. It
was shown that the anomalous exponents are nonuniversal as a result
of their dependence on a dimensionless parameter $u$, the ratio of
the velocity correlation time, and turnover time of scalar field.

In what follows we shall continue with the investigation of this
model from the point of view of the influence of the uniaxial
small-scale anisotropy on the anomalous scaling of the single-time
structure functions. In contradistinction with the studies of
\cite{Antonov99}, where the velocity was isotropic and the
large-scale anisotropy was introduced by the imposed linear mean
gradient, the uniaxial anisotropy in our model persists for all
scales, leading to nonuniversality of the anomalous exponents
through their dependence on the anisotropy parameters and ratio of
characteristic time scales. It can be consider as an additional step
to the construction of a more realistic model of anisotropic passive
advection.

The aim of the present paper is twofold. First of all we shall find
the dependence of the anomalous exponents on the anisotropy
parameters of the model and on the parameter $u$, therefore we shall
be able to answer the question whether the system with finite time
correlations of the velocity field with presence of small-scale
anisotropy is more anomalous, i.e., whether the corresponding
critical dimensions are less than those of the Kraichnan rapid
change model which was investigated in Ref.\,\cite{AdAnHnNo00}. The
answer on this question can be treated as the first step on the way
of investigating of the model with velocity field driven by the
stochastic Navier-Stokes equation which is more complicated form
mathematical point of view. The second aim is to analyze whether the
finite correlation time of velocity field can lead to more
complicated structure of critical dimensions than it was shown in
Ref.\,\cite{AdAnHnNo00} within the rapid-change model with
small-scale anisotropy.

The paper is organized as follows. In the first part of
Sec.\,\ref{sec2}, we give the precise formulation of used model. In
the second part, we give the field theoretic formulation of the
model and discuss corresponding diagrammatic technique. In
Sec.\,\ref{sec3}, we perform the ultraviolet (UV) renormalization of
the model, the renormalization constants are calculated in one-loop
approximation, and the corresponding RG equations are derived. In
Sec.\,\ref{sec4} we discuss the stability of possible scaling
regimes of the model which are governed by the corresponding
infrared (IR) fixed points. In Sec.\,\ref{sec5}, the renormalization
of needed composite operators is done and their critical dimensions
are found as functions of parameters of the model. Obtained results
are reviewed and discussed in Sec.\,\ref{sec6}.

\section{Field theoretic description of the model\label{sec2}}

The advection of a passive scalar field
$\theta(x)\equiv\theta(t,{\bf x})$ in an incompressible turbulent
environment is described by the stochastic equation
\begin{equation}
\partial_{t}\theta+v_{i}\partial_{i}\theta=\nu_{0}\Delta\theta+f\,,\label{eq:theta}
\end{equation}
where $\partial_{t}\equiv\partial/\partial t$,
$\partial_{i}\equiv\partial/\partial x_{i}$, $\nu_{0}$ is the
molecular diffusivity coefficient (in what follows, a subscript $0$
denotes bare parameters of unrenormalized theory),
$\Delta\equiv\partial^{2}$ is the Laplace operator, $v_{i}\equiv
v_{i}(x)$ is the $i$-th component of the divergence-free (owing to
the incompressibility) velocity field ${\bf v}(x)$, and $f\equiv
f(x)$ is an artificial Gaussian random noise with zero mean and
correlation function
\begin{equation}
D^{\theta}(x;x^{\prime})=\langle
f(x)f(x^{\prime})\rangle=\delta(t-t^{\prime})C({\bf r}/L),\,\,\,
{\bf r}={\bf x}-{\bf x^{\prime}},\label{eq:corelf}
\end{equation}
where the angular brackets $\langle...\rangle$ hereafter denote
average over the corresponding statistical ensemble and $L$ is an
integral scale related to the stirring. The random noise is
introduced to maintain the steady state of the system but the
detailed form of the function $C({\bf r}/L)$ in
Eq.\,(\ref{eq:corelf}) will be inessential in our consideration. The
only condition which must be satisfied by the function $C({\bf
r}/L)$ is that it must be finite and must decrease rapidly for $r\gg
L$. In the problems related to the genuine turbulence the velocity
field ${\bf v}(x)$ satisfies Navier-Stokes equation but, in what
follows, we shall work with a simplified model where we suppose that
the statistics of the velocity field is given in the form of a
Gaussian distribution with zero mean and pair correlation function
\cite{Antonov99,Antonov00}
\begin{eqnarray}D^v_{ij}(x;x^{\prime})&=&\langle
v_{i}(x)v_{j}(x^{\prime})\rangle = \int\frac{d\omega
d^{d}{\bf k}}{(2\pi)^{d+1}} P_{ij}({\bf k}) \nonumber \\
&\times& D^{v}(\omega, k) e^{-i[\omega(t-t^{\prime})-{\bf k} \cdot
({\bf x}-{\bf x^{\prime}})]},\label{eq:corelv}
\end{eqnarray}
with
\begin{equation}
D^{v}(\omega,k)=\frac{D_{0}k^{4-d-2\varepsilon-\eta}}
{(i\omega+u_{0}\nu_{0}k^{2-\eta})(-i\omega+u_{0}\nu_{0}k^{2-\eta})},\label{corrvelo}
\end{equation}
where $k=|{\bf k}|$ and a transverse projector $P_{ij}({\bf k})$
reflects vectorial nature of the solenoidal velocity field. In the
isotropic case it has the form of the simple transverse projector
\begin{equation}
P_{ij}({\bf k})=\delta_{ij}-\frac{k_i k_j}{k^2}\,.\label{projek}
\end{equation}
In the anisotropic case the transverse projector becomes more
complicated as it will be specified below (see also
Ref.\,\cite{AdAnHnNo00}). In Eq.\,(\ref{corrvelo}) $D_{0}\equiv
g_{0}\nu_{0}^{3}$ is a positive amplitude factor and introduced
parameter $g_{0}$  plays the role of the coupling constant of the
model. In addition, $g_{0}$ is a formal small parameter of the
ordinary perturbation theory. On the other hand, the parameter
$u_{0}$, introduced in the denominator of Eq.\,(\ref{corrvelo}),
gives the ratio of turnover time of scalar field and velocity
correlation time (see, e.g., Ref.\,\cite{Antonov99} for details).
The positive exponents $\varepsilon$ and $\eta$
($\varepsilon=O(\eta)$) are small RG expansion parameters. Thus, we
have a kind of double expansion model in the $\varepsilon-\eta$
plane around the origin $\varepsilon=\eta=0$. The coupling constant
$g_{0}$ and the exponent $\varepsilon$ control the behavior of the
equal-time pair correlation function of velocity field (mean square
velocity) or, equivalently, energy spectrum. On the other hand, the
parameter $u_{0}$ and the second exponent $\eta$ are related to the
frequency $\omega\simeq u_{0}\nu_{0}k^{2-\eta}$ which characterizes
the mode $k$ \cite{Antonov99,all3,ZhaGli92,ChFaLe96,Eyink96}. Thus,
in our notation, the value $\varepsilon=4/3$ corresponds to the
celebrated Kolmogorov \char`\"{}two-thirds law\char`\"{} for the
spatial statistics of the velocity field or, equivalently,
\char`\"{}five-thirds law\char`\"{} for the energy spectrum, and
$\eta=4/3$ corresponds to the Kolmogorov frequency. Simple
dimensional analysis shows that $g_{0}$ and $u_{0}$, which we
commonly term as charges, are related to the characteristic
ultraviolet (UV) momentum scale $\Lambda$ (or inner legth
$l\sim\Lambda^{-1}$) by the following relations
\begin{equation}
g_{0}\simeq\Lambda^{2\varepsilon+\eta},\qquad
u_{0}\simeq\Lambda^{\eta}.
\end{equation}

As was discussed in Introduction, in what follows, we shall take the
velocity statistics to be anisotropic at all scales. For that
purpose, we replace the ordinary transverse projector $P_{ij}({\bf
k})$ in Eq.\,(\ref{eq:corelv}) with the general uniaxially
anisotropic transverse tensor structure (see, e.g.,
Ref.\,\cite{AdAnHnNo00}):
\begin{equation}
T_{ij}({\bf k}) = a(\psi)P_{ij}({\bf k}) + b(\psi) P_{is}({\bf k})
n_s n_t P_{tj}({\bf k}), \label{eq:generaltij}
\end{equation}
where $n_i$ is the $i$-th component of the unit vector ${\bf n}$
$({\bf n}^{2}=1)$ which determines the distinguished direction of
uniaxial anisotropy and $\psi$ is the angle between the vectors
${\bf k}$ and ${\bf n}$, so that ${\bf n}\cdot {\bf k}=k\,\cos\psi$.
It is well known that functions $a(\psi)$ and $b(\psi)$ can be
decomposed into the $d$-dimensional generalization of the Legendre
polynomials which are known as the Gegenbauer polynomials
\cite{GradRizi}, namely,
\begin{equation}
a(\psi)=\sum_{l=0}^{\infty}a_{l}P_{2l}(\cos\psi),\qquad
b(\psi)=\sum_{l=0}^{\infty}b_{l}P_{2l}(\cos\psi) \label{deco}
\end{equation}
(as was shown in Ref.\,\cite{AdAnHnNo00} the odd polynomials do not
affect the scaling behavior). The necessary condition to have
positively defined velocity correlator (\ref{eq:corelv}) leads to
the following inequalities for these functions \cite{AdAnHnNo00}:
\begin{equation}
a(\psi)>0,\qquad
a(\psi)+b(\psi)\sin^{2}\psi>0.\label{eq:inequal}
\end{equation}
But in practical calculations it is impossible to work with the
general tensor structure as is defined in
Eq.\,(\ref{eq:generaltij}). The reason is, at least, because it
contains infinite number of parameters $a_i$ and $b_j$ in the
corresponding decomposition (\ref{deco}). Therefore, in what
follows, we shall work with the simplest special case of the general
uniaxial anisotropic transverse projector, namely,
\begin{equation}
T_{ij}({\bf k})= \left(1+\alpha_{1}\frac{{\bf n\cdot k}}{k^{2}}
\right)P_{ij}({\bf k}) + \alpha_{2}P_{is}({\bf k}) n_s n_t
P_{tj}({\bf k})\,,\label{eq:Tijk}
\end{equation}
which is sufficient for investigation of principal properties of the
uniaxial anisotropy (see the corresponding discussion in
Ref.\,\cite{AdAnHnNo00}). In this case, the inequalities
(\ref{eq:inequal}) reduce into the requirements $\alpha_{1}>-1,
\alpha_2>-1$. This special case represents nicely all main features
of the general model (\ref{eq:generaltij}). This can be seen from
the analysis given in Ref.\,\cite{AdAnHnNo00}.

Let us briefly discuss two special limits of the considered model
(\ref{eq:corelv}), (\ref{corrvelo}) (see also
Ref.\,\cite{Antonov99}). They will be also studied in what follows.
The first of them is the so-called rapid-change model limit when
$u_0\rightarrow \infty$ and $g_0^{\prime}\equiv g_0/u_0^2=$ const
\begin{equation}
D^v(\omega, k)\rightarrow g_0^{\prime} \nu_0 k^{-d-2\varepsilon +
\eta},
\end{equation}
and the second is the so-called quenched (time-independent or
frozen) velocity field limit, which is defined by $u_0\rightarrow 0$
and $g_0^{\prime\prime}\equiv g_0/u_0=$ const
\begin{equation}
D^v(\omega, k)\rightarrow g_0^{\prime\prime} \nu_0^2 \pi
\delta(\omega) k^{-d+2-2\varepsilon},
\end{equation}
which is similar to the well-known models of random walks in a
random environment with long-range correlations; see, e.g.,
Refs.\,\cite{Bouchaud,Honkonen}.

Using the well-known Martin-Siggia-Rose mechanism \cite{Martin} (see
also, e.g., Refs.\,\cite{ZinnJustin,Vasiliev}) the stochastic
problem (\ref{eq:theta})-(\ref{corrvelo}) can be treated as a field
theory with action functional
\begin{eqnarray}
S(\theta,\theta^{\prime},{\bf v}) &=& -\frac{1}{2} \int
dt_1\,d^d{\bf x_1}\,dt_2\,d^d{\bf x_2} \nonumber \\ && v_i(t_1,{\bf
x_1}) [D_{ij}^v(t_1,{\bf
x_1};t_2,{\bf x_2})]^{-1} v_j(t_2,{\bf x_2}) \nonumber  \\
&+& \frac{1}{2} \int dt_1\,d^d{\bf x_1}\,dt_2\,d^d{\bf x_2}
\nonumber \\ && \theta^{\prime}(t_1,{\bf x_1}) D^{\theta}(t_1,{\bf
x_1};t_2,{\bf x_2}) \theta^{\prime}(t_2,{\bf x_2}) \nonumber \\
&+& \int dt\,d^d{\bf x}\,\, \theta^{\prime}\left[-\partial_t -
v_i\partial_i+\nu_0\Delta \right]\theta\,,\label{eq:Sucinok}
\end{eqnarray}
where $\theta^{\prime}$ is an auxiliary scalar field, and
$D^{\theta}$ and $D^{v}$ are correlators (\ref{eq:corelf}) and
(\ref{eq:corelv}), respectively. In action (\ref{eq:Sucinok}) all
required summations over the vector indices are understood. The
second and the third integral in Eq.\,(\ref{eq:Sucinok}) represent
the DeDominicis-Janssen-type action for the stochastic problem
(\ref{eq:theta}), (\ref{eq:corelf}) at fixed ${\bf v}$, and the
first integral represents the Gaussian averaging over ${\bf v}$.

\begin{figure}
\begin{center}\includegraphics[%
  width=8cm]{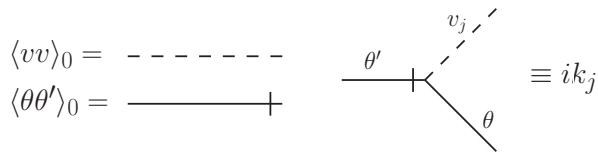}\end{center}
\caption{(Left) Graphical representation of needed propagators of
the model. (Right) The triple (interaction) vertex of the model.
Momentum $\mathbf{k}$ is flowing into the vertex via the auxiliary
field $\theta^{\prime}$. \label{fig1}}
\end{figure}

Model (\ref{eq:Sucinok}) corresponds to a standard Feynman
diagrammatic technique with the bare propagators
$\langle\theta\theta^{\prime}\rangle_{0}$ and $\langle
v_{i}v_{j}\rangle_{0}$ (in the time-momentum representation)
\begin{eqnarray}
\langle\theta(t,{\bf k})\theta^{\prime}(t^{\prime},{\bf k})\rangle_{0}
& = & \theta(t-t^{\prime})e^{-\nu_{0}k^{2}(t-t^{\prime})},\\
\langle v_{i}(t,{\bf k}) v_{j}(t^{\prime},{\bf k}) \rangle_{0} & = &
\frac{D_{0}}{2u_{0}k^{1+2\varepsilon}} \nonumber
\\ &\times& e^{-u_{0}\nu_{0}k^{2-\eta}(t-t^{\prime})}P_{ij}({\bf k}),
\end{eqnarray}
where $\theta(t-t^{\prime})$ is the step function, or (in the
frequency-momentum representation)
\begin{eqnarray}
\langle \theta(\omega,{\bf k}) \theta^{\prime}(\omega,{\bf k})\rangle_{0}
& = & \frac{1}{-i\omega+\nu_{0} k^{2}},\label{proptheta} \\
\langle v_{i}v_{j}\rangle_{0} & = & T_{ij}({\bf k}) D^{v}(\omega,k),
\end{eqnarray}
where $D^{v}(\omega,k)$ is given directly by Eq.\,(\ref{corrvelo}).
In the Feynman diagrams these propagators are represented by the
lines which are shown in Fig.\,\ref{fig1} (the end with a slash in
the propagator $\langle\theta\theta^{\prime}\rangle_{0}$ corresponds
to the field $\theta^{\prime}$, and the end without a slash
corresponds to the field $\theta$). The triple vertex (or
interaction vertex)
$-\theta^{\prime}v_{j}\partial_{j}\theta=\theta^{\prime}v_{j}V_{j}\theta$,
where $V_{j}=ik_{j}$ (in the momentum-frequency representation), is
present in Fig.\,\ref{fig1}, where momentum ${\bf k}$ is flowing
into the vertex via the auxiliary field $\theta^{\prime}$.

In the presence of anisotropy to have a multiplicatively
renormalized model  it is also necessary to introduce new
counterterm of the form $\theta^{\prime}({\bf n} \cdot {\bf
\partial})^{2} \theta$,
which is absent in the unrenormalized action functional
(\ref{eq:Sucinok}). It means that  the model given by action
(\ref{eq:Sucinok}) in its original formulation is not
multiplicatively renormalizable, and in order to use the standard RG
technique it is necessary to extend the model by adding the new
contribution to the unrenormalized action (\ref{eq:Sucinok}). The
extended action is
\begin{eqnarray}
S(\theta,\theta^{\prime},{\bf v}) &=& -\frac{1}{2} \int
dt_1\,d^d{\bf x_1}\,dt_2\,d^d{\bf x_2} \nonumber \\ && v_i(t_1,{\bf
x_1}) [D_{ij}^v(t_1,{\bf
x_1};t_2,{\bf x_2})]^{-1} v_j(t_2,{\bf x_2}) \nonumber  \\
&+& \frac{1}{2} \int dt_1\,d^d{\bf x_1}\,dt_2\,d^d{\bf x_2}
\label{eq:Saniz} \\ && \theta^{\prime}(t_1,{\bf x_1})
D^{\theta}(t_1,{\bf
x_1};t_2,{\bf x_2}) \theta^{\prime}(t_2,{\bf x_2}) \nonumber \\
&& \hspace{-1.7cm}+  \int dt\,d^d{\bf x}\,\,
\theta^{\prime}\left[-\partial_t - v_i\partial_i+\nu_0\Delta
+\chi_{0}\nu_{0}({\bf n}\cdot {\bf
\partial})^{2}\right]\theta\,,\nonumber
\end{eqnarray}
Here $\chi_{0}$ is a new dimensionless unrenormalized parameter. The
stability of the system implies the positivity of the total viscous
contribution $\nu_{0}k^{2}+\chi_{0}\nu_{0}(\mathbf{nk})^{2}$, which
leads to the inequality $\chi_{0}>-1$. Its {}``physical'' value is
zero, but this fact does not hinder the use of the RG technique, in
which it is first assumed to be arbitrary, and equality $\chi_{0}=0$
is imposed as the initial condition in solving the equations for
invariant variables. Below we shall see that the zero value of
$\chi_{0}$ corresponds to certain nonzero value of its renormalized
analog.

For the action (\ref{eq:Saniz}), the bare propagator in
Eq.\,(\ref{proptheta}) is replaced with
\begin{equation}
\langle\theta\theta'\rangle_{0} =
\frac{1}{-\mathrm{i}\omega+\nu_{0}k^{2}+\chi_{0}\nu_{0}(\mathbf{nk})^{2}}.\label{eq:propthetachi}
\end{equation}

The formulation of the problem through the action functional
(\ref{eq:Saniz}) replaces the statistical averages of random
quantities in the stochastic problem defined by
Eqs.\,(\ref{eq:theta}) and (\ref{eq:corelv}) with equivalent
functional averages with weight $\exp S(\Phi)$, where
$\Phi=\{\theta,\theta^{\prime},{\bf v}\}$. The generating
functionals of the total Green functions G(A) and connected Green
functions W(A) are then defined by the functional integral
\begin{equation}
G(A)=e^{W(A)}=\int {\cal D}\Phi \,\, e^{S(\Phi) +
A\Phi},\label{green}
\end{equation}
where $A(x)=\{A^{\theta},A^{\theta^{\prime}},{\bf A^{v}}\}$
represents a set of arbitrary sources for the set of fields $\Phi$,
${\cal D}\Phi \equiv {\cal D}\theta{\cal D}\theta^{\prime}{\cal
D}{\bf v}$ denotes the measure of functional integration, and the
linear form $A\Phi$ is defined as
\begin{equation}
A\Phi= \int d\,x
[A^{\theta}(x)\theta(x)+A^{\theta^{\prime}}(x)\theta^{\prime}(x) +
A_i^{v}(x) v_i(x)].\label{form}
\end{equation}

\section{Renormalization group
analysis\label{sec:Renormalization-group-analysis}}\label{sec3}

Using the standard analysis of canonical dimensions leads to the
information about possible UV divergences in the model (see, e.g.,
Refs.\,\cite{Vasiliev,ZinnJustin}). The dynamical model
(\ref{eq:Saniz}) belongs to the class of the so-called two-scale
models \cite{Vasiliev,AdAnVa96,AdAnVa99}, i.e., to the class of
models for which the canonical dimension of some quantity $F$ is
given by two numbers, namely, the momentum dimension $d^k_F$ and the
frequency dimension $d^{\omega}_F$. To find the dimensions of all
quantities it is convenient to use the standard normalization
conditions $d^k_k=-d^k_x=1, d^{\omega}_{\omega}=-d^{\omega}_t=1,
d^{\omega}_k=d^{\omega}_x=d^k_{\omega}=d^k_t=0$, and the requirement
that each term of the action functional must be dimensionless
separately with respect to the momentum and frequency dimensions.
The total canonical dimension $d_F$ is then defined as $d_F=d^k_F+2
d^{\omega}_F$ (it is related to the fact that $\partial_t \propto
\nu_0 \partial^2$ in the free action (\ref{eq:Saniz}) with choice of
zero canonical dimension for $\nu_0$). In the framework of the
theory of renormalization  the total canonical dimension in
dynamical models plays the same role as the momentum dimension does
in static models.

\begin{table}
\caption{\label{table1} Canonical dimensions of the fields and
parameters of the model under consideration.}
\begin{ruledtabular}
\begin{tabular}{ccccccccc}
$F$ & ${\bf v}$ & $\theta$ & $\theta^{\prime}$ & $m, \Lambda, \mu$ &
$\nu_0, \nu$ & $g_0$ & $u_0$ & $g, u, \chi_0,\chi$ \\
\hline $d^k_F$ & -1 & 0 & $d$ & 1 & -2 & $2 \varepsilon +\eta$ &
$\eta$ & 0 \\
$d^{\omega}_F$ & 1 & -1/2 & 1/2 & 0 & 1 & 0 & 0 & 0 \\
$d_F$ & 1 & -1 & $d+1$ & 1 & 0 & $2 \varepsilon +\eta$ & $\eta$ & 0 \\
\end{tabular}
\end{ruledtabular}
\end{table}

The canonical dimensions of our model are present in Table
\ref{table1}, where also the canonical dimensions of the
renormalized parameters are shown.

The necessity to work with the model based on the action
(\ref{eq:Saniz}) instead of the action (\ref{eq:Sucinok}) is given
by the following consideration. The model (\ref{eq:Sucinok}) is
logarithmic at $\varepsilon=\eta=0$ (the coupling constants $g_{0}$
and $u_0$ are dimensionless); therefore, in the framework of the
minimal substraction (MS) scheme \cite{ZinnJustin}, which is always
used in what follows, possible UV divergences in the correlation
functions have the form of poles in $\varepsilon,\eta$, and their
linear combinations. It is well known that the superficial
divergences can be present only in the 1-irreducible Green functions
for which the corresponding total canonical dimensions are a
nonnegative integer. Detail analysis of the possible divergences was
done, e.g., in Ref.\,\cite{AdAnHnNo00}, therefore we shall not
repeat it here. This analysis shows that superficially divergent
function of our model is only function $\langle \theta^{\prime}
\theta \rangle_{1-ir}$. From the action functional
(\ref{eq:Sucinok}) one immediately obtains that the corresponding
counterterms, which are needed to remove these divergences, must be
proportional to two symbols $\partial$ and, in the isotropic case,
it is reduced to the structure $\theta^{\prime} \triangle \theta$.
However, in the anisotropic case, it is necessary to introduce
possible anisotropic counterterm $\theta^{\prime} ({\bf n}\cdot {\bf
\partial})^{2} \theta$ which is not present in the original action
(\ref{eq:Sucinok}) but which is generated during calculations. This
is the reason why our starting action is the action given in
Eq.\,(\ref{eq:Saniz}).

After this extension the model has become multiplicatively
renormalizable. It means that all divergences can be removed by the
counterterms of the forms $\theta'\Delta \theta$ and $\theta'({\bf
n}\cdot {\bf \partial})^{2}\theta$ \cite{Antonov99,AdAnHnNo00}. This
can be explicitly expressed in the multiplicative renormalization of
the parameters $g_{0},u_{0}$, $\nu_{0}$, and $\chi_{0}$ in the form
\begin{equation} \nu_{0}=\nu Z_{\nu},
g_{0}=g\mu^{2\varepsilon+\eta}Z_{g}, u_{0}=u\mu^{\eta}Z_{u},
\chi_{0}=\chi Z_{\chi}.\label{zetka}\end{equation}
 Here the dimensionless parameters $g,u,\nu$, and $\chi$ are the
renormalized counterparts of the corresponding bare ones, $\mu$ is
the renormalization mass (a scale setting parameter), an artefact of
the dimensional regularization. Quantities
$Z_{i}=Z_{i}(g,u,\chi;d;\epsilon,\eta)$ are the so-called
renormalization constants and, in general, they contain poles in
linear combinations of $\epsilon$ and $\eta.$

The renormalized action functional has the following form:
\begin{eqnarray}
S_R(\theta,\theta^{\prime},{\bf v}) &=& -\frac{1}{2} \int
dt_1\,d^d{\bf x_1}\,dt_2\,d^d{\bf x_2} \nonumber \\ && v_i(t_1,{\bf
x_1}) [D_{ij}^v(t_1,{\bf
x_1};t_2,{\bf x_2})]^{-1} v_j(t_2,{\bf x_2}) \nonumber  \\
&+& \frac{1}{2} \int dt_1\,d^d{\bf x_1}\,dt_2\,d^d{\bf x_2}
\label{eq:Srenorm} \\ && \theta^{\prime}(t_1,{\bf x_1})
D^{\theta}(t_1,{\bf
x_1};t_2,{\bf x_2}) \theta^{\prime}(t_2,{\bf x_2}) \nonumber \\
&& \hspace{-2.0cm}+  \int dt\,d^d{\bf x}\,\,
\theta^{\prime}\left[-\partial_t - v_i\partial_i+\nu Z_1\Delta +\chi
\nu Z_2 ({\bf n}\cdot {\bf
\partial})^{2}\right]\theta\,,\nonumber
\end{eqnarray}
By comparison of the renormalized action (\ref{eq:Srenorm}) with
definitions of the renormalization constants $Z_{i}$,
$i=g,u,\nu,\chi$, which are given in Eqs.\,(\ref{zetka}), we come to
the relations among them:
\begin{equation}
Z_{\nu}=Z_{1},\, Z_{\chi}=Z_{2}Z_{1}^{-1},\, Z_{g}=Z_{1}^{-3},\,
Z_{u}=Z_{1}^{-1}.\label{zetka1}
\end{equation}
%The third and fourth
%relations are consequences of the absence of the renormalization of
%the term with $D_{v}:$\begin{equation}
%g_{0}\nu_{0}^{3}=g\mu^{2\epsilon+\eta}\nu^{3},\qquad
%u_{0}\nu_{0}=u\mu^{\eta}\nu\label{D0}\end{equation}
% in renormalized action (\ref{eq:Srenorm}).

\begin{figure}
\begin{center}\includegraphics[%
  width=6cm]{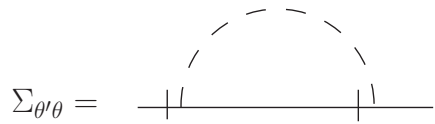}\end{center}
\caption{The one-loop diagram that contribute to the self-energy
operator $\Sigma_{\theta^{\prime}\theta}$. \label{fig2}}
\end{figure}

The issue of interest is, in particular, the behavior of response
functions, e.g., $\langle\theta(x)\theta'(x')\rangle$, correlation
functions
$\langle\theta(x_{1})\theta(x_{2})...\theta(x_{n})\rangle$, and the
equal-time structure functions
\begin{equation}
S_{N}(r)\equiv\langle[\theta(t,\mathbf{x})-\theta(t,\mathbf{x'})]^{N}\rangle,\qquad
r=|\mathbf{x}-\mathbf{x'}|\label{struc}
\end{equation}
in the inertial range specified by the inequalities $l\sim 1/\Lambda
\ll r \ll L=1/m$ ($l$ is an internal length). In the field theoretic
formulation of our stochastic problem the angular brackets
$\langle\ldots\rangle$ mean functional average over fields
$\theta,\theta',\mathbf{v}$ with weight $\exp(S_{R})$. Independence
of the original unrenormalized model of the scale-setting parameter
$\mu$ of the renormalized model yields the RG differential equations
for the renormalized correlation functions of the fields, e.g.,
\begin{equation}
[{\mathcal{{D}}}_{\mu}+\sum_{i=g,\chi,u} \beta_{i} \partial_{i}-
\gamma_{\nu} {\mathcal{{D}}}_{\nu}]
\langle\theta(\mathbf{x},t)\theta(\mathbf{x^{\prime}},t^{\prime})\rangle_{R}=0.
\label{RGoper}
\end{equation}
Here ${\mathcal{{D}}}_{x}\equiv x\partial_{x}$ stands for any
variable $x$ and the RG functions (the $\beta$ and $\gamma$
functions) are given by the well known definitions
\cite{ZinnJustin,Vasiliev}. In our case, using the relations
(\ref{zetka1}) for the renormalization constants, they acquire the
following form:
\begin{eqnarray} \gamma_{i} & \equiv &
\mathcal{{D}}_{\mu}\ln Z_{i}\label{eq:gammai}
\end{eqnarray}
for any renormalization constant $Z_{i}$, and
\begin{eqnarray}
\beta_{g} & \equiv & \mathcal{{D}}_{\mu}g=g(-2\varepsilon-\eta+3\gamma_{1}),\label{betag}\\
\beta_{u} & \equiv & \mathcal{{D}}_{\mu}u=u(-\eta+\gamma_{1}),\label{betau}\\
\beta_{\chi} & \equiv &
\mathcal{{D}}_{\mu}\chi=\chi(\gamma_{1}-\gamma_{2}).\label{betachi}
\end{eqnarray}
The renormalization constants $Z_{1}$ and $Z_{2}$ are determined by
the requirement that the one-particle irreducible Green function
$\langle \theta^{\prime} \theta\rangle_{1-ir}$ must be UV finite
when is written in the renormalized variables. In our case this
means that it has no singularities in the limit
$\varepsilon,\eta\rightarrow0$. The one-particle irreducible Green
function $\langle \theta^{\prime} \theta\rangle_{1-ir}$ is related
to the self-energy operator $\Sigma_{\theta^{\prime}\theta}$, which
is expressed via Feynman graphs,  by the Dyson equation. In
frequency-momentum representation it has the following form:
\begin{equation}
\langle \theta^{\prime}\theta \rangle_{1-ir}=-i\omega+\nu_0 p^2
+\nu_0\chi_0({\bf n}\cdot{\bf p})^2 -
\Sigma_{\theta^{\prime}\theta}(\omega, p).\label{Dyson}
\end{equation}
Thus $Z_{1}$ and $Z_{2}$ are found from the requirement that the UV
divergences are canceled in Eq.\,(\ref{Dyson}) after the
substitution $\nu_{0}=\nu Z_{\nu}$, $\chi_{0}=\chi Z_{\chi}$. This
determines $Z_{1}$ and $Z_{2}$ up to an UV finite contribution,
which is fixed by the choice of the renormalization scheme. In the
MS scheme all the renormalization constants have the form: 1 +
\textit{poles in $\varepsilon,\eta$ and their linear combinations}.
In one-loop approximation the self-energy operator
$\Sigma_{\theta^{\prime}\theta}$ is defined by Feynman diagram which
is shown in Fig.\,\ref{fig2}.

It can be shown that in one-loop calculations it is enough to work
with $\eta=0$ (see, e.g., Refs.\,\cite{Antonov99,Antonov00,
AdAnHo02} for details). This possibility essentially simplifies the
evaluations of all quantities. Then the divergent part of the
diagram given in Fig.\,\ref{fig2} has only poles in $\varepsilon$.
Its explicit analytical form is given as follows (in renormalized
parameters and within one-loop approximation):
\begin{eqnarray}
\Sigma_{\theta'\theta}(p) & = &
-\frac{S_{d}}{(2\pi)^{d}}\frac{g\nu}{2u(1+u)}
\frac{1}{d(d+2)}\frac{1}{\epsilon} \label{sigma} \\ &&\times
\left[p^{2}A+({\bf n}\cdot {\bf p})^{2}B\right], \nonumber
\end{eqnarray}
with
\begin{eqnarray}
A &=& (1+\alpha_{1})d(d+2)\,_{2}F_{1}\left(1;\frac{1}{2};\frac{d}{2};\frac{-\chi}{1+u}\right) \nonumber \\
 &+&
 %\hspace{-1.1cm}
 (\alpha_{2}-\alpha_{1}d-1)(d+2)
 \,_{2}F_{1}\left(1;\frac{1}{2};1+\frac{d}{2};\frac{-\chi}{1+u}\right) \nonumber \\
 &+&
 %\hspace{-0.9cm}
 (\alpha_{1}-\alpha_{2})(d+1)
 \,_{2}F_{1}\left(1;\frac{1}{2};2+\frac{d}{2};\frac{-\chi}{1+u}\right),\label{aaa} \\
B &=& -(1+\alpha_{1})d(d+2)\,_{2}F_{1}\left(1;\frac{1}{2};\frac{d}{2};\frac{-\chi}{1+u}\right) \nonumber \\
 && \hspace{-1.1cm} -[\alpha_{1}(1-2d)+\alpha_{2}-d](d+2)
 \,_{2}F_{1}\left(1;\frac{1}{2};1+\frac{d}{2};\frac{-\chi}{1+u}\right)\nonumber \\
 && \hspace{-0.4cm} -(\alpha_{1}-\alpha_{2})d (d+1)
 \,_{2}F_{1}\left(1;\frac{1}{2};2+\frac{d}{2};\frac{-\chi}{1+u}\right).
 \label{bbb}
\end{eqnarray}
where $S_{d}=2\pi^{d/2}/\Gamma(d/2)$ denotes surface of the
$d$-dimensional unit sphere and  ${_{2}F_{1}}(a,b,c,z)=1+\frac{a\,
b}{c\cdot1}z+\frac{a(a+1)b(b+1)}{c(c+1)\cdot1\cdot2}z^{2}+\ldots$
represents the corresponding hypergeometric function.

In the end, the renormalization constants $Z_{1}$ and $Z_{2}$ are
given as follows
\begin{eqnarray}
Z_{1} & = & 1-\frac{\bar{g}}{2u(1+u)}\frac{1}{d(d+2)}\frac{A}{\epsilon}, \label{z11}\\
Z_{2} & = &
1-\frac{\bar{g}}{2u(1+u)}\frac{1}{d(d+2)\chi}\frac{B}{\epsilon},
\label{z22}
\end{eqnarray}
 where we have introduced new notation $\bar{g}=gS_{d}/(2\pi)^{d}$.

Now using the definition of the anomalous dimensions $\gamma_{1}$
and $\gamma_{2}$ in Eq.(\ref{eq:gammai}) one comes to the following
expressions:
\begin{eqnarray}
\gamma_{1} & = & \frac{\bar{g}}{2u(1+u)d(d+2)}A,\label{gama11} \\
\gamma_{2} & = & \frac{\bar{g}}{2u(1+u)d(d+2)\chi}B. \label{gama22}
\end{eqnarray}
In the next section we shall use these results for investigation of
possible scaling regimes of the model.

\section{Fixed points and scaling regimes\label{sec4}}

Possible scaling regimes of a renormalized model are directly given
by the infrared (IR) stable fixed points of the corresponding system
of the RG equations \cite{ZinnJustin,Vasiliev}. The fixed point of
the RG equations is defined by $\beta$-functions, namely, by
requirement of their vanishing. In our model the coordinates
$g_{*},u_{*},\chi_{*}$ of all possible fixed points are found from
the system of three equations
\begin{equation}
\beta_{g}(g_{*},u_{*},\chi_{*})=\beta_{u}(g_{*},u_{*},\chi_{*})=\beta_{\chi}(g_{*},u_{*},\chi_{*})=0.
\end{equation}
The $\beta$-functions $\beta_{g}$, $\beta_{u}$, and $\beta_{\chi}$
are defined in Eqs. (\ref{betag}), (\ref{betau}), and
(\ref{betachi}). To investigate the IR stability of a fixed point it
is enough to analyze the eigenvalues of the matrix $\Omega$ of the
first derivatives:
\begin{equation} \Omega_{ij}=\left(\begin{array}{ccc}
\partial\beta_{g}/\partial g & \partial\beta_{g}/\partial u & \partial\beta_{g}/\partial\chi\\
\partial\beta_{u}/\partial g & \partial\beta_{u}/\partial u & \partial\beta_{u}/\partial\chi\\
\partial\beta_{\chi}/\partial g & \partial\beta_{\chi}/\partial u & \partial\beta_{\chi}/\partial\chi\end{array}\right).
\end{equation}
The IR asymptotic behavior is governed by the IR stable fixed
points, i.e., those for which real parts of all eigenvalues are
nonnegative.

First of all, we shall study the rapid-change model limit:
$u\rightarrow\infty$. In this regime, it is convenient to make
transformation to new variables, namely, $w\equiv1/u$, and
$g^{\prime}\equiv g/u^{2}$ \cite{Antonov99}, with the corresponding
changes in the $\beta$ functions:
\begin{eqnarray}
\beta_{g^{\prime}} & = & g^{\prime}(-2\varepsilon+\eta+\gamma_{1}),\\
\beta_{w} & = & w(\eta-\gamma_{1}),
\end{eqnarray} while
$\beta_{\chi}$ is unchanged, i.e., it is given by
Eq.\,(\ref{betachi}). In this notation the anomalous dimensions
$\gamma_{1}$ and $\gamma_{2}$ acquire the following form:
\begin{eqnarray}
\gamma_{1} & = & \frac{\bar{g}^{\prime}}{2(1+w)d(d+2)}A^{\prime},\label{gama11w0} \\
\gamma_{2} & = &
\frac{\bar{g}^{\prime}}{2(1+w)d(d+2)\chi}B^{\prime},
\label{gama22w0}
\end{eqnarray}
where again $\bar{g}^{\prime}=g^{\prime}S_{d}/(2\pi)^{d}$, and
$A^{\prime}$ and $B^{\prime}$ acquire the form
\begin{eqnarray}
A^{\prime} & = & (1+\alpha_{1})d(d+2)\,_{2}F_{1}\left(1;\frac{1}{2};\frac{d}{2};\frac{-\chi w}{1+w}\right) \label{aw0}\\
&& \hspace{-0.3cm} +(\alpha_{2}-\alpha_{1}d-1)(d+2)\,_{2}F_{1}\left(1;\frac{1}{2};1+\frac{d}{2};\frac{-\chi w}{1+w}\right) \nonumber \\
&  & +(\alpha_{1}-\alpha_{2})(d+1)\,_{2}F_{1}\left(1;\frac{1}{2};2+\frac{d}{2};\frac{-\chi w}{1+w}\right),\nonumber\\
B^{\prime} & = & -(1+\alpha_{1})d(d+2)\,_{2}F_{1}\left(1;\frac{1}{2};\frac{d}{2};\frac{-\chi w}{1+w}\right) \label{bw0} \\
&& \hspace{-1.1cm} -[\alpha_{1}(1-2d)+\alpha_{2}-d](d+2)\,_{2}F_{1}\left(1;\frac{1}{2};1+\frac{d}{2};\frac{-\chi w}{1+w}\right)\nonumber \\
&&
-(\alpha_{1}-\alpha_{2})(d+1)d\,_{2}F_{1}\left(1;\frac{1}{2};2+\frac{d}{2};\frac{-\chi
w}{1+w}\right).\nonumber
\end{eqnarray}

\begin{figure}
\begin{center}\includegraphics[%
  bb=13bp 45bp 570bp 650bp,
  clip,
  width=7.5cm]{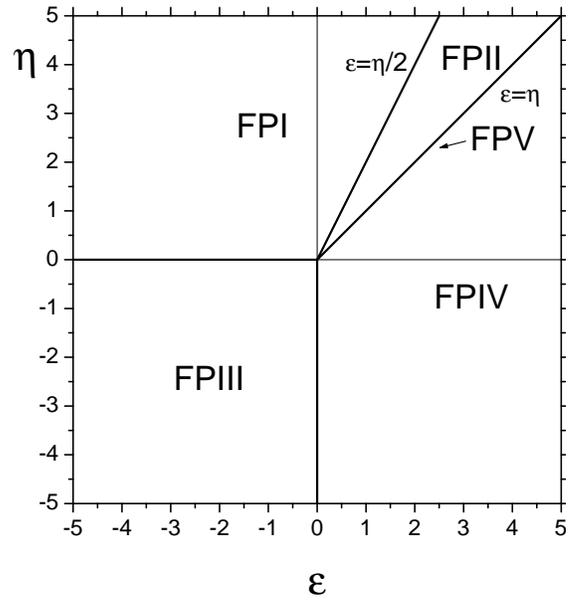}\end{center}
\caption{The "phase" diagram of the fixed points of the model (see
the text for details).\label{fig3}}
\end{figure}

In the rapid-change model limit $w\rightarrow0$
($u\rightarrow\infty$) we are coming to the result of
Refs.\,\cite{AdAnHnNo00} with the anomalous dimensions $\gamma_{1}$
and $\gamma_{2}$ of the form
\begin{eqnarray}
\gamma_{1} &=& \lim_{w\rightarrow0}\frac{\bar{g}^{\prime}}{2(1+w)d(d+2)}A^{\prime}  \\
&=&\frac{\bar{g}^{\prime}}{2d(d+2)}[(d-1)(d+2)+\alpha_{1}(d+1)+\alpha_{2}], \nonumber \\
\gamma_{2} &=&
\lim_{w\rightarrow0}\frac{\bar{g}^{\prime}}{2(1+w)d(d+2)\chi}B^{\prime}\nonumber \\
&=&
\frac{\bar{g}^{\prime}}{2d(d+2)\chi}[-2\alpha_{1}+(d^{2}-2)\alpha_{2}].
\end{eqnarray}
For completeness we shall briefly discuss this spacial case. In this
limit we have two fixed points denoted as FPI and FPII. The first
fixed point is trivial, namely
\begin{equation}
\mathrm{FPI}:\qquad w_{*}=g_{*}^{\prime}=0,
\end{equation}
with arbitrary $\chi_{*}$ and $\gamma_{1}^{*}=0$,
$\gamma_{2}^{*}=0$. The corresponding \char`\"{}stability
matrix\char`\"{} is triangular with diagonal elements (eigenvalues):
\begin{equation}
\lambda_{1}=-2\varepsilon+\eta,\qquad\lambda_{2}=\eta,\qquad\lambda_{3}=0.\end{equation}
The region of the IR stability is shown in Fig.\,\ref{fig3}. The
second point is defined as
\begin{eqnarray}
\mathrm{FPII}:\quad w_{*} & = & 0,\\
\bar{g}_{*}^{\prime} & = & \frac{2d(d+2)(2\varepsilon-\eta)}{(d+2)(d-1)+\alpha_{1}(d+1)+\alpha_{2}},\\
\chi_{*} & = &
\frac{-2\alpha_{1}+\alpha_{2}(d^{2}-2)}{(d+2)(d-1)+\alpha_{1}(d+1)+\alpha_{2}}.
\end{eqnarray}
with $\gamma_{1}^{*}=\gamma_{2}^{*}=2\varepsilon-\eta$. The
triangular matrix $\Omega$ has the following eigenvalues (diagonal
elements)
\begin{equation}
\lambda_{1}=2\varepsilon-\eta,\qquad\lambda_{2}=2\varepsilon-\eta,\qquad\lambda_{3}=-2\varepsilon+2\eta.
\end{equation}
The region of the IR stability of this fixed point is shown in
Fig.\,\ref{fig3}.

Now let us analyze the \char`\"{}frozen regime\char`\"{} with frozen
velocity field. It is mathematically obtained from the model under
consideration in the limit $u\rightarrow0$. To study this transition
it is appropriate to change the variable $g$ to the new variable
$g^{\prime\prime}\equiv g/u$ \cite{Antonov99}. Then the $\beta_g$
function is transformed to the following one:
\begin{equation}
\beta_{g^{\prime\prime}} =
g^{\prime\prime}(-2\varepsilon+2\gamma_{1}),\label{betagu0}
\end{equation}
while $\beta_{u}$ and $\beta_{\chi}$ functions are not changed,
i.e., they are the same as the initial ones given by
Eqs.\,(\ref{betau}) and (\ref{betachi}). In this notation the
anomalous dimensions $\gamma_{1}$ and $\gamma_{2}$ have the form
\begin{eqnarray}
\gamma_{1} & = & \frac{\bar{g}^{\prime\prime}}{2(1+u)d(d+2)}A,\\
\gamma_{2} & = &
\frac{\bar{g}^{\prime\prime}}{2(1+u)d(d+2)\chi}B,
\end{eqnarray}
where, as always,
$\bar{g}^{\prime\prime}=g^{\prime\prime}S_{d}/(2\pi)^{d}$. In the
limit $u\rightarrow0$ the anomalous dimensions $\gamma_{1}$ and
$\gamma_{2}$ acquire the following form:
\begin{eqnarray}
\gamma_{1} & = & \frac{\bar{g}^{\prime\prime}}{2d(d+2)}A^{\prime\prime},\\
\gamma_{2} & = &
\frac{\bar{g}^{\prime\prime}}{2d(d+2)\chi}B^{\prime\prime}.
\end{eqnarray}
where $A^{\prime\prime}$ and $B^{\prime\prime}$ are given as follows
\begin{eqnarray}
A^{\prime\prime} & = & (1+\alpha_{1})d(d+2)\,_{2}F_{1}\left(1;\frac{1}{2};\frac{d}{2};-\chi\right)\nonumber\\
\hspace{-0.5cm} && + (\alpha_{2}-\alpha_{1}d-1)(d+2)\,_{2}F_{1}\left(1;\frac{1}{2};1+\frac{d}{2};-\chi\right) \nonumber\\
&  & +(\alpha_{1}-\alpha_{2})(d+1)\,_{2}F_{1}\left(1;\frac{1}{2};2+\frac{d}{2};-\chi\right), \label{adva}\\
B^{\prime\prime} & = & -(1+\alpha_{1})d(d+2)\,_{2}F_{1}\left(1;\frac{1}{2};\frac{d}{2};-\chi\right)\nonumber \\
&& \hspace{-1.0cm} -[\alpha_{1}(1-2d)+\alpha_{2}-d](d+2)\,_{2}F_{1}\left(1;\frac{1}{2};1+\frac{d}{2};-\chi\right)\nonumber\\
&&
-(\alpha_{1}-\alpha_{2})(d+1)d\,_{2}F_{1}\left(1;\frac{1}{2};2+\frac{d}{2};-\chi\right).
 \label{bdva}
\end{eqnarray}
The system of $\beta$ functions (\ref{betau}), (\ref{betachi}), and
(\ref{betagu0}) exhibits two fixed points, denoted as FPIII and
FPIV, related to the corresponding two scaling regimes. One of them
is again trivial, namely,
\begin{equation}
\mathrm{FPIII}:\qquad u_{*}=g_{*}^{\prime\prime}=0,\end{equation}
 with arbitrary $\chi_{*}$ and $\gamma_{1}^{*}=\gamma_{2}^{*}=0$.
The eigenvalues of the corresponding matrix $\Omega$ are
\begin{equation}
\lambda_{1}=-2\varepsilon,\qquad\lambda_{2}=-\eta,\qquad\lambda_{3}=0.\end{equation}
 Thus this regime is IR stable only if both parameters $\varepsilon$,
and $\eta$ are negative simultaneously as can be seen in
Fig.\ref{fig3}. The second, non-trivial, point is
\begin{eqnarray}
\mathrm{FPIV}:\qquad u_{*} & = & 0,\\
\bar{g}_{*}^{\prime\prime} & = &
\frac{2d(d+2)\varepsilon}{A_{*}^{\prime\prime}},
\end{eqnarray}
where $A_{*}^{\prime\prime}$ is $A^{\prime\prime}$ given in
Eq.\,(\ref{adva}) taken at fixed point, i.e.,  $\chi$ is replaced by
$\chi_{*}$ which is given only implicitly by the equation
\begin{equation}
\chi_* A_{*}^{\prime\prime} - B_{*}^{\prime\prime} = 0,
\label{podchi}
\end{equation}
where $B_{*}^{\prime\prime}$ is $B^{\prime\prime}$ given in
Eq.\,(\ref{bdva}) taken at the fixed point.

Straightforward analysis shows that to have
$\bar{g}_{*}^{\prime\prime}>0$ together with $\chi_*>-1$ one must
suppose $\varepsilon>0$. It is the only condition related to the
coordinates of the fixed point.

The IR stability of the fixed point is again given by the $\Omega$
matrix, namely, by the positive values of real parts of its
eigenvalues. It is triangular in this case, thus its eigenvalues are
given directly by the diagonal elements. The eigenvalues are
\begin{eqnarray}
\lambda_{1} & = & 2\varepsilon,\\
\lambda_{2} & = & \varepsilon-\eta,\\
\lambda_{3} & = & \chi_* \left( \frac{\partial \gamma_1}{\partial
\chi}-\frac{\partial \gamma_2}{\partial \chi}\right)_*.
\end{eqnarray}
Here $\lambda_{3}$ has rather complicated explicit form but it can
be numerically shown that $\lambda_{3}$ is always positive for
$\alpha_{1,2}>-1, \varepsilon>0$, and $d>0$. The region of stability
of this fixed point is shown in Fig.\,\ref{fig3}.

Now let us turn to the most interesting scaling regime with finite
value of the fixed point for the variable $u$. By short analysis one
immediately concludes that the system of equations
\begin{eqnarray}
\beta_{g} & = & g(-2\varepsilon-\eta+3\gamma_{1})=0,\\
\beta_{u} & = & u(-\eta+\gamma_{1})=0,\\
\beta_{\chi} & = & \chi(\gamma_{1}-\gamma_{2})=0. \label{chiii}
\end{eqnarray}
can be fulfilled simultaneously for finite values of $g$ and $u$
only when the parameter $\varepsilon$ is equal to $\eta$:
$\varepsilon=\eta$. In this case the function $\beta_{g}$ is
proportional to function $\beta_{u}$. As a result we have not one
fixed point but a set of fixed points $g_{*}$, $\chi_{*}$ that
depend on arbitrary parameter $u_{*}>0$. The value of the fixed
point for the variable $g$ in one-loop approximation is given as
follows (we denote it as FPV)
\begin{equation}
\mathrm{FPV}:
g_{*}=\frac{2u(1+u_{*})d(d+2)\varepsilon}{A_{*}}\label{eq:FPVg}
\end{equation}
where $A_{*}$ is $A$ from Eq.\,(\ref{aaa}) with $u$ and $\chi$
replaced by $u_{*}$ and $\chi_{*}$, respectively. On the other hand,
$\chi_{*}$ is again known only implicitly and it can be obtained
from Eq.\,(\ref{chiii}) which is equivalent to the condition
\begin{equation}
\gamma_{1}^{*}=\gamma_{2}^{*}, \label{eq:FPVchi}
\end{equation}
where $\gamma_{1}^{*}$, $\gamma_{2}^{*}$ are $\gamma_{1}$,
$\gamma_{2}$ given by Eqs.\,(\ref{gama11}) and (\ref{gama22}) where
$g$, $u$ are replaced by $g_{*}$ and $u_{*}$ respectively.

The eigenvalues of the corresponding stability matrix are

\begin{eqnarray}
\lambda_{1} & = & 0\\
\lambda_{2,3} & = & \frac{1}{2}[C \pm \sqrt{C^2-4 D}],
\end{eqnarray}
where
\begin{eqnarray*}
C&=&3\varepsilon+\chi_*\partial_{\chi}(\gamma_{1}-\gamma_{2})|_*+
u_*\partial_{u}\gamma_{1}|_* \,, \\
D&=&
3\varepsilon\chi_*\partial_{\chi}(\gamma_{1}-\gamma_{2})|_*\nonumber
\\ && -\chi_*
u_*(\partial_{u}\gamma_{1}\,\partial_{\chi}\gamma_{2}-\partial_{\chi}\gamma_{1}\,\partial_{u}\gamma_{2})|_*
\end{eqnarray*}
where $|_*$ means that the quantity must be taken at the fixed
point.
%$\partial_{x}=\left.\frac{\partial}{\partial
%x}\right|_{x=x_{*}}$.
It can be shown numerically that for any positive values of $u_{*}$
and for all possible values of the anisotropy parameters
$\alpha_{1,2}$ the eigenvalues $\lambda_{2}$ and $\lambda_{3}$ are
always greater then zero. Therefore, the corresponding fixed point
is IR stable and satisfy stability condition. It corresponds to the
line $\varepsilon=\eta$ in Fig.\,\ref{fig3}, where the regions of
stability for all possible fixed points are shown.

As was already mentioned (see the previous section) the issue of
interest are especially multiplicatively renormalizable equal-time
two-point quantities $G(r)$ (see also, e.g.,
Ref.\,\cite{Antonov99}). Examples of such quantities are the
equal-time structure functions in the inertial interval as they were
defined in Eq.\,(\ref{struc}). The IR scaling behavior of the
function $G(r)$ (for $r/l\gg 1$ and any fixed $r/L$)
\begin{equation}
G(r)\simeq \nu_0^{d^{\omega}_G} l^{-d_G} (r/l)^{-\Delta_G} R(r/L)
\label{frscaling}
\end{equation}
is related to the existence of IR stable fixed points of the RG
equations (see above). In Eq.\,(\ref{frscaling}) $d^{\omega}_G$ and
$d_G$ are corresponding canonical dimensions of the function $G$
(the canonical dimensions of the model are given in
Sec.\,\ref{sec:Renormalization-group-analysis}), $R(r/L)$ is the
so-called scaling function, which cannot be determined by the RG
equation (see, e.g., Ref.\,\cite{Vasiliev}), and $\Delta_G$ is the
critical dimension defined as
\begin{equation}
\Delta_G=d_G^k+\Delta_{\omega} d_G^{\omega} + \gamma_G^*.
\end{equation}
Here $\gamma_G^*$ is the fixed point value of the anomalous
dimension $\gamma_G\equiv \mu \partial_{\mu} \ln Z_G$, where $Z_G$
is the renormalization constant of the multiplicatively
renormalizable quantity $G$, i.e., $G=Z_G G^R$ \cite{Antonov00}, and
$\Delta_{\omega}=2-\gamma_{\nu}^*$ is the critical dimension of the
frequency with $\gamma_{\nu}^*=\gamma_{1}^*$ which is defined in
Eq.\,(\ref{gama11}) and $\gamma_{1}^*$ means that $\gamma_{1}$ is
taken at the corresponding fixed point. From above discussion of the
possible scaling regimes we have
\begin{equation}
\gamma_{\nu}^*\equiv \xi =\left\{ \begin{array}{cc}
2\varepsilon-\eta & \mathrm{for \,\, FPII}
\\ \varepsilon  & \,\, \mathrm{for \,\,FPIV} \\ \varepsilon=\eta  & \mathrm{for \,\,FPV}\end{array} \right\}\,. \label{xi}
\end{equation}
We are working only in one-loop approximation but the anomalous
dimension $\gamma_{\nu}^*$ is already exact for all fixed points at
one-loop level \cite{Antonov99,ChHnJuJuMaRe06ab}, i.e., it has no
loop corrections of higher order, therefore the critical dimensions
of frequency $\omega$ and of fields $\Phi\equiv\{{\bf v}, \theta,
\theta^{\prime}\}$  are also found exactly at one-loop level
approximation \cite{Antonov99}. In our notation they read
\begin{equation}
\Delta_{\omega}=2-\gamma_{\nu}^* = \left\{
\begin{array}{cc} 2-2\varepsilon+\eta & \mathrm{for \,\, FPII}
\\ 2-\varepsilon  & \,\, \mathrm{for \,\,FPIV} \\ 2-\varepsilon=2-\eta  &
\mathrm{for \,\,FPV}\end{array} \right\}\,.  \label{deltao}
\end{equation}
and
\begin{equation}
\Delta_{{\bf
v}}=1-\gamma^*_{\nu},\,\,\,\Delta_{\theta}=-1+\gamma^*_{\nu}/2,\,\,\,\Delta_{\theta^{\prime}}=d+1-\gamma^*_{\nu}/2.
\end{equation}

\input epsf
   \begin{figure}[t]
     \vspace{0.3cm}
       \begin{center}
       \leavevmode
       \epsfxsize=3.6cm
       \epsffile{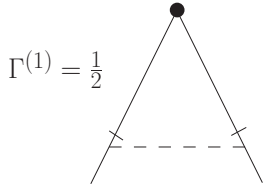}
   \end{center}
\vspace{0cm} \caption{Graphical representation of the one-loop
correction to $\Gamma_N$ in Eq.\,(\ref{Gamma2}).  \label{fig4}}
\end{figure}

The renormalized function $G^R$ must satisfy the RG equation of the
form
\begin{equation}
({\cal D}_{RG}+\gamma_G)G^R(r)=0,
\end{equation}
with operator ${\cal D}_{RG}$ given explicitly in
Eq.\,(\ref{RGoper}), namely,
\begin{equation}
{\cal D}_{RG}\equiv {\mathcal{{D}}}_{\mu}+\sum_{i=g,\chi,u}
\beta_{i} \partial_{i}- \gamma_{\nu} {\mathcal{{D}}}_{\nu}.
\end{equation}
The difference between the functions $G$ and $G^R$ is only in the
normalization, choice of parameters (bare or renormalized), and
related to this choice the form of the perturbation theory (in $g_0$
or in $g$). The existence of a nontrivial IR stable fixed point
means that in the IR asymptotic region $ r/l \gg 1$ and any fixed
$r/L$ the function $G(r)$ takes on the self-similar form given in
Eq.\,(\ref{frscaling}). As was already mentioned the scaling
function $R(r/L)$ is not determined by the RG equation itself. The
dependence of the scaling functions on the argument $r/L$ in the
region $r/L \ll 1$ can be studied using the well-known Wilson
operator product expansion (OPE)
\cite{ZinnJustin,Vasiliev,AdAnVa96,AdAnVa99}. It shows that, in the
limit $r/L\to 0$, the function $R(r/L)$ can be written in the
following asymptotic form:
\begin{equation}
R(r/L) = \sum_{i} C_{F_i}(r/L)\, (r/L)^{\Delta_{F_i}}, \label{ope}
\end{equation}
where $C_{F_i}$ are coefficients regular in $r/L$. In general, the
summation is implied over certain  renormalized composite operators
$F_i$  with critical dimensions $\Delta_{F_i}$. In the case under
consideration the leading contribution is given by operators $F_i$
having the form $F[N,p]=\partial_{i_1} \theta \cdots \partial_{i_p}
\theta (\partial_i \theta \partial_i \theta)^n$ with $N=p+2 n$. In
the next section we shall consider them in detail, where the
complete one-loop calculation of the critical dimensions  of the
composite operators $F_N$ will be presented for arbitrary values of
$N$, $d$, $u$, and $\alpha_{1,2}$.

\begin{figure}[t]
 \vspace{-0.8cm}
\includegraphics[width=70mm]{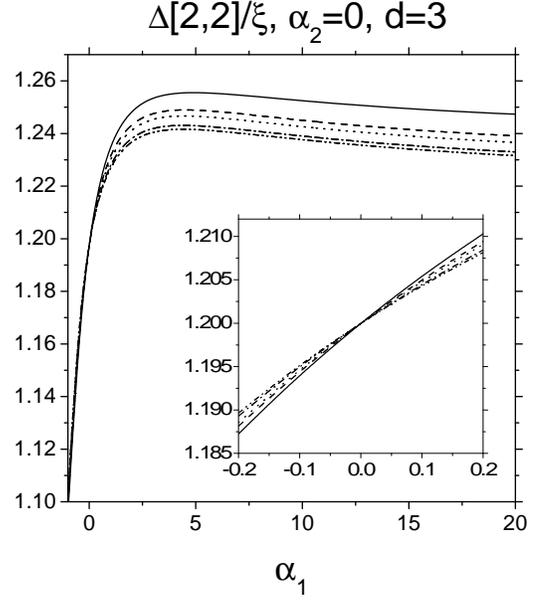}% Here is how to import EPS art
\vspace{-1.0cm} \caption{\label{fig5} Dependence of the critical
dimension $\Delta[2,2]/\xi$ on anisotropy parameter $\alpha_1$
($\alpha_2=0$) for different fixed point values of the parameter
$u$: $u^*=0$ (frozen limit) - solid line, $u^*=0.5$ - dash line,
$u^*=1$ - dot line, $u^*=5$ - dash dot line, $u^*=\infty$
(rapid-change model limit) - dash dot dot line. The small figure
shows details that are not visible in basic figure.}
\end{figure}

\begin{figure}[t]
 \vspace{-0.8cm}
\includegraphics[width=70mm]{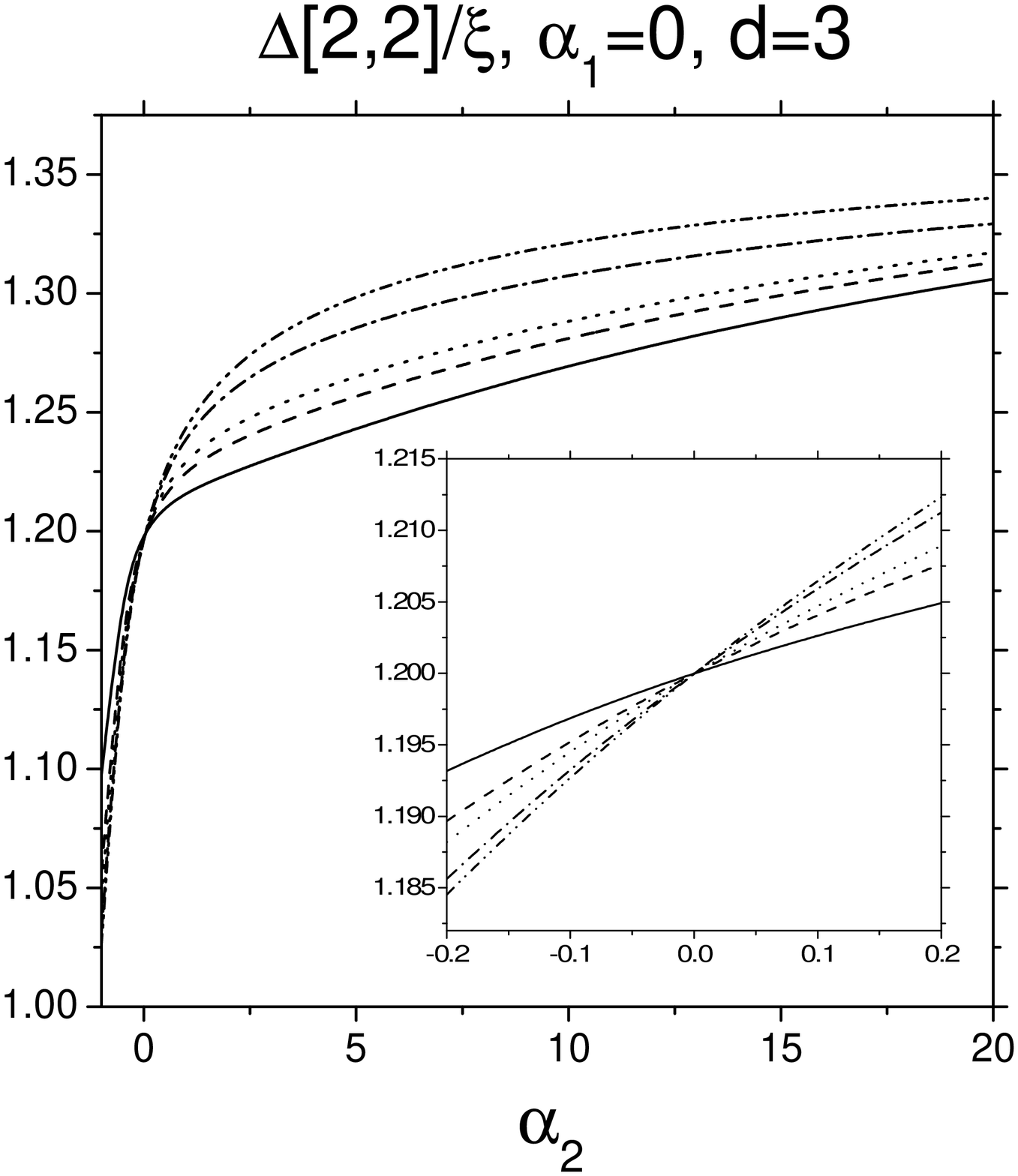}% Here is how to import EPS art
\vspace{-1.2cm} \caption{\label{fig6} Dependence of the critical
dimension $\Delta[2,2]/\xi$ on anisotropy parameter $\alpha_2$
($\alpha_1=0$) for different fixed point values of the parameter $u$
(for notation see the caption in Fig.\,\ref{fig5}).}
\end{figure}

\section{Critical Dimensions of Composite Operators and Anomalous Scaling \label{sec5}}

\subsection{Operator product expansion}

According to the OPE \cite{ZinnJustin,Vasiliev,AdAnVa96,AdAnVa99},
the equal-time product $F_1(x^{\prime})F_2(x^{\prime\prime})$ of two
renormalized composite operators \footnote{By definition we use the
term "composite operator" for any local monomial or polynomial
constructed from primary fields and their derivatives at a single
point $x\equiv(t, {\bf x})$. Constructions $\theta^n(x)$ and
$[\partial_i \theta(x)
\partial_i \theta(x)]^n$ are typical examples.} at ${\bf x}=({\bf x^{\prime}}+{\bf
x^{\prime\prime}})/2=const$ and ${\bf r}={\bf x^{\prime}}-{\bf
x^{\prime\prime}} \rightarrow 0$ can be written in the following
form:
\begin{equation}
F_1(x^{\prime})F_2(x^{\prime\prime})=\sum_i C_{F_i}({\bf r})
F_i({\bf x},t), \label{fff}
\end{equation}
where the summation is taken over all possible renormalized local
composite operators $F_i$ allowed by symmetry with definite critical
dimensions $\Delta_{F_i}$, and the functions $C_{F_i}$ are the
corresponding Wilson coefficients regular in $L^{-2}$. The
renormalized correlation function $\langle
F_1(x^{\prime})F_2(x^{\prime\prime}) \rangle$ can now be found by
averaging Eq.\,(\ref{fff}) with the weight $\exp S^R$ with $S^R$
from Eq.\,(\ref{eq:Srenorm}). The quantities $\langle F_i \rangle$
appear on the right-hand side and their asymptotic behavior in the
limit $L^{-1} \rightarrow 0$ is then found from the corresponding RG
equations and has the form $\langle F_i \rangle \varpropto
L^{-\Delta_{F_i}}$.

\begin{figure}[t]
 \vspace{-0.8cm}
\includegraphics[width=70mm]{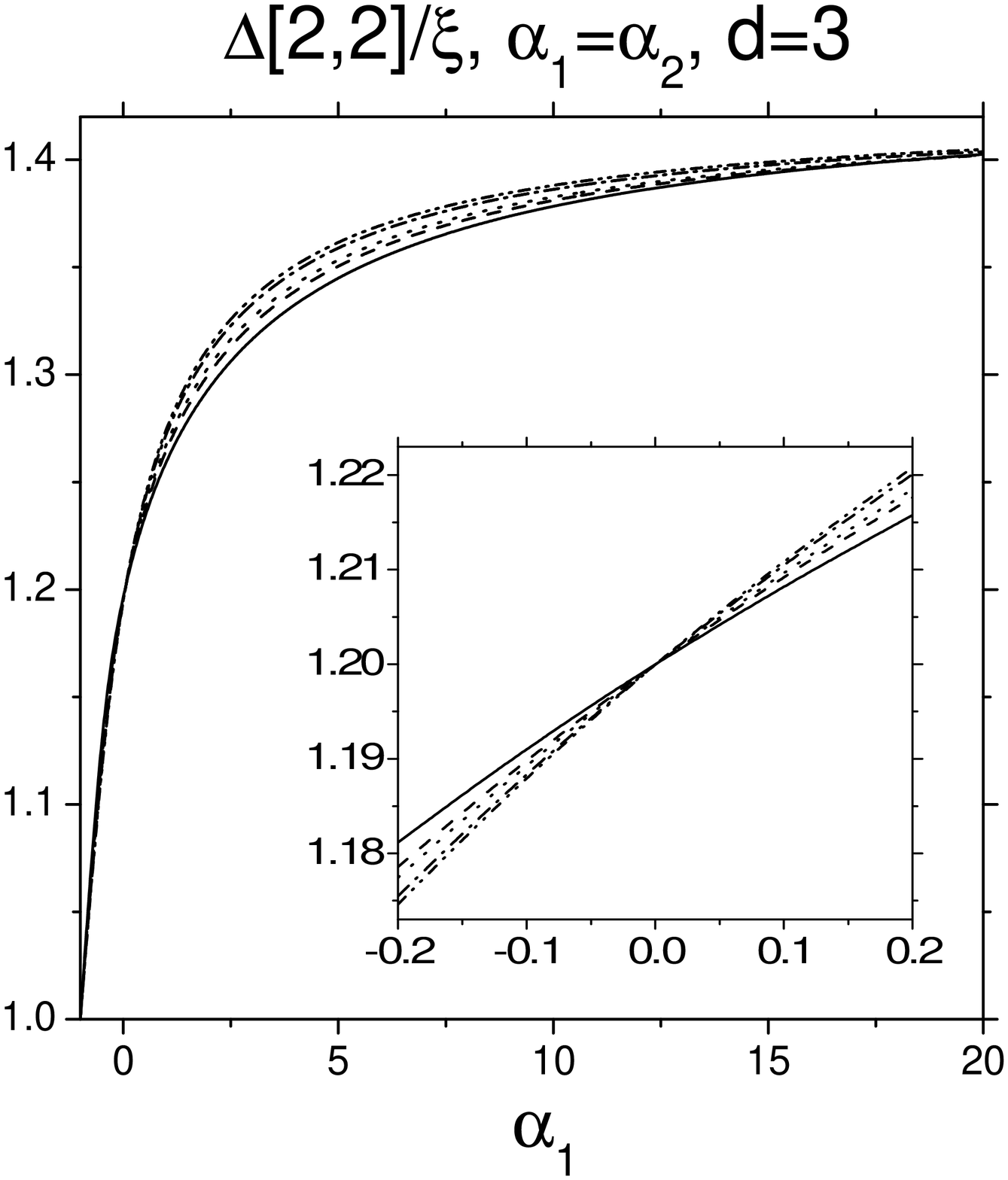}% Here is how to import EPS art
\vspace{-1.2cm} \caption{\label{fig7} Dependence of the critical
dimension $\Delta[2,2]/\xi$ on anisotropy parameter
$\alpha_1=\alpha_2$ for different fixed point values of the
parameter $u$ (for notation see the caption in Fig.\,\ref{fig5}).}
\end{figure}

\begin{figure}[t]
 \vspace{-0.8cm}
\includegraphics[width=70mm]{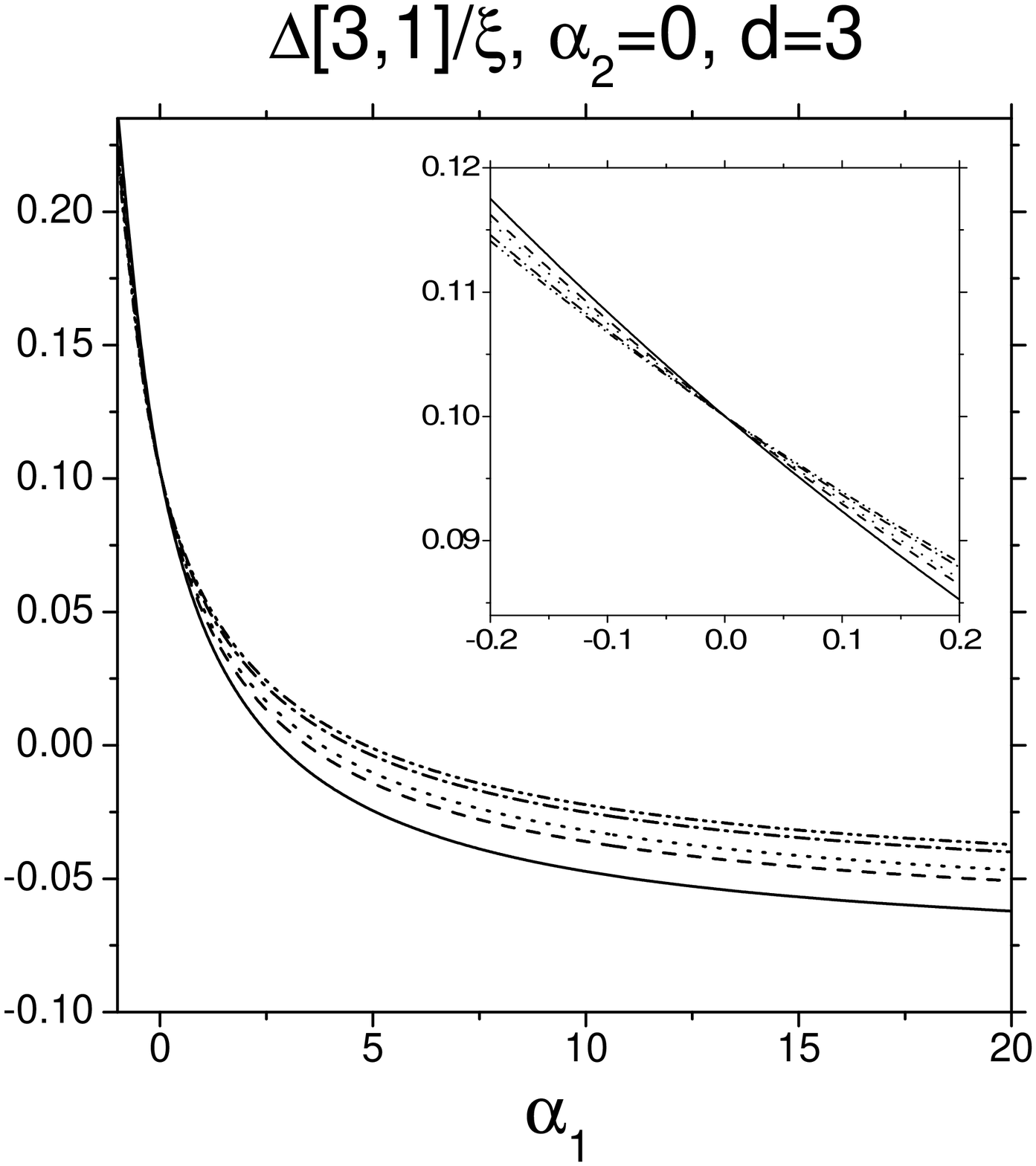}% Here is how to import EPS art
\vspace{-1.2cm} \caption{\label{fig8} Dependence of the critical
dimension $\Delta[3,1]/\xi$ on anisotropy parameter $\alpha_1$
($\alpha_2=0$) for different fixed point values of the parameter $u$
(for notation see the caption in Fig.\,\ref{fig5}).}
\end{figure}

\begin{figure}[t]
 \vspace{-0.8cm}
\includegraphics[width=70mm]{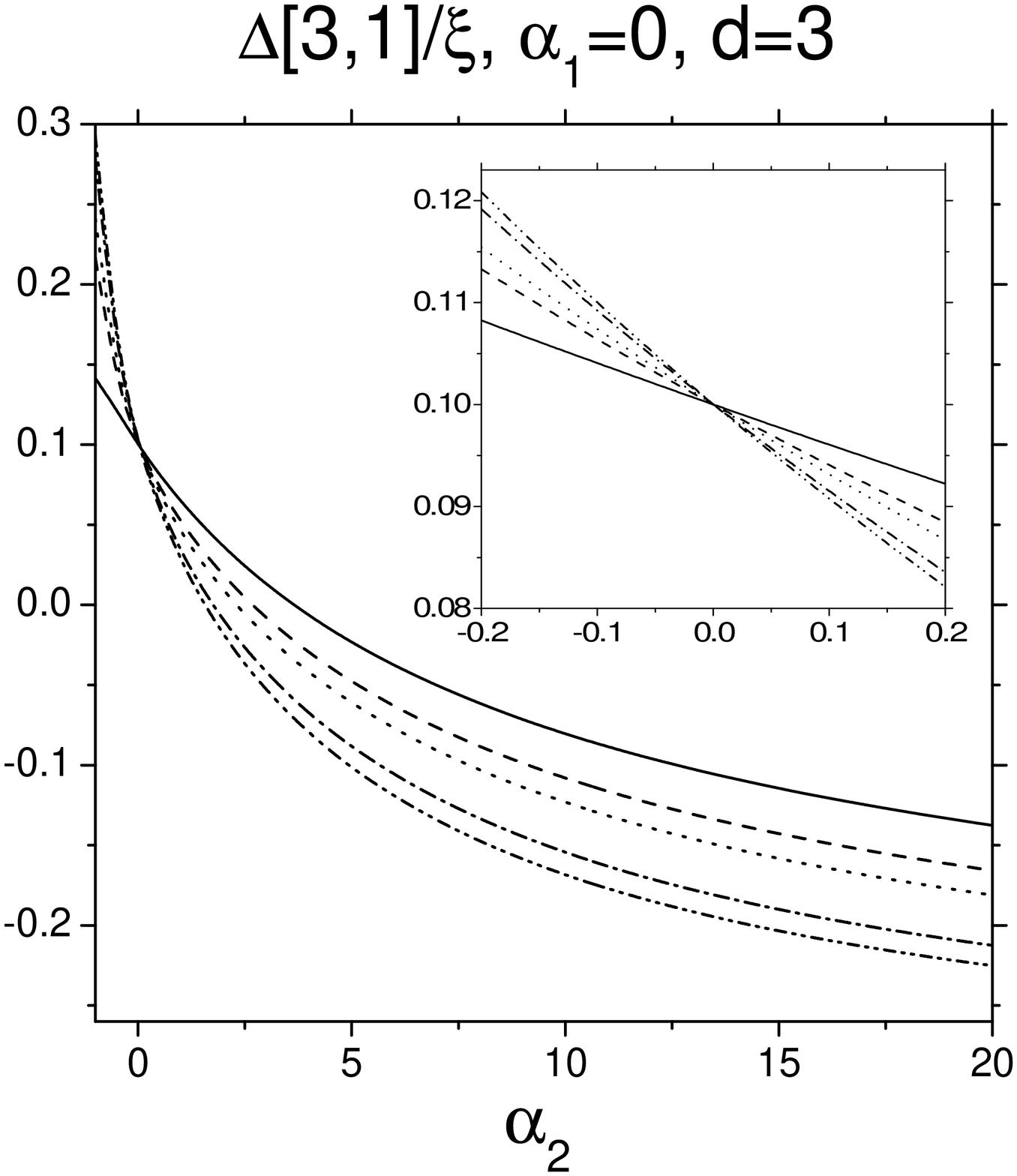}% Here is how to import EPS art
\vspace{-1.2cm} \caption{\label{fig9} Dependence of the critical
dimension $\Delta[3,1]/\xi$ on anisotropy parameter $\alpha_2$
($\alpha_1=0$) for different fixed point values of the parameter $u$
(for notation see the caption in Fig.\,\ref{fig5}).}
\end{figure}

\begin{figure}[t]
 \vspace{-0.8cm}
\includegraphics[width=70mm]{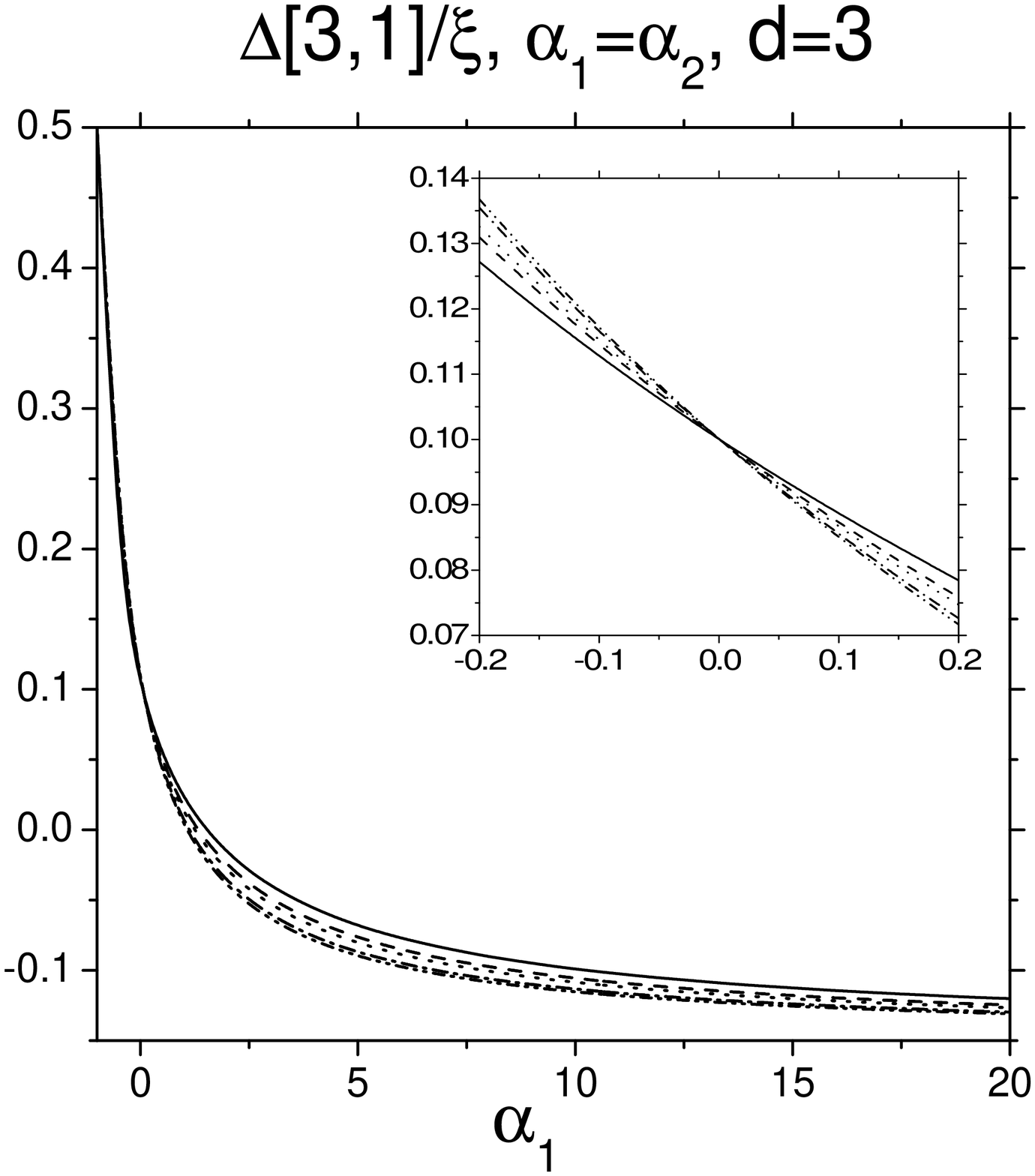}% Here is how to import EPS art
\vspace{-1.2cm} \caption{\label{fig10} Dependence of the critical
dimension $\Delta[3,1]/\xi$ on anisotropy parameter
$\alpha_1=\alpha_2$ for different fixed point values of the
parameter $u$ (for notation see the caption in Fig.\,\ref{fig5}).}
\end{figure}

From the OPE (\ref{fff}) one can find that the scaling function
$R(r/L)$ in the representation (\ref{frscaling}) for the correlation
function $F_1(x^{\prime})F_2(x^{\prime\prime})$ has the form given
in Eq.\,(\ref{ope}), where the coefficients $C_{F_i}$ are regular in
$(r/L)^2$.

\begin{figure}[t]
 \vspace{-0.8cm}
\includegraphics[width=70mm]{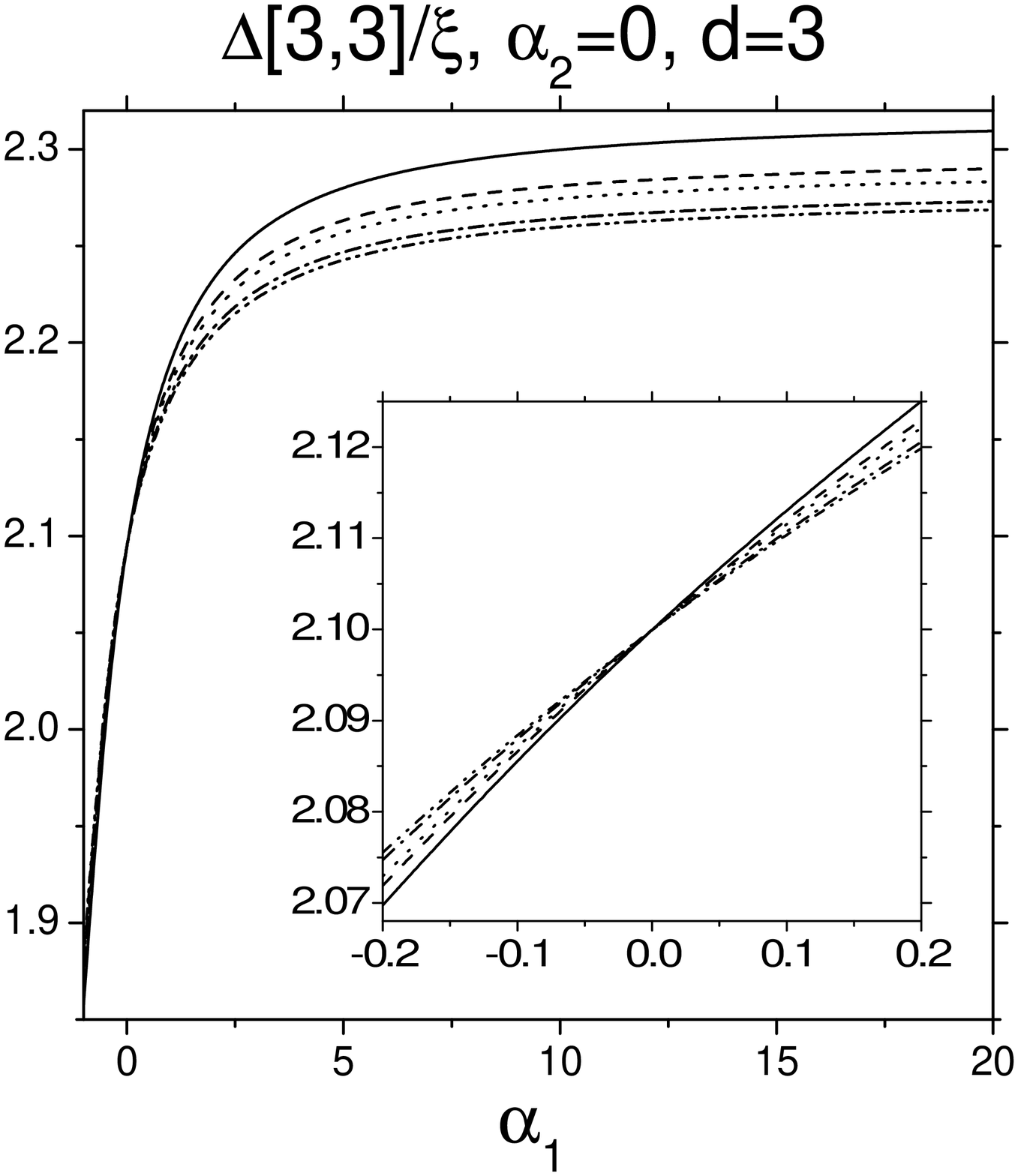}% Here is how to import EPS art
\vspace{-1.2cm} \caption{\label{fig11} Dependence of the critical
dimension $\Delta[3,3]/\xi$ on anisotropy parameter $\alpha_1$
($\alpha_2=0$) for different fixed point values of the parameter $u$
(for notation see the caption in Fig.\,\ref{fig5}).}
\end{figure}

\begin{figure}[t]
 \vspace{-0.8cm}
\includegraphics[width=70mm]{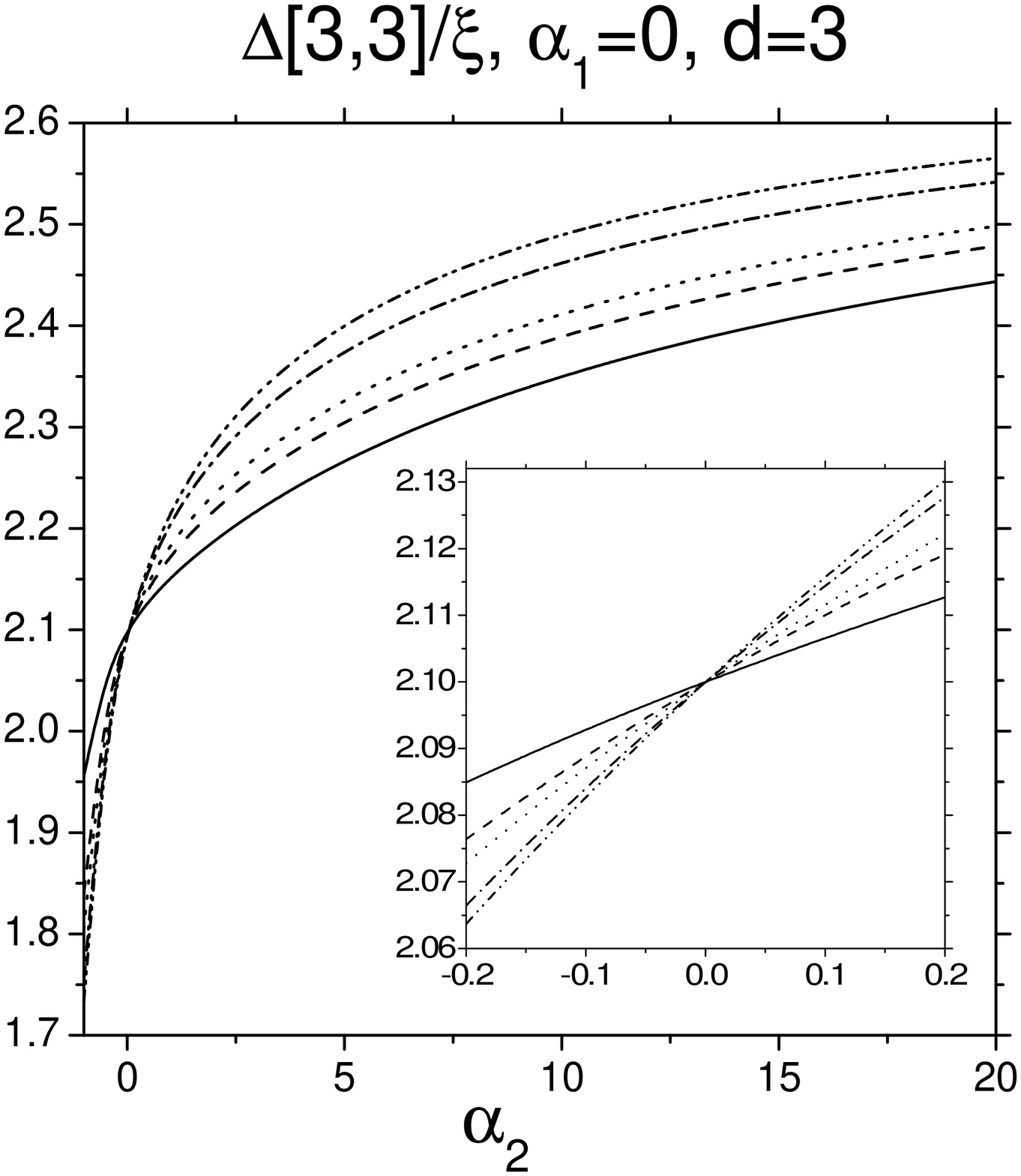}% Here is how to import EPS art
\vspace{-1.2cm} \caption{\label{fig12} Dependence of the critical
dimension $\Delta[3,3]/\xi$ on anisotropy parameter $\alpha_2$
($\alpha_1=0$) for different fixed point values of the parameter $u$
(for notation see the caption in Fig.\,\ref{fig5}).}
\end{figure}

\begin{figure}[t]
 \vspace{-0.8cm}
\includegraphics[width=70mm]{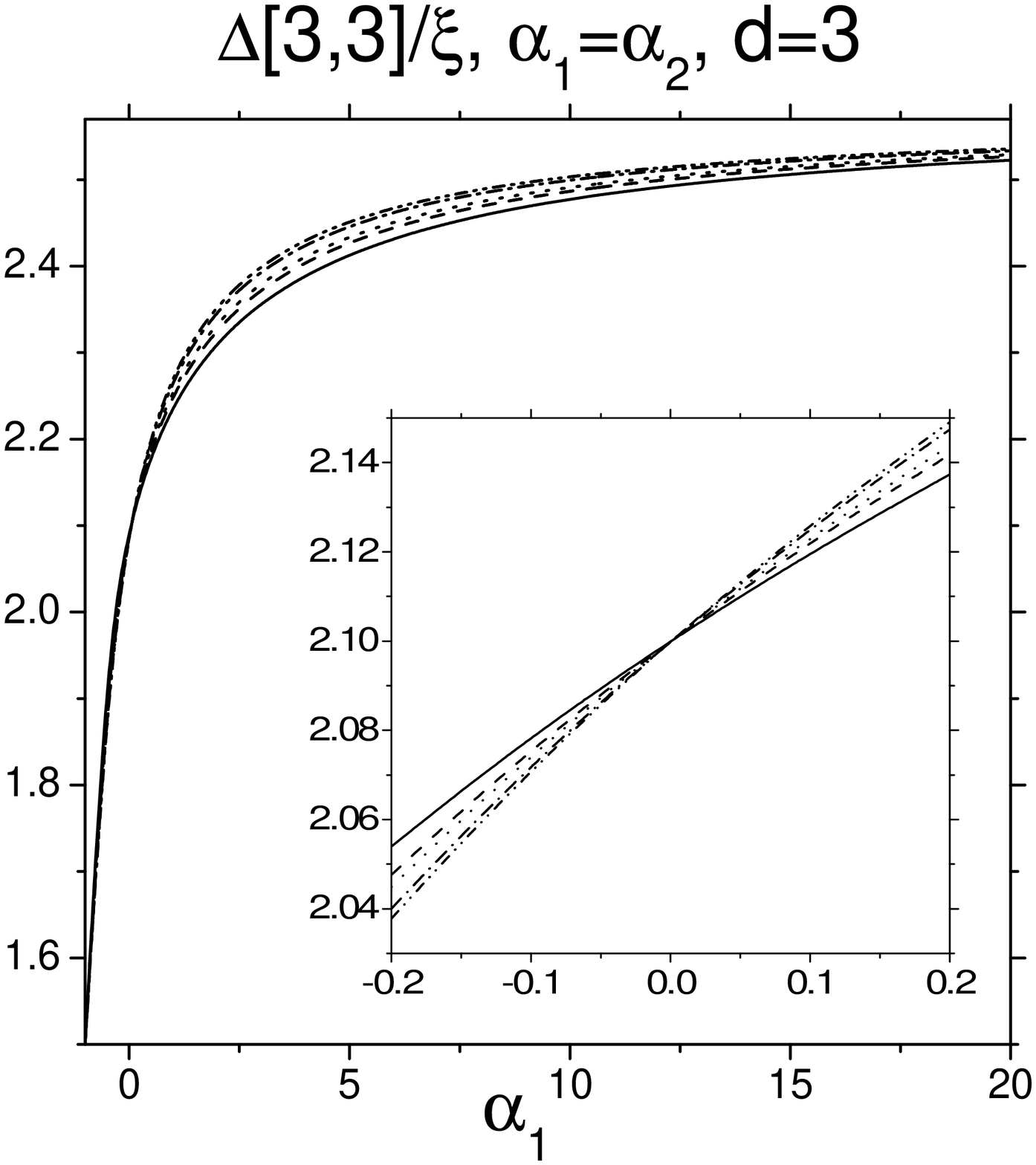}% Here is how to import EPS art
\vspace{-1.2cm} \caption{\label{fig13} Dependence of the critical
dimension $\Delta[3,3]/\xi$ on anisotropy parameter
$\alpha_1=\alpha_2$ for different fixed point values of the
parameter $u$ (for notation see the caption in Fig.\,\ref{fig5}).}
\end{figure}

It is well known
%(see, e.g., Refs.\,\cite{Vasiliev,AdAnVa96,AdAnVa99})
that the specific feature of the turbulence models is the existence
of operators with negative critical dimensions (the so-called
"dangerous" operators)
\cite{AdAnVa98,AdAn98,AdAnVa96,AdAnVa99,Vasiliev}. Their presence in
the OPE determines the IR behavior of the scaling functions and
leads to their singular dependence on $L$ when $r/L \rightarrow0$.
At this point the turbulence models are crucially different from the
models of critical phenomena, where the leading contribution to the
representation (\ref{frscaling}) is given by the simplest operator
$F=1$ with the dimension $\Delta_F=0$, and the other operators
determine only the corrections that vanish for $r/L \rightarrow0$.
If the spectrum of the dimensions $\Delta_{F_i}$ for a given scaling
function is bounded from below, the leading term of its behavior for
$r/L \rightarrow 0$ is given by the minimal dimension. As was
discussed in Ref.\,\cite{Antonov99}, the model under consideration
belongs to this case for small enough values of the exponents
$\varepsilon, \eta$.

In what follows, we shall concentrate on the equal-time structure
functions of the scalar field as defined in Eq.\,(\ref{struc}). The
representation (\ref{frscaling}) is valid with the dimensions
$d_G^{\omega}=-N/2$, $d_G=-N$, and
$\Delta_G=N\Delta_{\theta}=N(-1+\gamma_{\nu}^*/2)$. In general, not
only do the operators which are present in the corresponding Taylor
expansion are entering into the OPE but also all possible operators
that admix to them in renormalization. In the present anisotropic
model the leading contribution of the Taylor expansion for the
structure functions (\ref{struc}) is given by the tensor composite
operators constructed solely of the scalar gradients
\begin{equation}
F[N,p]\equiv \partial_{i_1}\theta \cdots \partial_{i_p}\theta
(\partial_{i}\theta
\partial_{i} \theta)^n, \label{composite}
\end{equation}
where $N=p+2n$ is the total number of the fields $\theta$ entering
into the operator and $p$ is the number of the free vector indices
(see, e.g., Ref.\,\cite{AdAnHnNo00}).

\begin{figure}[t]
 \vspace{-0.8cm}
\includegraphics[width=70mm]{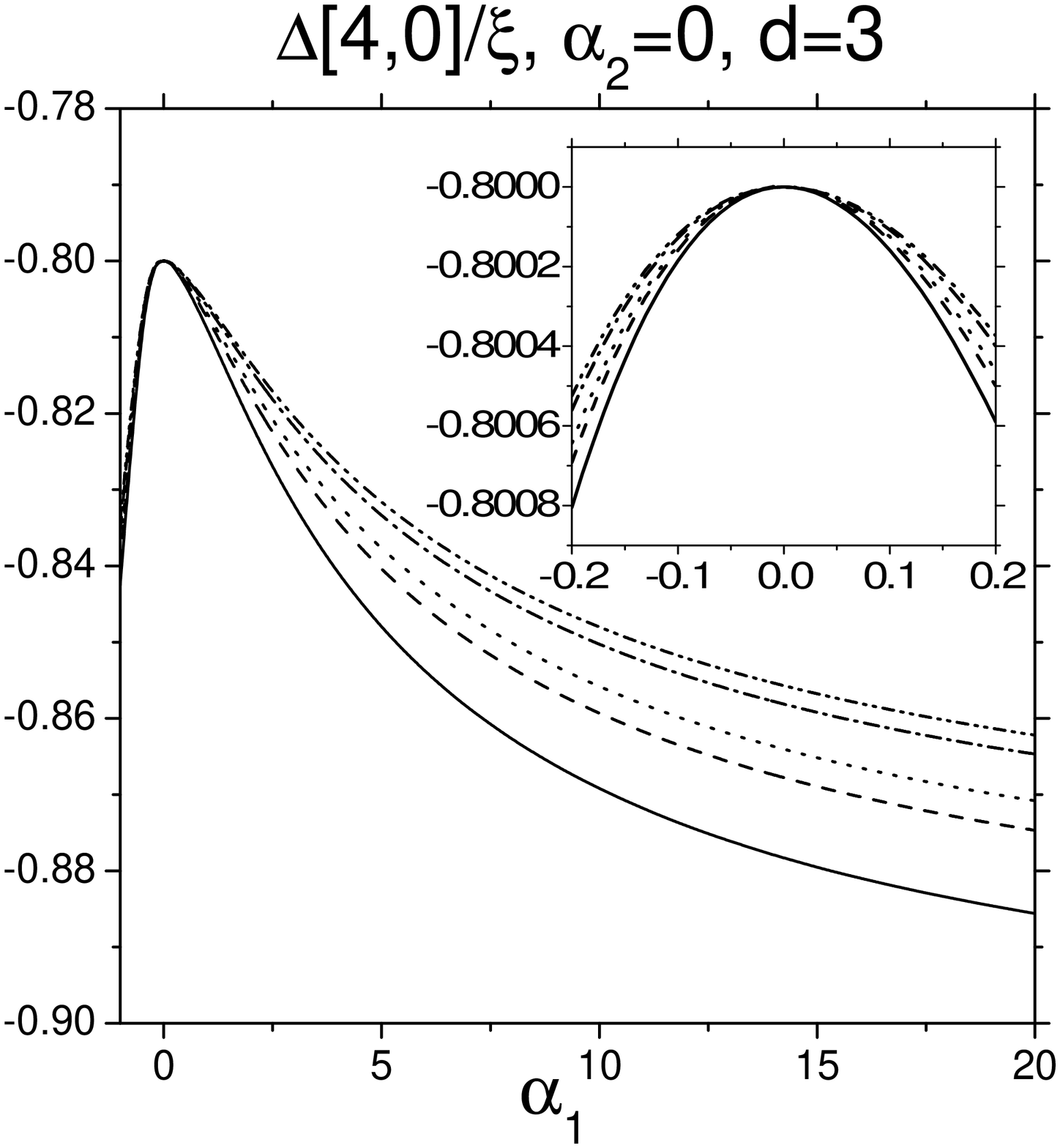}% Here is how to import EPS art
\vspace{-1.2cm} \caption{\label{fig14} Dependence of the critical
dimension $\Delta[4,0]/\xi$ on anisotropy parameter $\alpha_1$
($\alpha_2=0$) for different fixed point values of the parameter $u$
(for notation see the caption in Fig.\,\ref{fig5}).}
\end{figure}

\begin{figure}[t]
 \vspace{-0.8cm}
\includegraphics[width=70mm]{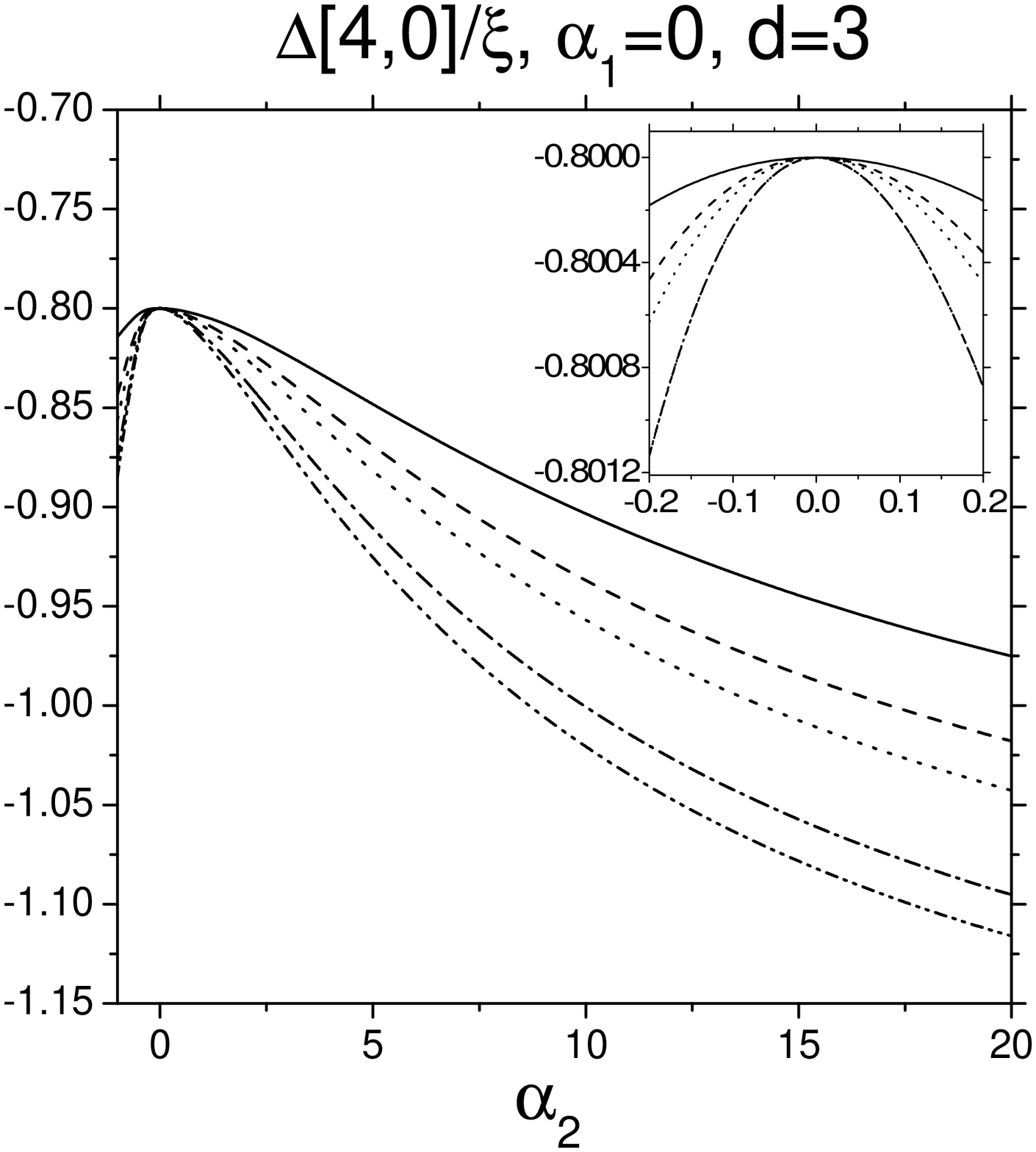}% Here is how to import EPS art
\vspace{-1.2cm} \caption{\label{fig15} Dependence of the critical
dimension $\Delta[4,0]/\xi$ on anisotropy parameter $\alpha_2$
($\alpha_2=0$) for different fixed point values of the parameter $u$
(for notation see the caption in Fig.\,\ref{fig5}).}
\end{figure}

\begin{figure}[t]
 \vspace{-0.8cm}
\includegraphics[width=70mm]{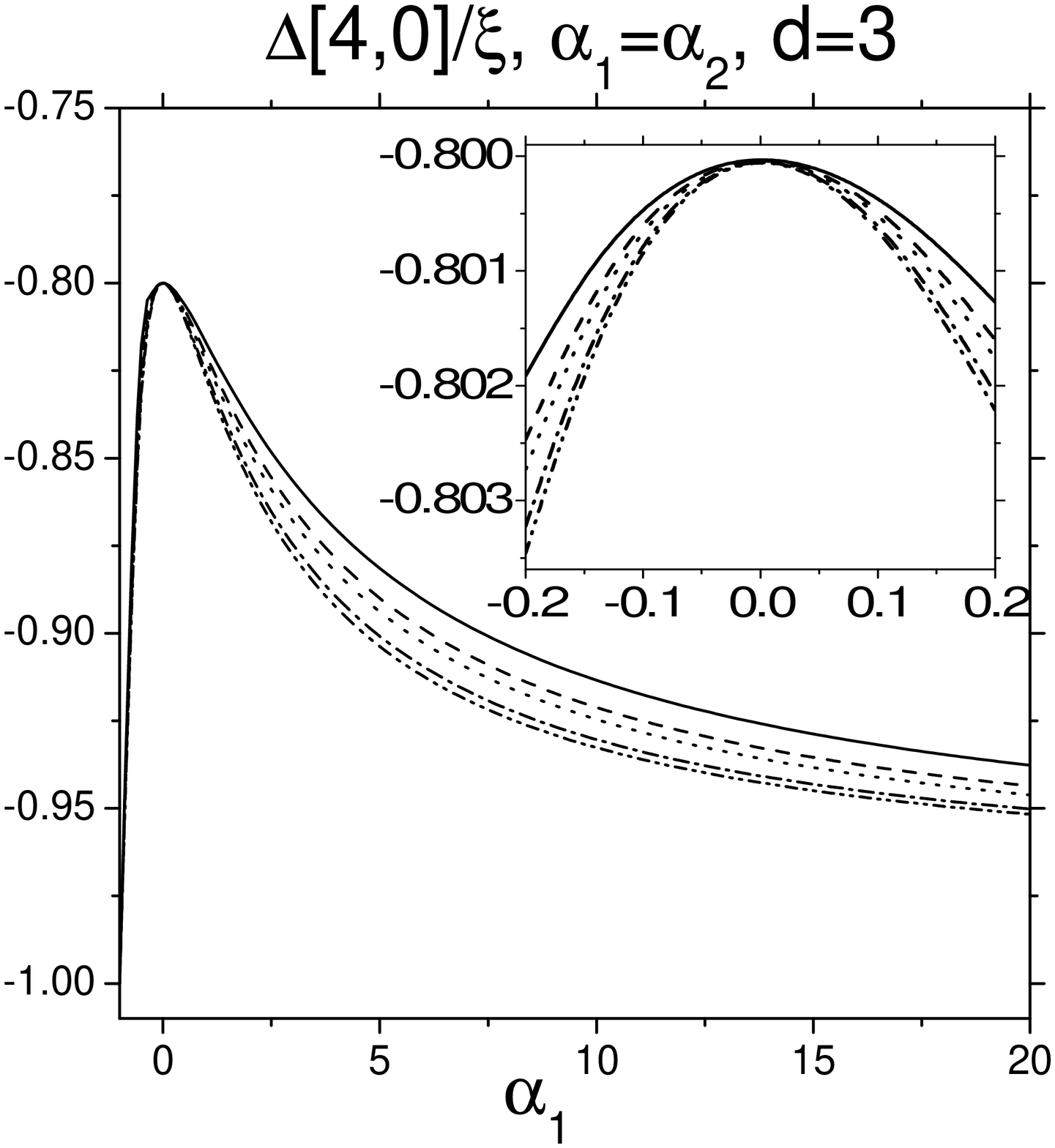}% Here is how to import EPS art
\vspace{-1.2cm} \caption{\label{fig16} Dependence of the critical
dimension $\Delta[4,0]/\xi$ on anisotropy parameter
$\alpha_1=\alpha_2$ for different fixed point values of the
parameter $u$ (for notation see the caption in Fig.\,\ref{fig5}).}
\end{figure}

\subsection{ Composite operators $F[N,p]$:
renormalization and critical dimensions}

Let us briefly discuss renormalization of the composite operators
(\ref{composite}). A complete and detailed discussion of the
renormalization of the composite operators is given in
Ref.\,\cite{AdAnBaKaVa01}. Therefore, we shall discuss  only basic
moments necessary to present explicit expressions for composite
operators.

The necessity of additional renormalization of the composite
operators (\ref{composite}) is related to the fact that the
coincidence of the field arguments in Green functions containing
them leads to additional UV divergences. These divergences must be
removed by special kind of renormalization procedure which can be
found, e.g., in Refs.\,\cite{ZinnJustin,Vasiliev,Collins}, where
their renormalization is studied in general. As for the
renormalization of composite operators in the models of turbulence
it is discussed in Refs.\,\cite{AdAnVa96,AdAnVa99}. Besides,
typically, the composite operators are mixed under renormalization.
Therefore, let us briefly discuss this issue \cite{Vasiliev}.

\begin{figure}[t]
 \vspace{-0.8cm}
\includegraphics[width=70mm]{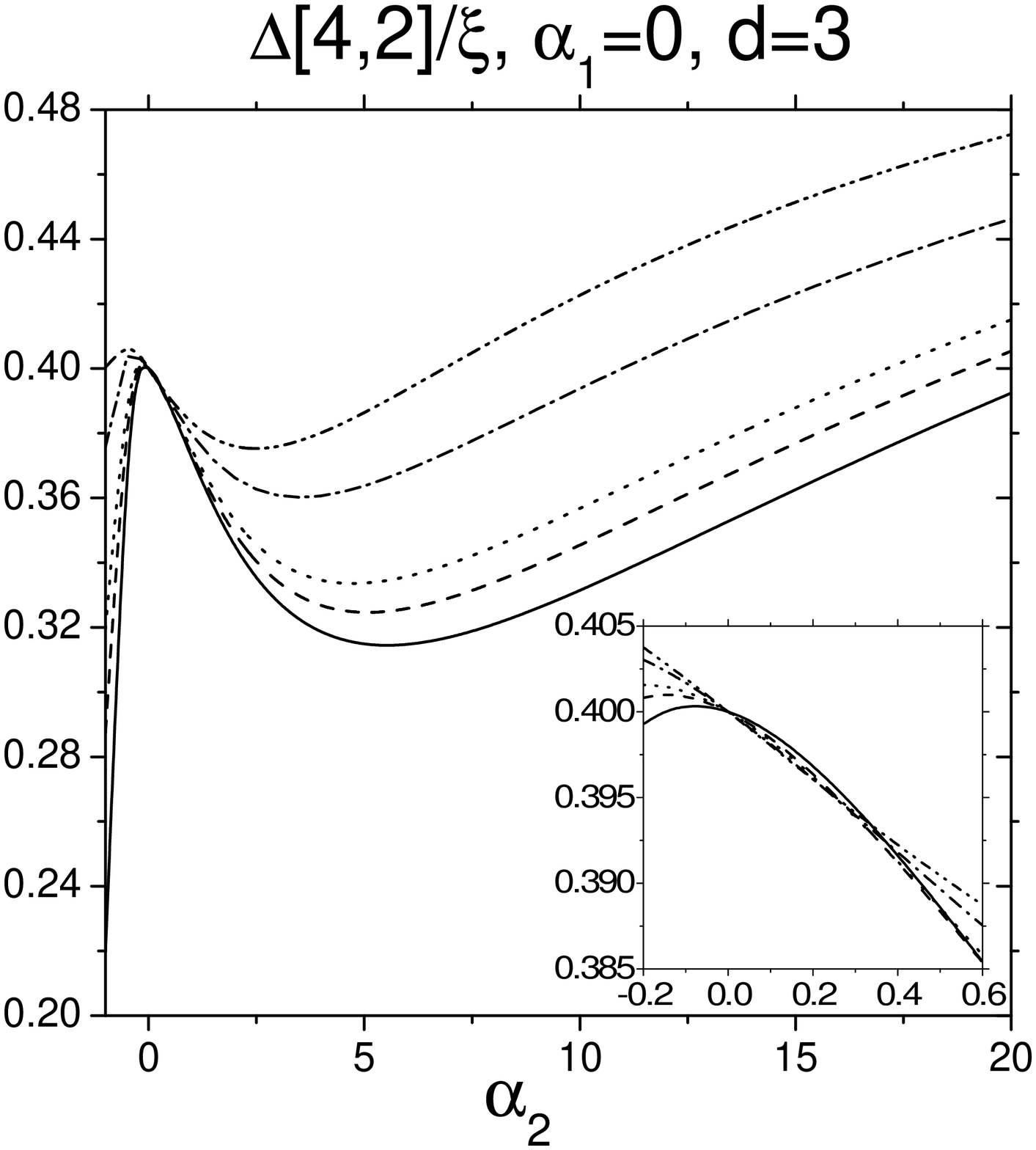}% Here is how to import EPS art
\vspace{-1.2cm} \caption{\label{fig17} Dependence of the critical
dimension $\Delta[4,2]/\xi$ on anisotropy parameter $\alpha_1$
($\alpha_2=0$) for different fixed point values of the parameter $u$
(for notation see the caption in Fig.\,\ref{fig5}).}
\end{figure}

\begin{figure}[t]
 \vspace{-0.8cm}
\includegraphics[width=70mm]{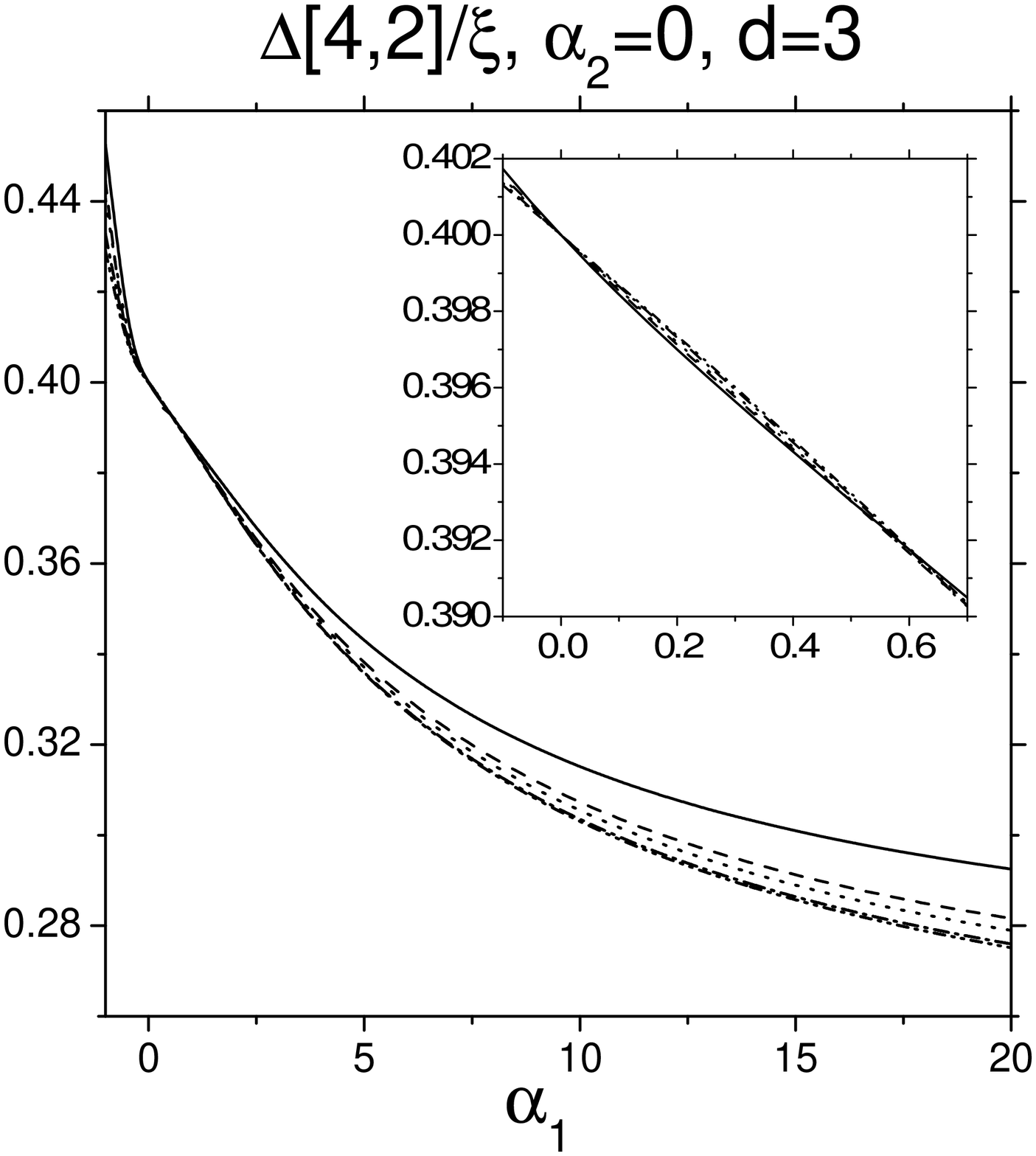}% Here is how to import EPS art
\vspace{-1.2cm} \caption{\label{fig18} Dependence of the critical
dimension $\Delta[4,2]/\xi$ on anisotropy parameter $\alpha_2$
($\alpha_2=0$) for different fixed point values of the parameter $u$
(for notation see the caption in Fig.\,\ref{fig5}).}
\end{figure}

\begin{figure}[t]
 \vspace{-0.8cm}
\includegraphics[width=70mm]{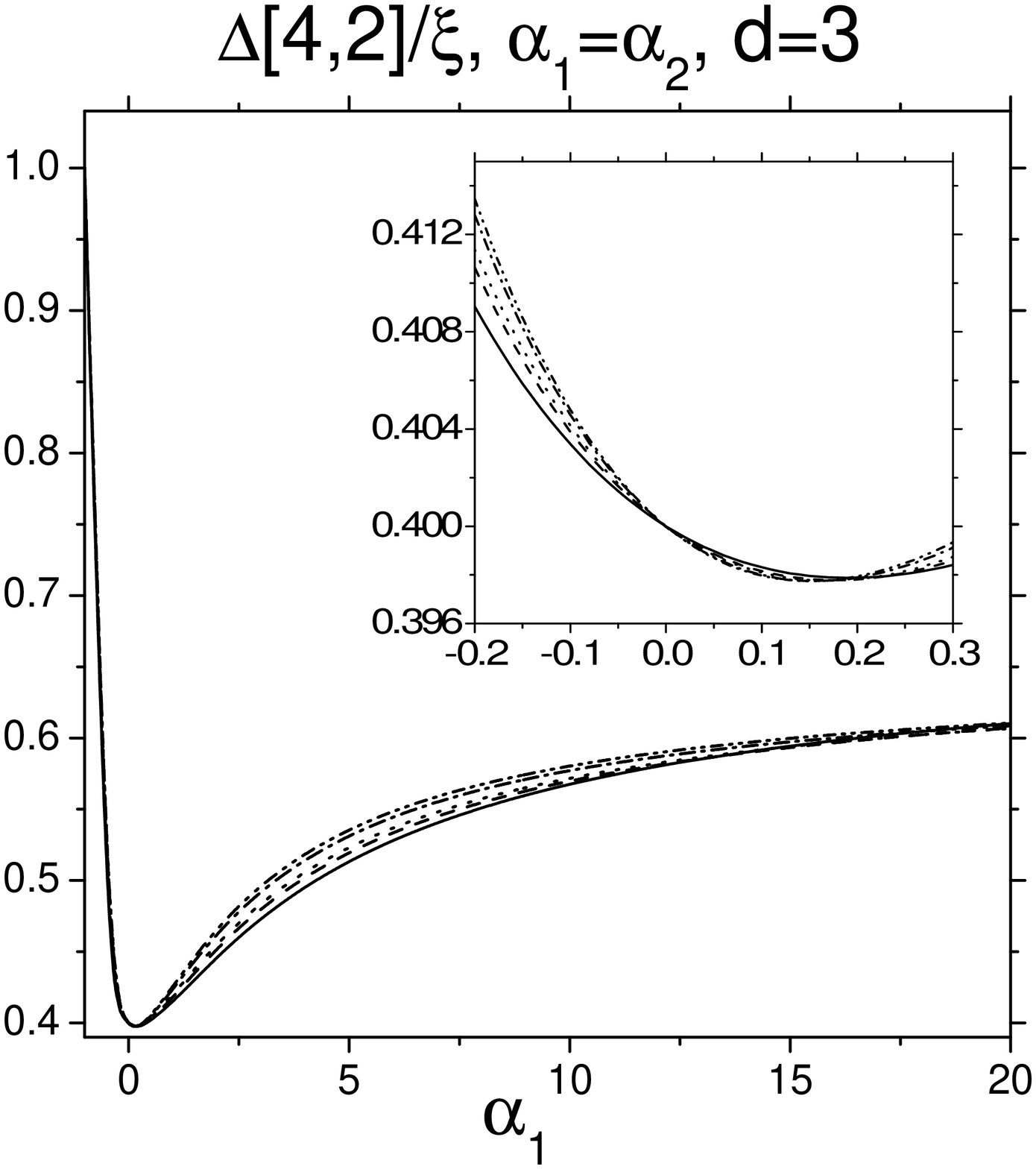}% Here is how to import EPS art
\vspace{-1.2cm} \caption{\label{fig19} Dependence of the critical
dimension $\Delta[4,2]/\xi$ on anisotropy parameter
$\alpha_1=\alpha_2$ for different fixed point values of the
parameter $u$ (for notation see the caption in Fig.\,\ref{fig5}).}
\end{figure}

Let $F\equiv\{F_{\alpha}\}$ be a closed set of composite operators
which are mixed only with each other in renormalization. Then the
renormalization matrix $Z_F\equiv\{Z_{\alpha \beta}\}$ and the
matrix of corresponding anomalous dimensions $\gamma_F \equiv
\{\gamma_{\alpha\beta}\}$ for this set are given as follows
\begin{equation}
F_{\alpha}=\sum_{\beta}Z_{\alpha\beta}F_{\beta}^{R},
\qquad\gamma_{F}=Z_{F}^{-1} \tilde{D}_{\mu}Z_{F}.\label{2.2}
\end{equation}
Renormalized composite operators are subject to the following RG
differential equations
\begin{equation}
({\mathcal{{D}}}_{\mu}+\sum_{i=g,\chi,u} \beta_{i}
\partial_{i}-\gamma_{\nu}{\mathcal{{D}}}_{\nu})F_{\alpha}^{R}
=-\sum_{\beta}\gamma_{\alpha\beta}F_{\beta}^{R},
\end{equation}
which lead to the following matrix of critical dimensions
$\Delta_{F}\equiv\{\Delta_{\alpha\beta}\}$
\begin{equation}
\Delta_{F}=d_{F}^{k}+\Delta_{\omega}d_{F}^{\omega}+\gamma_{F}^{*},
\qquad\Delta_{\omega}=2-\gamma_{\nu}^*,\label{32B}
\end{equation}
where $d_{F}^{k}$ a $d_{F}^{\omega}$ are diagonal matrices of
corresponding canonical dimensions and $\gamma_{F}^{*}$ is the
matrix of anomalous dimensions (\ref{2.2}) taken at the fixed point.
In the end, the critical dimensions of the set of operators
$F\equiv\{ F_{\alpha}\}$ are given by the eigenvalues of the matrix
$\Delta_F$. The so-called "basis" operators that possess definite
critical dimensions have the form
\begin{equation}
F_{\alpha}^{bas}=\sum_{\beta}U_{\alpha\beta}F_{\beta}^{R}\,\,,\label{2.5}
\end{equation}
where the matrix $U_{F}=\{ U_{\alpha\beta}\}$ is such that
$\Delta'_{F}=U_{F}\Delta_{F}U_{F}^{-1}$ is diagonal.

\begin{figure}[t]
 \vspace{-0.8cm}
\includegraphics[width=70mm]{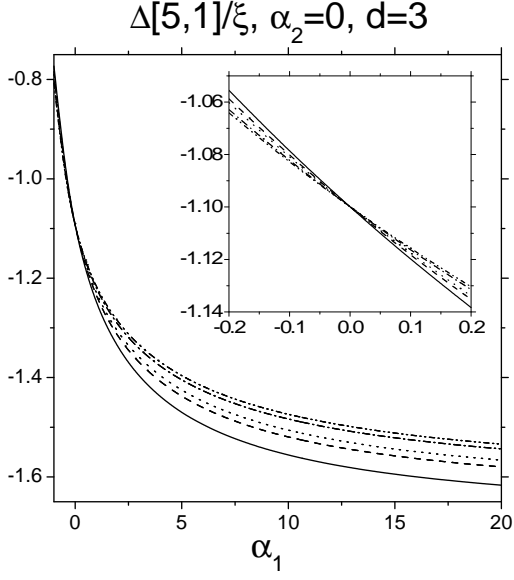}% Here is how to import EPS art
\vspace{-1.2cm} \caption{\label{fig20} Dependence of the critical
dimension $\Delta[5,1]/\xi$ on anisotropy parameter $\alpha_1$
($\alpha_2=0$) for different fixed point values of the parameter $u$
(for notation see the caption in Fig.\,\ref{fig5}).}
\end{figure}

\begin{figure}[t]
 \vspace{-0.8cm}
\includegraphics[width=70mm]{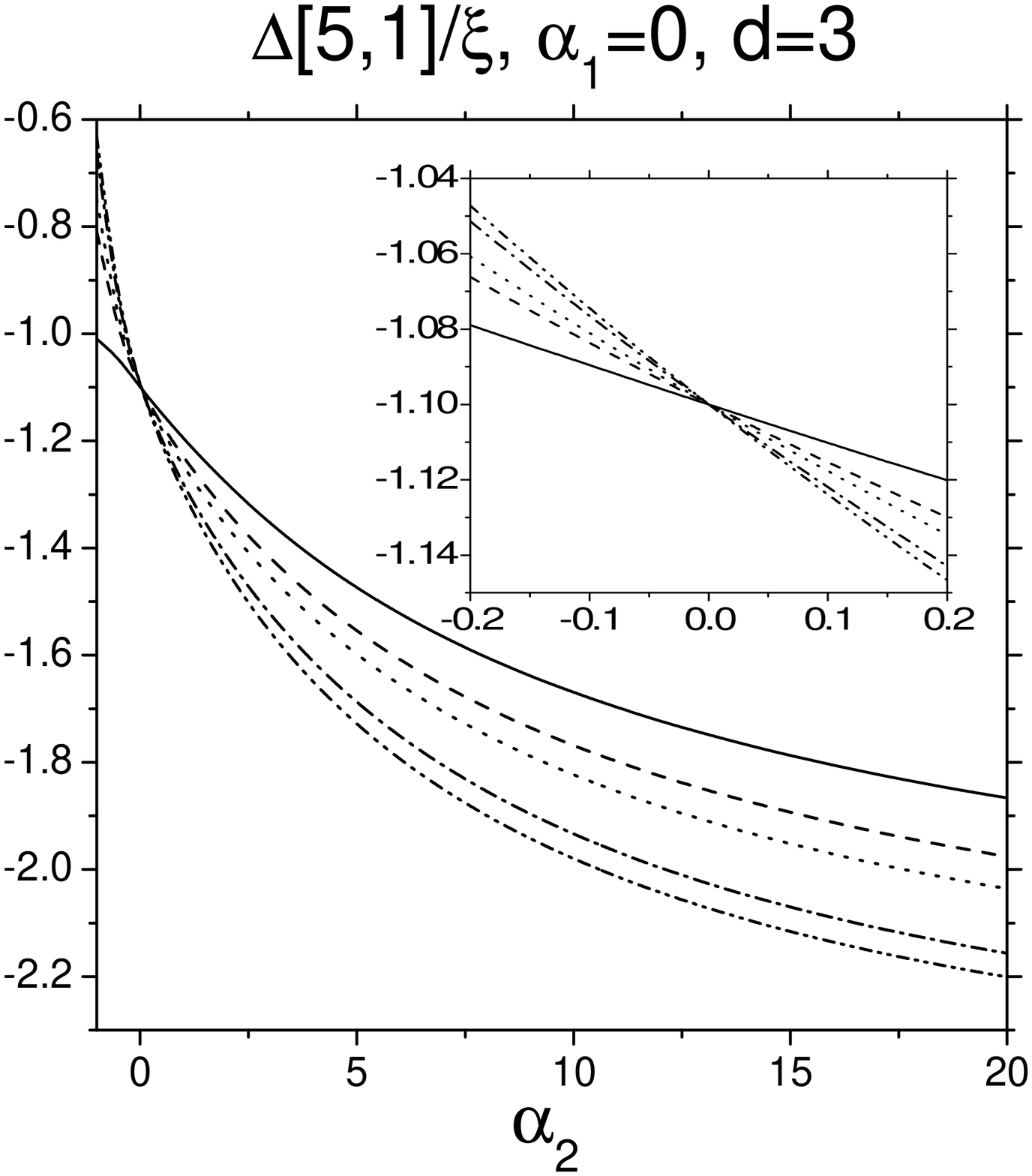}% Here is how to import EPS art
\vspace{-1.2cm} \caption{\label{fig21} Dependence of the critical
dimension $\Delta[5,1]/\xi$ on anisotropy parameter $\alpha_2$
($\alpha_1=0$) for different fixed point values of the parameter $u$
(for notation see the caption in Fig.\,\ref{fig5}).}
\end{figure}

\begin{figure}[t]
 \vspace{-0.8cm}
\includegraphics[width=70mm]{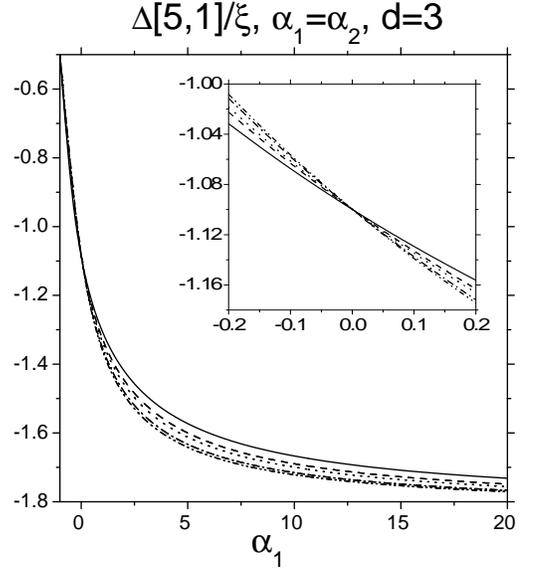}% Here is how to import EPS art
\vspace{-1.2cm} \caption{\label{fig22} Dependence of the critical
dimension $\Delta[5,1]/\xi$ on anisotropy parameter
$\alpha_1=\alpha_2$ for different fixed point values of the
parameter $u$ (for notation see the caption in Fig.\,\ref{fig5}).}
\end{figure}

As was already mentioned, in what follows, the central role is
played by the tensor composite operators
$\partial_{i_{1}}\theta\cdots\partial_{i_{p}}\theta\,(\partial_{i}\theta\partial_{i}\theta)^{n}$,
constructed solely of the scalar gradients. It is convenient to deal
with the scalar operators obtained by contracting the tensors with
the appropriate number of the uniaxial anisotropy vectors
$\mathbf{n}$ \cite{AdAnHnNo00},
\begin{equation}
F[N,p]\equiv[(\mathbf{n \cdot
\partial})\theta]^{p}(\partial_{i}\theta\partial_{i}\theta)^{n},\quad
N\equiv2n+p.\label{Fnp}
\end{equation}
Detail analysis shows that the composite operators (\ref{Fnp}) with
different $N$ are not mixed in renormalization, and therefore the
corresponding renormalization matrix
$Z_{[N,p][N^{\prime},p^{\prime}]}$ is in fact block-diagonal, i.e.,
$Z_{[N,p][N^{\prime},p^{\prime}]}=0$ for $N^{\prime}\neq N$
\cite{AdAnHnNo00}.

In the isotropic case, as well as in the case when large-scale
anisotropy is present, the elements $Z_{[N,p]\,[N,p']}$ vanish for
$p<p'$, thus the block $Z_{[N,p]\,[N,p']}$ is in fact triangular
along with the corresponding blocks of the matrices $U_{F}$ and
$\Delta_{F}$ from Eqs.\,(\ref{2.5}) and (\ref{32B}). In the
isotropic case it can be diagonalized by changing to irreducible
operators (scalars, vectors, and traceless tensors), but even for
nonzero imposed gradient its eigenvalues are the same as in the
isotropic case. Therefore, the inclusion of large-scale anisotropy
does not affect critical dimensions of the operators (\ref{Fnp}). On
the other hand, in the case of small-scale anisotropy, the operators
with different values of $p$ mix heavily in renormalization, and the
matrix $Z_{[N,p]\,[N,p']}$ is neither diagonal nor triangular here
and one can write
\begin{equation}
F[N,p]=\sum_{l=0}^{\lfloor N/2 \rfloor} Z_{[N,p]\,[N,N-2l]}
F^R[N,N-2l]\,, \label{Fnl}
\end{equation}
where $\lfloor N/2 \rfloor$ means the integer part of the $N/2$.
Therefore, each block of renormalization constants with given $N$ is
an $(\lfloor N/2 \rfloor+1)\times(\lfloor N/2 \rfloor+1)$ matrix. Of
course, the matrix of critical dimensions (\ref{32B}), whose
eigenvalues at IR stable fixed point are the critical dimensions
$\Delta[N,p]$ of the set of operators $F[N,p]$, has also dimension
$(\lfloor N/2 \rfloor+1)\times(\lfloor N/2 \rfloor+1)$.

\begin{figure}[t]
 \vspace{-0.8cm}
\includegraphics[width=70mm]{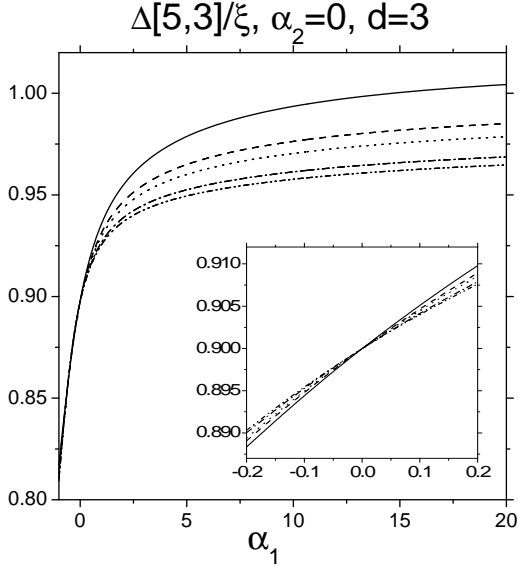}% Here is how to import EPS art
\vspace{-1.2cm} \caption{\label{fig23} Dependence of the critical
dimension $\Delta[5,3]/\xi$ on anisotropy parameter $\alpha_1$
($\alpha_2=0$) for different fixed point values of the parameter $u$
(for notation see the caption in Fig.\,\ref{fig5}).}
\end{figure}

\begin{figure}[t]
 \vspace{-0.8cm}
\includegraphics[width=70mm]{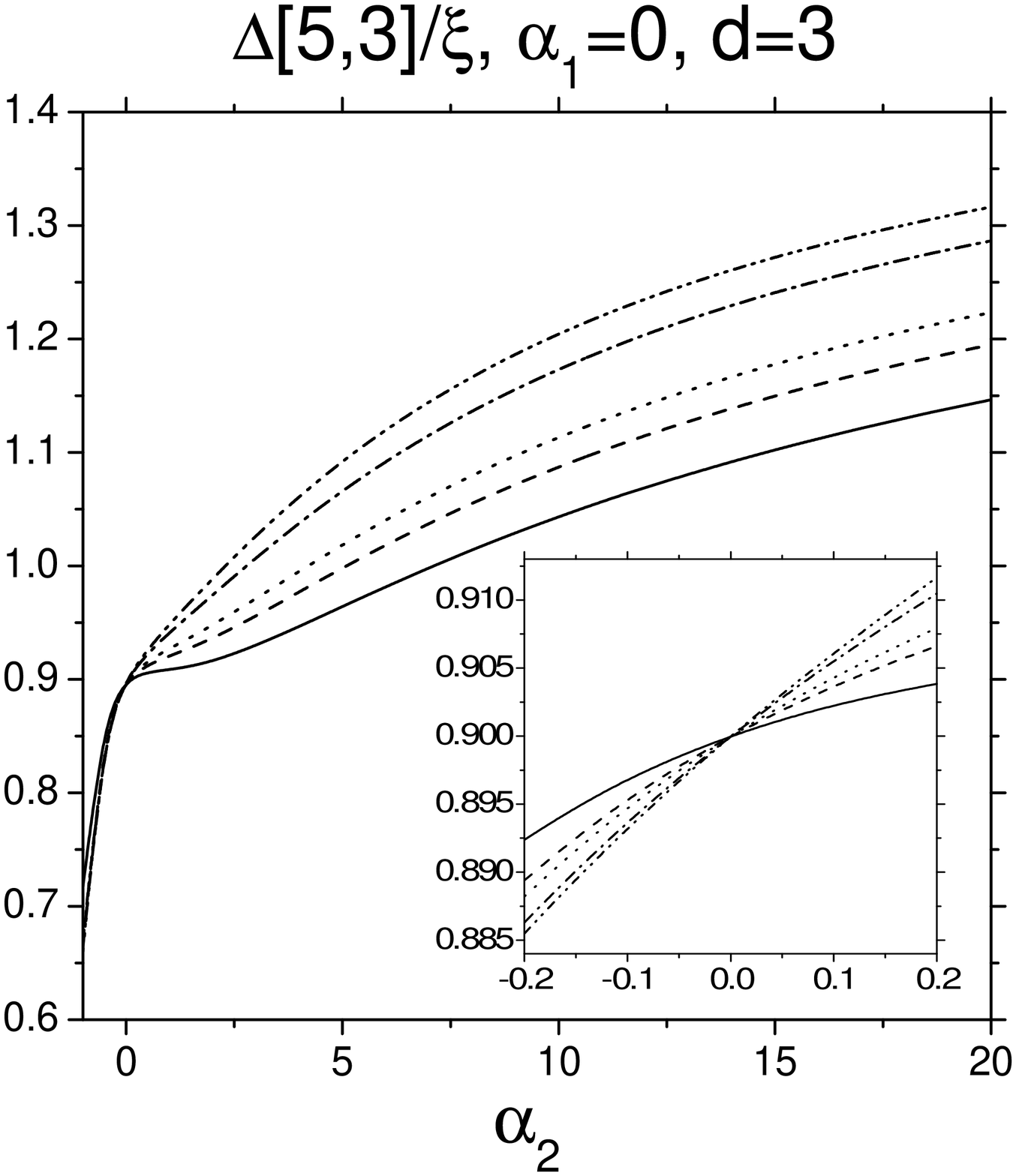}% Here is how to import EPS art
\vspace{-1.2cm} \caption{\label{fig24} Dependence of the critical
dimension $\Delta[5,3]/\xi$ on anisotropy parameter $\alpha_2$
($\alpha_1=0$) for different fixed point values of the parameter $u$
(for notation see the caption in Fig.\,\ref{fig5}).}
\end{figure}

\begin{figure}[t]
 \vspace{-0.8cm}
\includegraphics[width=70mm]{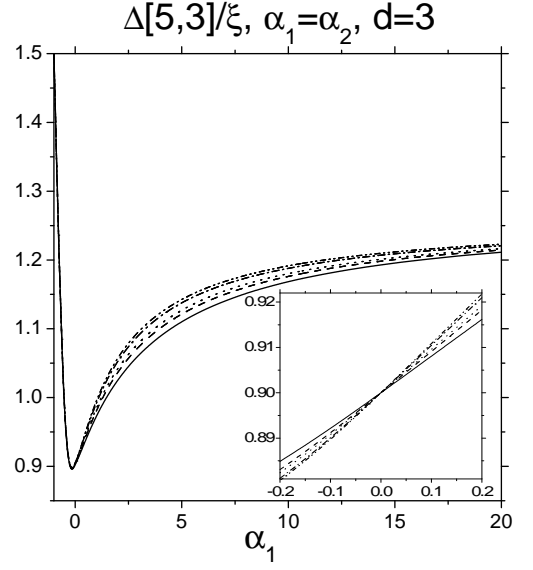}% Here is how to import EPS art
\vspace{-1.2cm} \caption{\label{fig25} Dependence of the critical
dimension $\Delta[5,3]/\xi$ on anisotropy parameter
$\alpha_1=\alpha_2$ for different fixed point values of the
parameter $u$ (for notation see the caption in Fig.\,\ref{fig5}).}
\end{figure}

Now let us turn to the calculation of the renormalization constants
$Z_{[N,p]\,[N,p']}$ in the one-loop approximation in our model. We
shall proceed as in Refs.\,\cite{AdAnHnNo00,Antonov99}. Let
$\Gamma(x;\theta)$ be the generating functional of the 1-irreducible
Green functions with one composite operator $F[N,p]$ from Eq.
(\ref{Fnp}) and any number of fields $\theta$. We shall be
interested in the $N$-th term of the expansion of $\Gamma(x;\theta)$
in $\theta$, which we denote $\Gamma_{N}(x;\theta)$; it has the form
\begin{eqnarray}
\Gamma_{N}(x;\theta)&=&\frac{1}{N!}\int dx_{1}\cdots\int
dx_{N}\,\theta(x_{1})\cdots\theta(x_{N}) \nonumber \\
&&\times\langle
F[N,p](x)\theta(x_{1})\cdots\theta(x_{N})\rangle_{\textrm{1-ir}},\label{Gamma1}
\end{eqnarray}
and in the one-loop approximation it is given as
\begin{equation}
\Gamma_{N}=F[N,p]+\Gamma^{(1)}\,, \label{Gamma2}
\end{equation}
where $\Gamma^{(1)}$ is given by the analytical calculation of the
diagram in Fig.\,\ref{fig4}, and the first term in
Eq.\,(\ref{Gamma2}) represents "tree" approximation (see also
Ref.\,\cite{AdAnHnNo00}).

\begin{figure}[t]
 \vspace{-0.8cm}
\includegraphics[width=70mm]{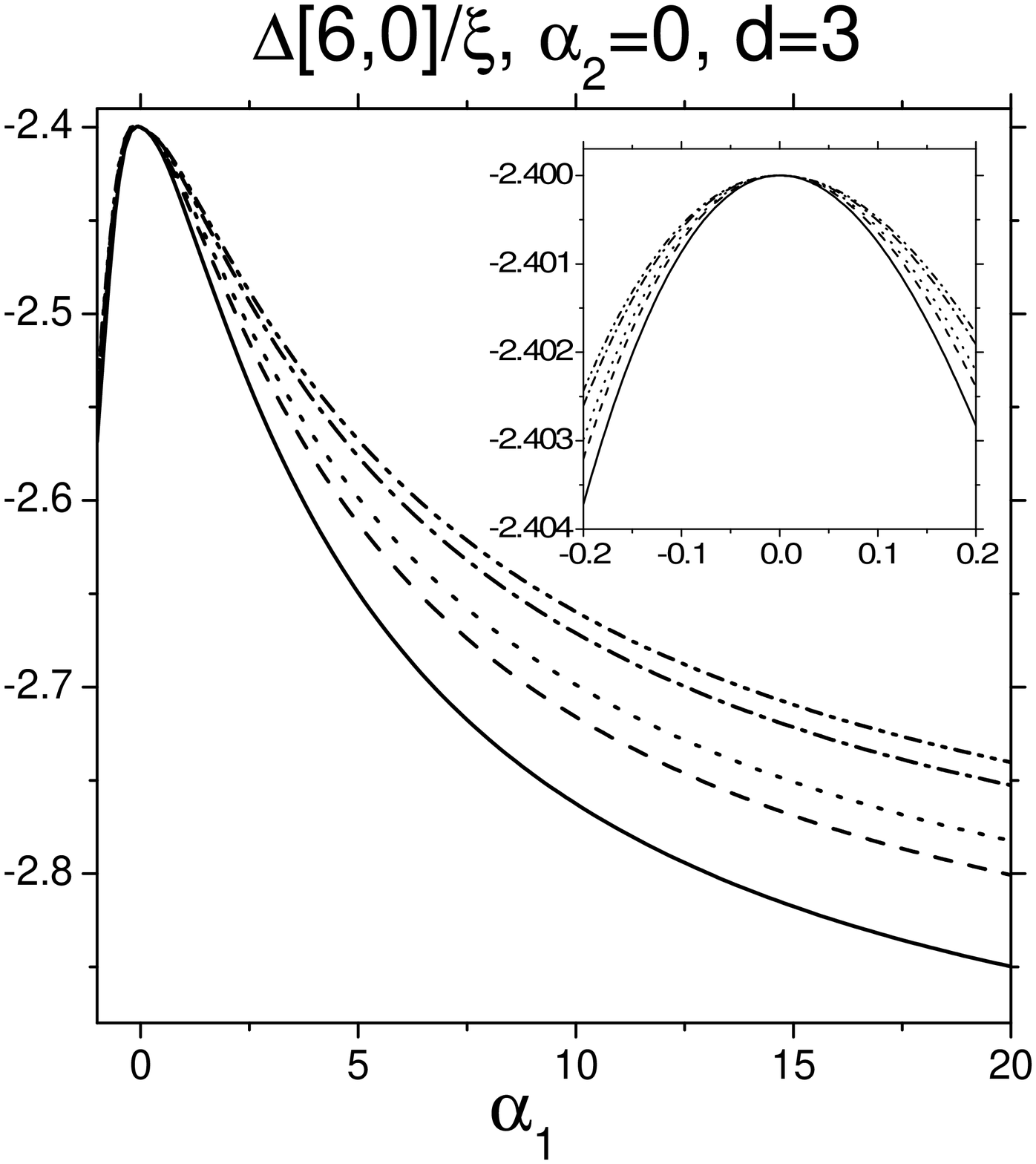}% Here is how to import EPS art
\vspace{-1.2cm} \caption{\label{fig26} Dependence of the critical
dimension $\Delta[6,0]/\xi$ on anisotropy parameter $\alpha_1$
($\alpha_2=0$) for different fixed point values of the parameter $u$
(for notation see the caption in Fig.\,\ref{fig5}).}
\end{figure}

\begin{figure}[t]
 \vspace{-0.8cm}
\includegraphics[width=70mm]{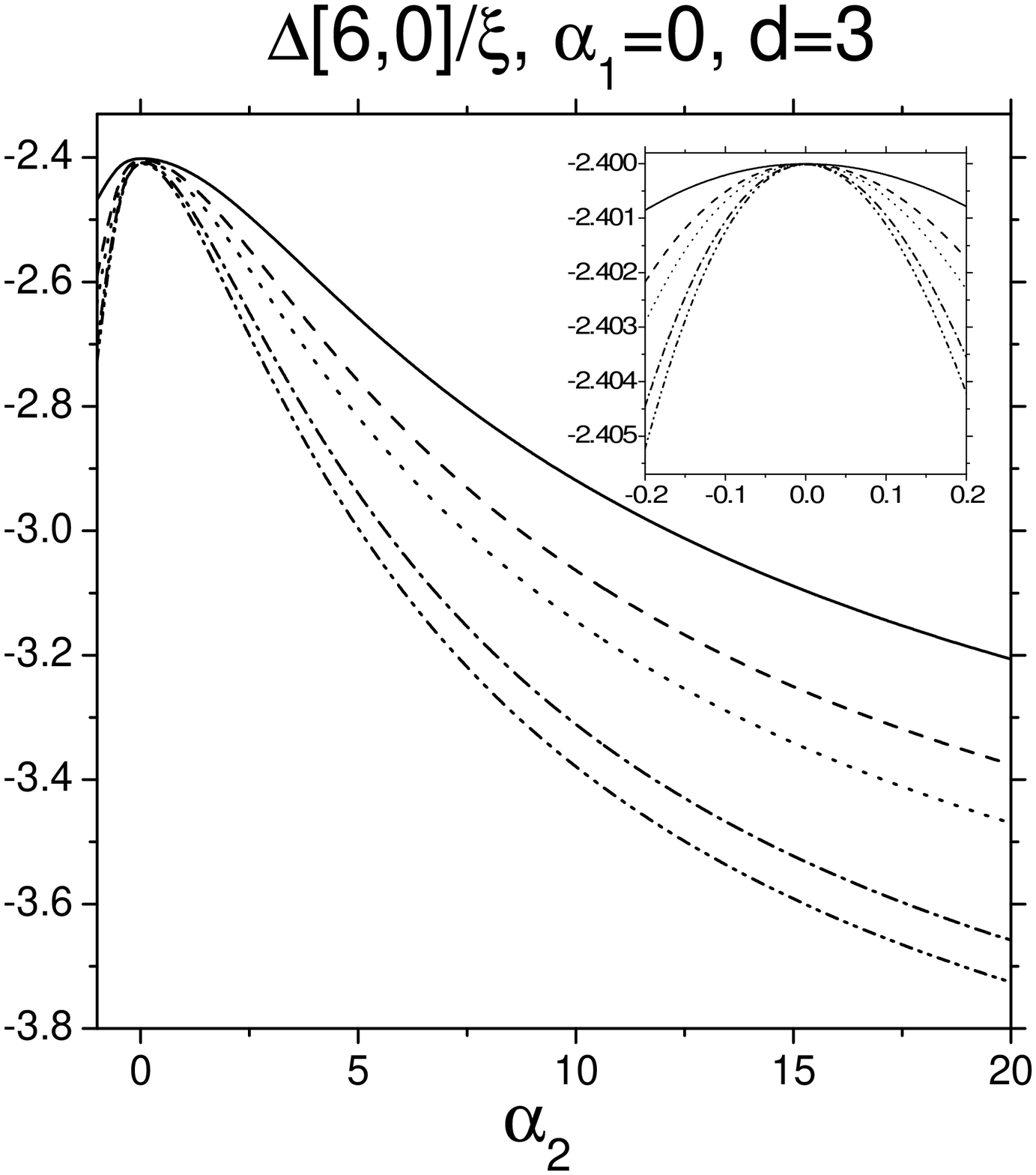}% Here is how to import EPS art
\vspace{-1.2cm} \caption{\label{fig27} Dependence of the critical
dimension $\Delta[6,0]/\xi$ on anisotropy parameter $\alpha_2$
($\alpha_1=0$) for different fixed point values of the parameter $u$
(for notation see the caption in Fig.\,\ref{fig5}).}
\end{figure}

\begin{figure}[t]
 \vspace{-0.8cm}
\includegraphics[width=70mm]{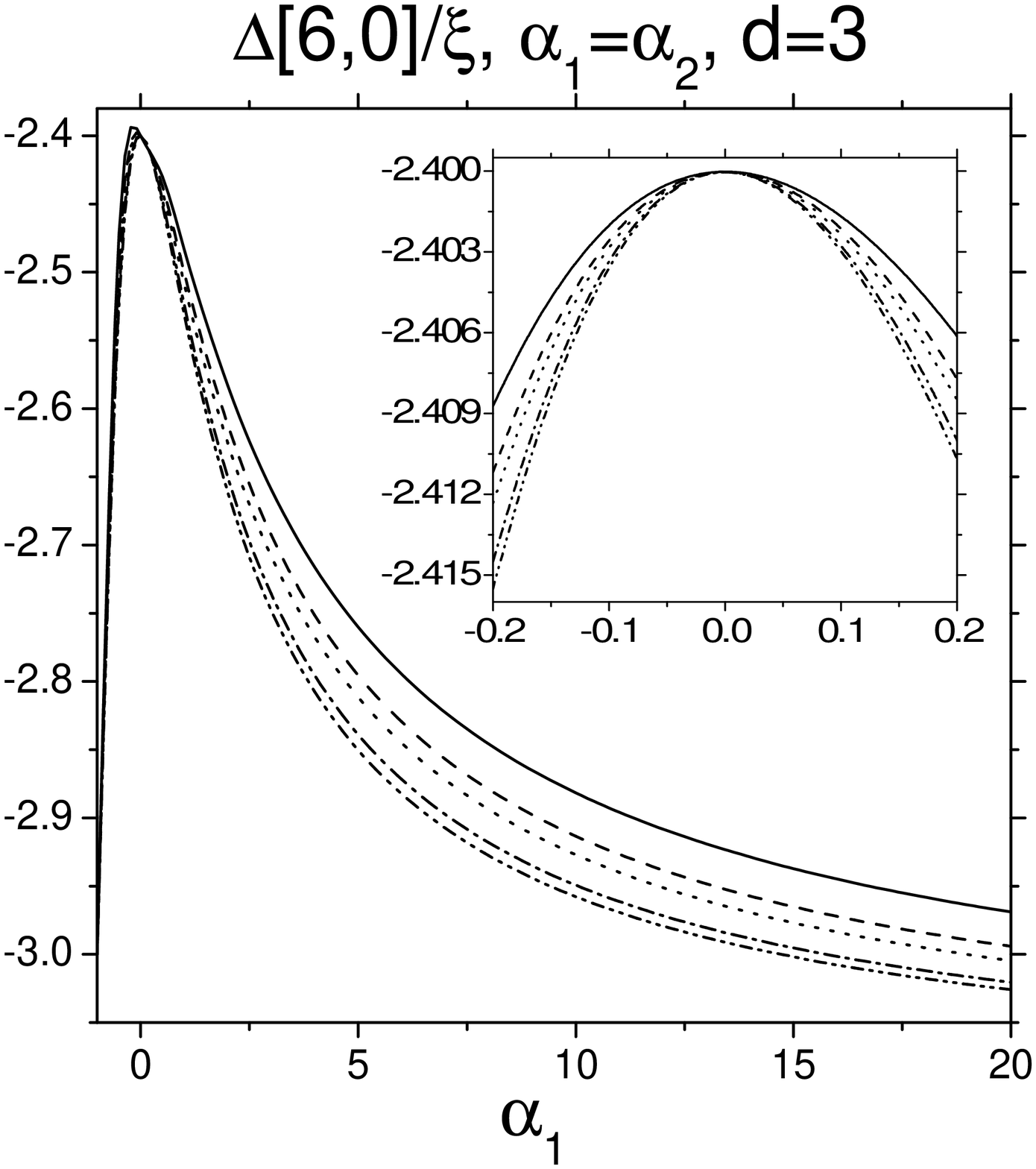}% Here is how to import EPS art
\vspace{-1.2cm} \caption{\label{fig28} Dependence of the critical
dimension $\Delta[6,0]/\xi$ on anisotropy parameter
$\alpha_1=\alpha_2$ for different fixed point values of the
parameter $u$ (for notation see the caption in Fig.\,\ref{fig5}).}
\end{figure}

The black circle with two attached lines in the diagram in
Fig.\,\ref{fig4} denotes the variational derivative $V(x;\,
x_{1},x_{2})\equiv\delta^{2}F[N,p]/{\delta\theta(x_{1})\delta\theta(x_{2})}$.
It can be represented in the following convenient form
\cite{AdAnHnNo00}
\begin{eqnarray}
V(x;\,x_{1},x_{2})&=&\partial_{i}\delta(x-x_{1})\,\partial_{j}\delta(x-x_{2})\,
\nonumber \\ && \times \frac{\partial^{2}}{\partial a_{i}\partial
a_{j}}\,\Bigl[(\mathbf{na})^{p}(a^{2})^{n}\Bigr],\label{Vertex}
\end{eqnarray}
where  a constant vector $a_{i}$ will be substituted with
$\partial_{i}\theta(x)$  after the differentiation. Analytical form
of the the diagram in Fig.\,\ref{fig4} (without the symmetry factor
$1/2$) is the following:
\begin{eqnarray}
&& \hspace{-0.7cm}\int dx_{1}\cdots\int dx_{4} V(x;\,
x_{1},x_{2})\langle\theta(x_{1})\theta'(x_{3})\rangle_{0} \nonumber
\\ && \hspace{-0.5cm} \times \langle\theta(x_{2})\theta'(x_{4})\rangle_{0}
 \langle
v_{k}(x_{3})v_{l}(x_{4})\rangle_{0}\partial_{k}\theta(x_{3})\partial_{l}\theta(x_{4}),\label{Coor}
\end{eqnarray}
where the bare propagators are given in Eqs.\,(\ref{eq:corelv}),
(\ref{eq:propthetachi}) and the derivatives are related to the
ordinary vertex factors shown in Fig.\,\ref{fig1}.

We are interested in the UV divergent part of the expression
(\ref{Coor}) which is needed for determination of the corresponding
renormalization constants. But the needed UV divergent part is
proportional to the polynomial built of $N$ gradients
$\partial_{i}\theta(x)$ at a single spacetime point $x$, and all of
them have been already extracted from the (\ref{Coor}), namely,
$N-2$ gradients are given by the vertex (\ref{Vertex}) and the
others $2$ gradients are given by the ordinary vertex factors in
Fig.\,\ref{fig1}. This important point from the view of calculations
allows us to replace the gradients with the constant vectors ${\bf
a}$. Therefore, in the end, the divergent part of expression
(\ref{Coor}) can be written in the following compact form:
\begin{equation}
a_{k}a_{l}\,\frac{\partial^{2}}{\partial a_{i}\partial
a_{j}}\,\Bigl[(\mathbf{na})^{p}(a^{2})^{n}\Bigr]\, X_{ij,\,
kl},\label{Coor2}
\end{equation}
with
\begin{eqnarray} X_{ij,\, kl} &\equiv& \int
dx_{3} \int
dx_{4}\,\,\partial_{i}\langle\theta(x)\theta'(x_{3})\rangle_{0}\,
\nonumber
\\ && \times \partial_{j}\langle\theta(x)\theta'(x_{4})\rangle_{0}\,\langle
v_{k}(x_{3})v_{l}(x_{4})\rangle_{0},\label{X}
\end{eqnarray}
or, in the momentum-frequency representation (suitable for the
further calculations), after integration over the frequency,
\begin{eqnarray}
\hspace{-0.4cm} X_{ij,\, kl}&=&\frac{D_0}{2u_0^{2}\nu_0^3}\,\int
\frac{d\mathbf{k}}{(2\pi)^{d}}\frac{k_{i}k_{j}}{(k^2+m^2)^{d/2+\varepsilon}}T_{kl}(\mathbf{k})
\nonumber
\\ &\times &\left(\frac{1}{k^{2}+\chi(\mathbf{nk})^{2}}-\frac{1}{k^{2}(1+u)+\chi(\mathbf{nk})^{2}}\right),\label{X2}
\end{eqnarray}
with $D_{0}$ from Eq.\,(\ref{corrvelo}) and $T_{kl}$ from
Eq.\,(\ref{eq:Tijk}) (we again use the possibility to work with
$\eta=0$ within one-loop approximation \cite{Antonov99,Antonov00}).
Expression (\ref{X2}) can be decomposed into some tensor structures
(see, e.g., Ref.\,\cite{AdAnHnNo00}) and after rather long but
direct calculations
%In the end, using all above defined expressions,
we are coming to the following result for the quantity
defined in Eq.\,(\ref{Coor2}):
\begin{eqnarray}
&& \hspace{-1cm}
\frac{S_{d}}{(2\pi)^{d}}\frac{g}{4u^{2}}\left(\frac{\mu}{m}\right)^{2\varepsilon}\frac{1}{\varepsilon}
\{ Q_{1}\, F[N,p-2]+Q_{2}\, F[N,p]\nonumber \\ && + Q_{3}\, F[N,p+2]
+ Q_{4}\, F[N,p+4] \},\label{X6}
\end{eqnarray}
where we have substituted the unrenormalized quantities with the
renormalized one, $a_{i}$ have been replaced with the gradients
$\partial_{i}\theta(x)$ (thus they again form the operators
$F[N,q]$, with $q=p-2,p,p+2,p+4$), and the following notation was
applied for the corresponding coefficients
\begin{equation}
Q_i=\sum_{j=0}^3 A_{ij} \left(H_j - \frac{1}{1+u} G_j \right),\quad
i=1,...,4, \label{qecka}
\end{equation}
where $H_j$ and $G_j$ are the hypergeometric functions of the
following form:
\begin{eqnarray}
H_j&=& _{2}F_{1}\left(\frac{1}{2}, 1; j+\frac{d}{2};-\chi\right),
\nonumber \\ G_j&=& _{2}F_{1}\left(\frac{1}{2}, 1;
j+\frac{d}{2};-\frac{\chi}{1+u}\right), \nonumber
\end{eqnarray}
with $j=0,...,3$, and coefficients $A_{ij}$ for $i=1,...,4$ and
$j=0,...,3$ are given in Appendix A.

\begin{figure}[t]
 \vspace{-0.8cm}
\includegraphics[width=70mm]{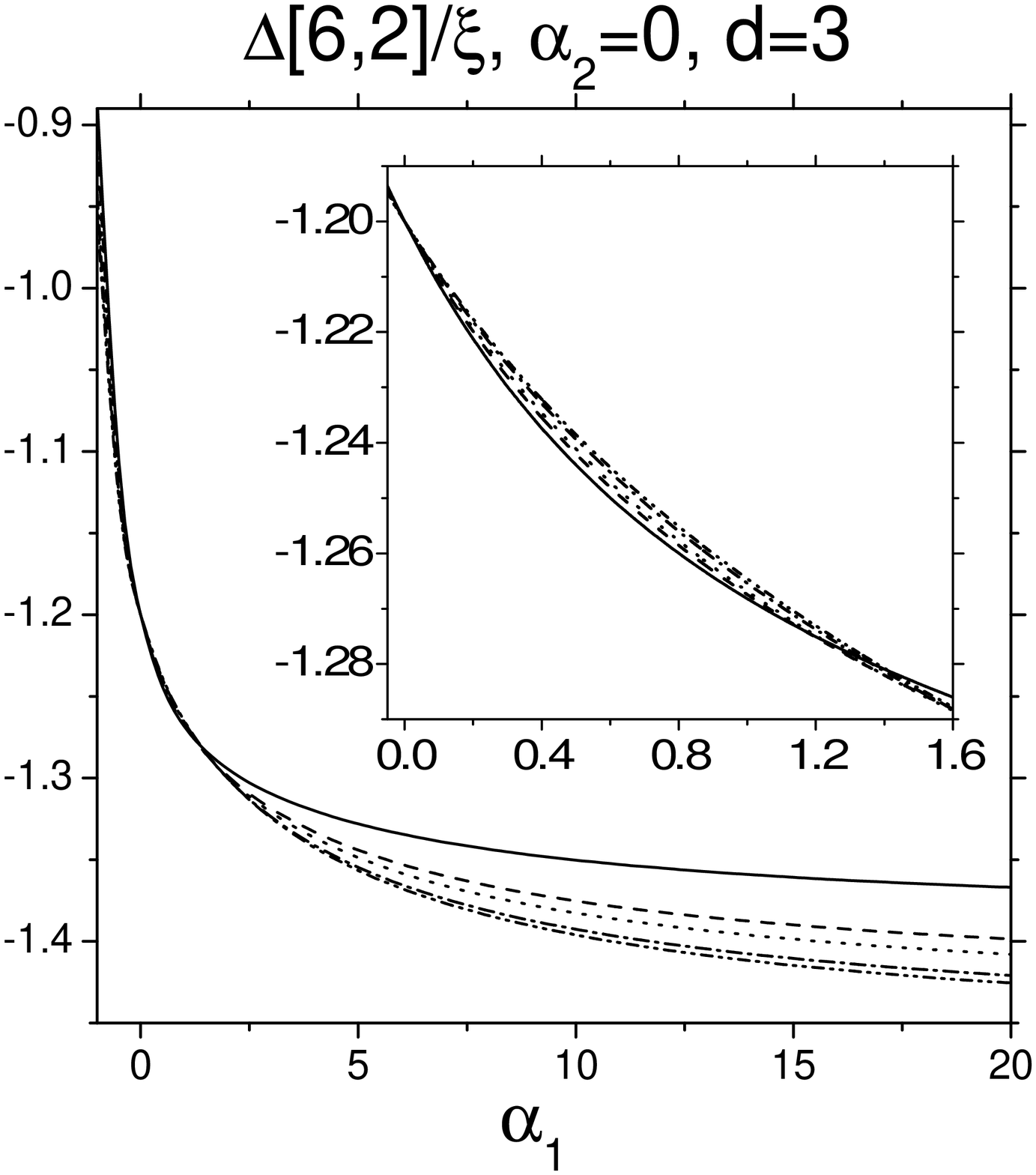}% Here is how to import EPS art
\vspace{-1.2cm} \caption{\label{fig29} Dependence of the critical
dimension $\Delta[6,2]/\xi$ on anisotropy parameter $\alpha_1$
($\alpha_2=0$) for different fixed point values of the parameter $u$
(for notation see the caption in Fig.\,\ref{fig5}).}
\end{figure}

\begin{figure}[t]
 \vspace{-0.8cm}
\includegraphics[width=70mm]{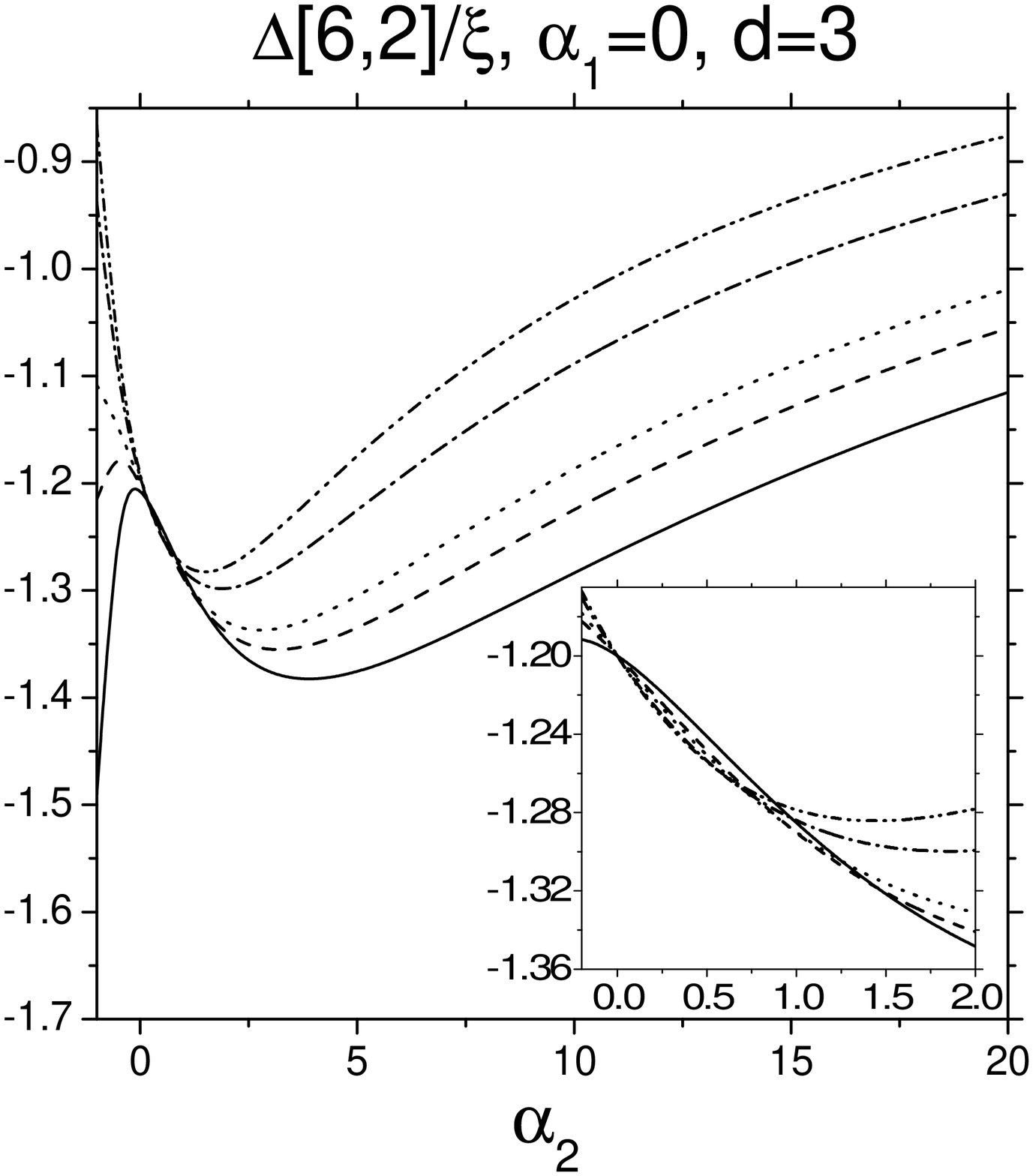}% Here is how to import EPS art
\vspace{-1.2cm} \caption{\label{fig30} Dependence of the critical
dimension $\Delta[6,2]/\xi$ on anisotropy parameter $\alpha_2$
($\alpha_1=0$) for different fixed point values of the parameter $u$
(for notation see the caption in Fig.\,\ref{fig5}).}
\end{figure}

\begin{figure}[t]
 \vspace{-0.8cm}
\includegraphics[width=70mm]{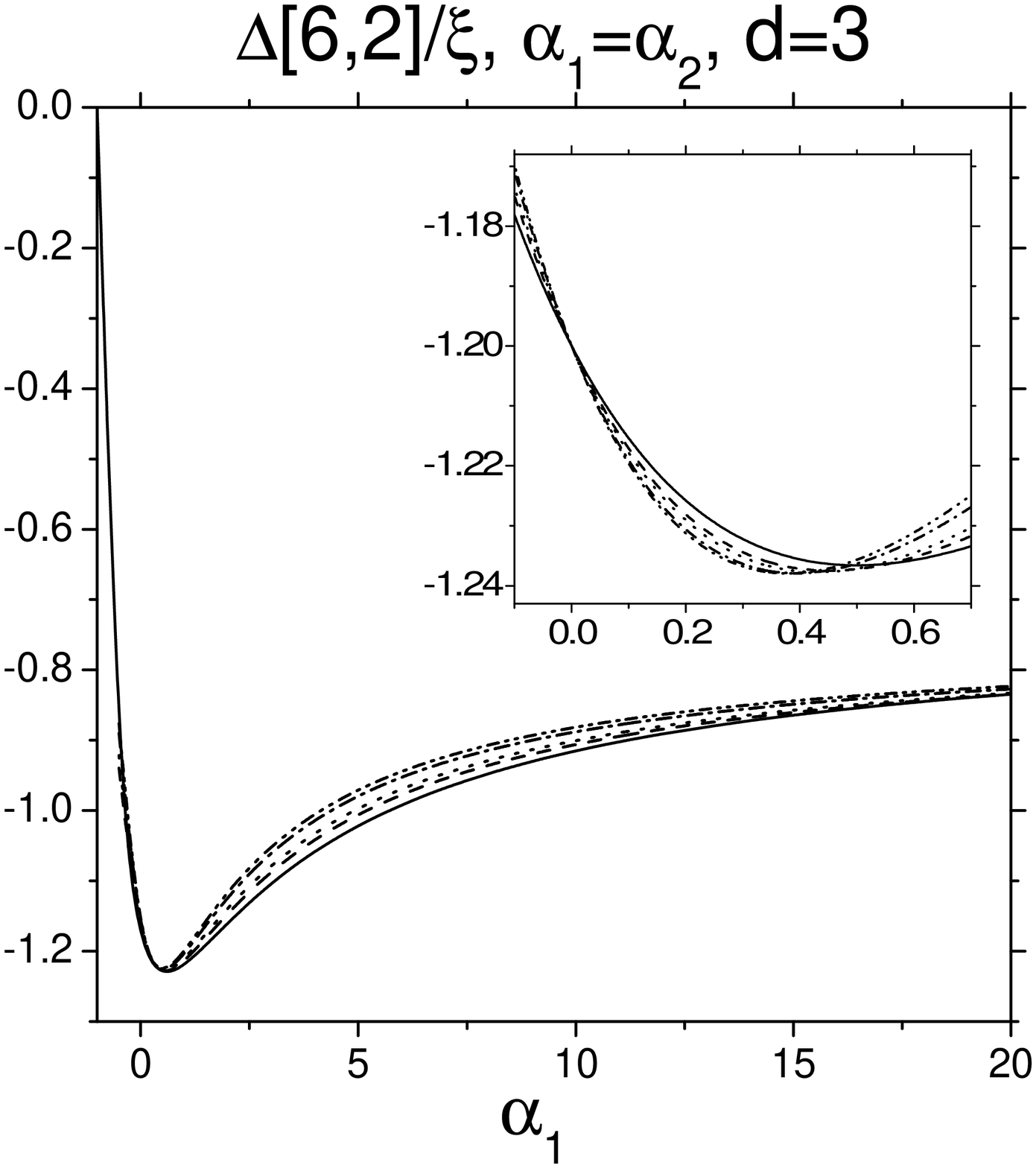}% Here is how to import EPS art
\vspace{-1.2cm} \caption{\label{fig31} Dependence of the critical
dimension $\Delta[6,2]/\xi$ on anisotropy parameter
$\alpha_1=\alpha_2$ for different fixed point values of the
parameter $u$ (for notation see the caption in Fig.\,\ref{fig5}).}
\end{figure}

Using the standard renormalization procedure the renormalization
constants $Z_{[N,p]\,[N,p']}$ defined in Eq.\,(\ref{Fnl}) are found
from the requirement that function (\ref{Gamma2}) is UV finite
(contains no poles in $\varepsilon$) when is written in renormalized
variables and with the replacement $F[N,p]\rightarrow F^{R}[N,p]$.
In the end, from Eqs. (\ref{Gamma2}) and (\ref{X6}) we have
\begin{eqnarray}
Z_{[N,p][N,p-2]} &=&  \frac{\bar{g}}{8u^{2}\varepsilon}\, Q_{1},
\\
Z_{[N,p][N,p]} &=& 1+\frac{\bar{g}}{8u^{2}\varepsilon}\, Q_{2},
\\
Z_{[N,p][N,p+2]} &=& \frac{\bar{g}}{8u^{2}\varepsilon}\,
Q_{3}, \\
Z_{[N,p][N,p+4]} &=& \frac{\bar{g}}{8u^{2}\varepsilon}\,
Q_{4},\label{Znp}
\end{eqnarray}
with coefficients $Q_{i}$ given in Eq. (\ref{qecka}). Using the
definition of the matrix of anomalous dimensions
$\gamma_{[N,p]\,[N',p']}$ given in Eq.\,(\ref{2.2}) we are coming to
the following result
\begin{eqnarray}
\gamma_{[N,p][N,p-2]} & = & -\frac{\bar{g}}{4u^{2}}Q_{1}, \nonumber \\
\gamma_{[N,p][N,p]} & = & -\frac{\bar{g}}{4u^{2}}Q_{2}, \nonumber \\
\gamma_{[N,p][N,p+2]}&=& -\frac{\bar{g}}{4u^{2}}Q_{1}, \label{Gnp} \\
\gamma_{[N,p][N,p+4]} & = & -\frac{\bar{g}}{4u^{2}}Q_{1}, \nonumber
\end{eqnarray}
and the desired matrix of critical dimensions (\ref{32B}) has the
form
\begin{equation}
\Delta_{[N,p][N,p']}=N\,\gamma_{\nu}^*/2+\gamma_{[N,p][N,p']}^{*},\label{Dnp}
\end{equation}
where the asterisk means that the quantities are taken at the
corresponding fixed point (see Sec.\,\ref{sec4}) and
$\gamma_{\nu}^*$ is given in Eq.\,(\ref{xi}). The nonzero one-loop
contribution to the matrix of critical dimension (\ref{Dnp}) is
represented by Eqs.\,(\ref{Gnp}) with $Q_i,i=1,...,4$ defined in
Eq.\,(\ref{qecka}). It means that the matrix elements of the matrix
$\gamma_{[N,p]\,[N',p']}$ other than given in Eq.\,(\ref{Gnp}) are
equal to zero. It can be seen immediately that the matrix of
critical dimensions depends on the anisotropy parameters $\alpha_1$
and $\alpha_2$ and, what is now more interesting and important here,
on the parameter $u$ (see below).

\begin{figure}[t]
 \vspace{-0.8cm}
\includegraphics[width=70mm]{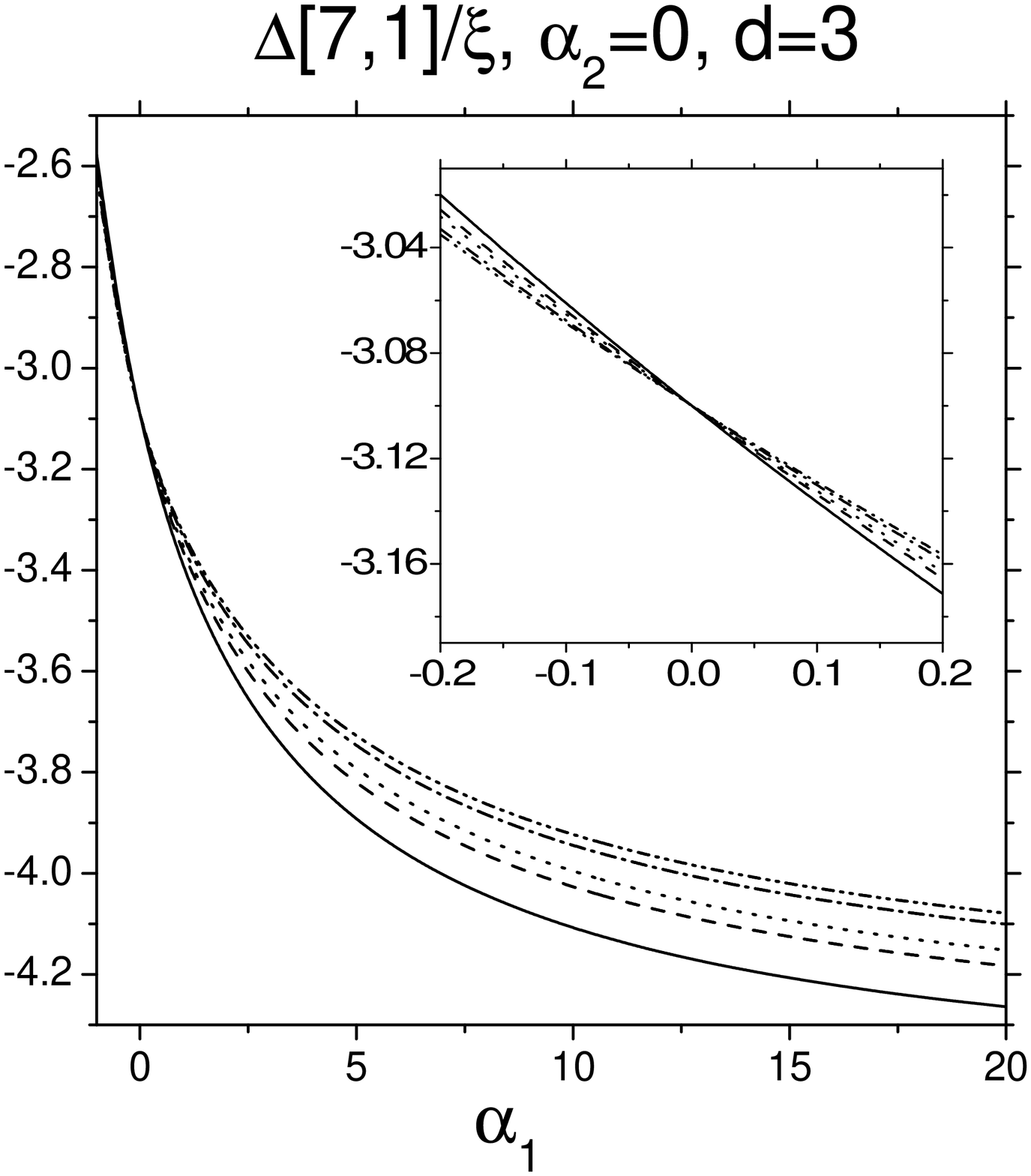}% Here is how to import EPS art
\vspace{-1.2cm} \caption{\label{fig32} Dependence of the critical
dimension $\Delta[7,1]/\xi$ on anisotropy parameter $\alpha_1$
($\alpha_2=0$) for different fixed point values of the parameter $u$
(for notation see the caption in Fig.\,\ref{fig5}).}
\end{figure}

\begin{figure}[t]
 \vspace{-0.8cm}
\includegraphics[width=70mm]{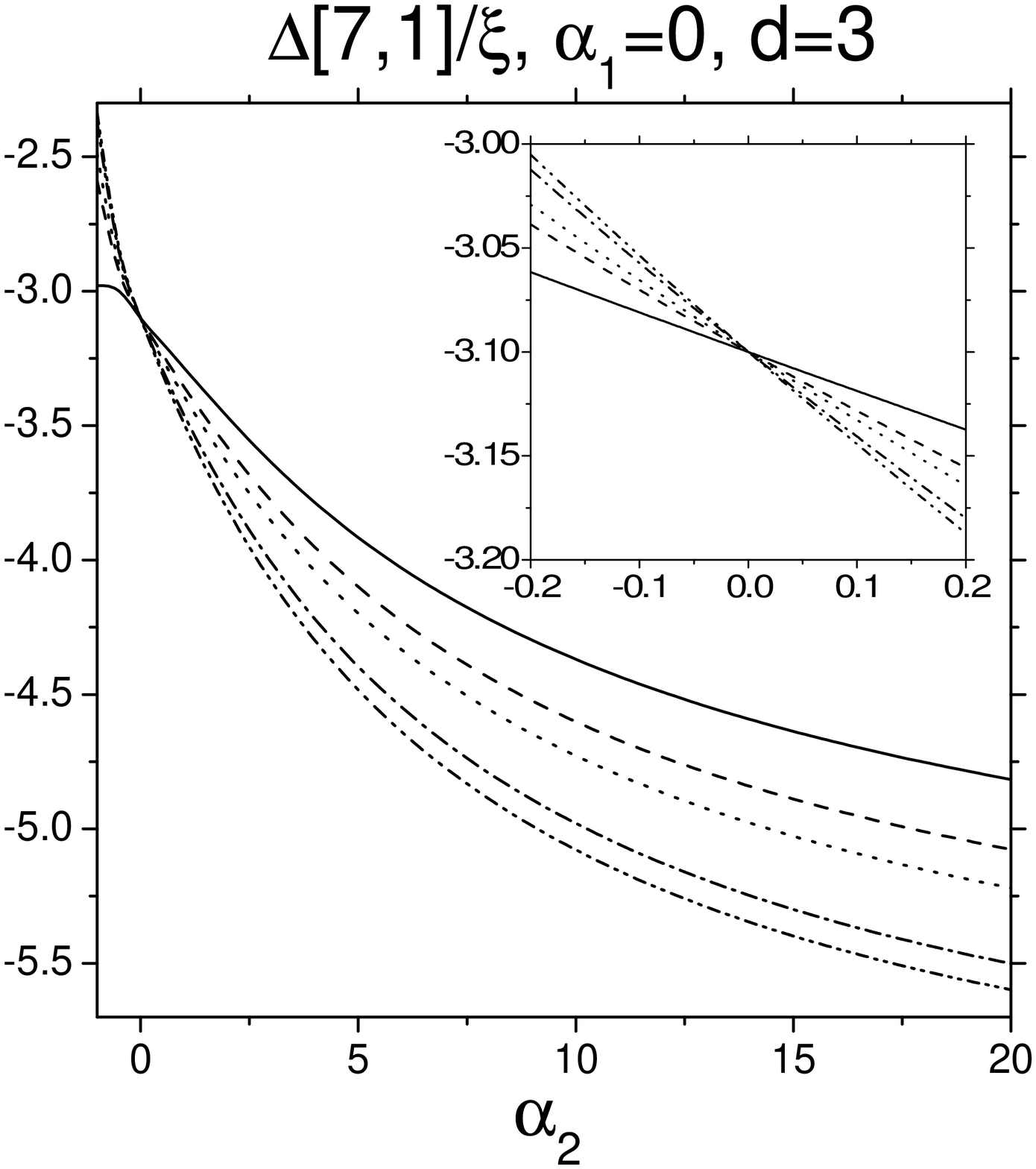}% Here is how to import EPS art
\vspace{-1.2cm} \caption{\label{fig33} Dependence of the critical
dimension $\Delta[7,1]/\xi$ on anisotropy parameter $\alpha_2$
($\alpha_1=0$) for different fixed point values of the parameter $u$
(for notation see the caption in Fig.\,\ref{fig5}).}
\end{figure}

In the end, the critical dimensions $\Delta[N,p]$ are given by the
eigenvalues of the matrix (\ref{Dnp}). The simplest situation occurs
in the isotropic limit with $\alpha_1=\alpha_2=0$ and,
correspondingly, $\chi^*=0$. In this case, one comes to the
triangular matrix, therefore its eigenvalues are given directly by
the diagonal elements. But more interesting is the fact that within
the isotropic model we have the same eigenvalues of the matrix of
critical dimensions for all fixed point values of $u^*$, i.e., the
eigenvalues are independent of $u$ at the fixed point, namely,
\begin{equation}
\Delta[N,p]=\left(\frac{N}{2}+\frac{p(p-1)-n(d-1)(d+N+p)}{(d-1)(d+2)}\right)\xi,\label{deltais}
\end{equation}
where $\xi$ is given in Eq.\,(\ref{xi}) (see, e.g.,
Ref.\,\cite{Antonov99} for details). As a result, it means that
within the one-loop approximation  there is no difference between
general model with finite time correlations and its two special
limits, namely, Kraichnan's rapid change limit and the frozen limit
of the model as for the anomalous behavior of the equal-time
structure functions (it, of course, also holds for the other
equal-time correlation functions).

\begin{figure}[t]
 \vspace{-0.8cm}
\includegraphics[width=70mm]{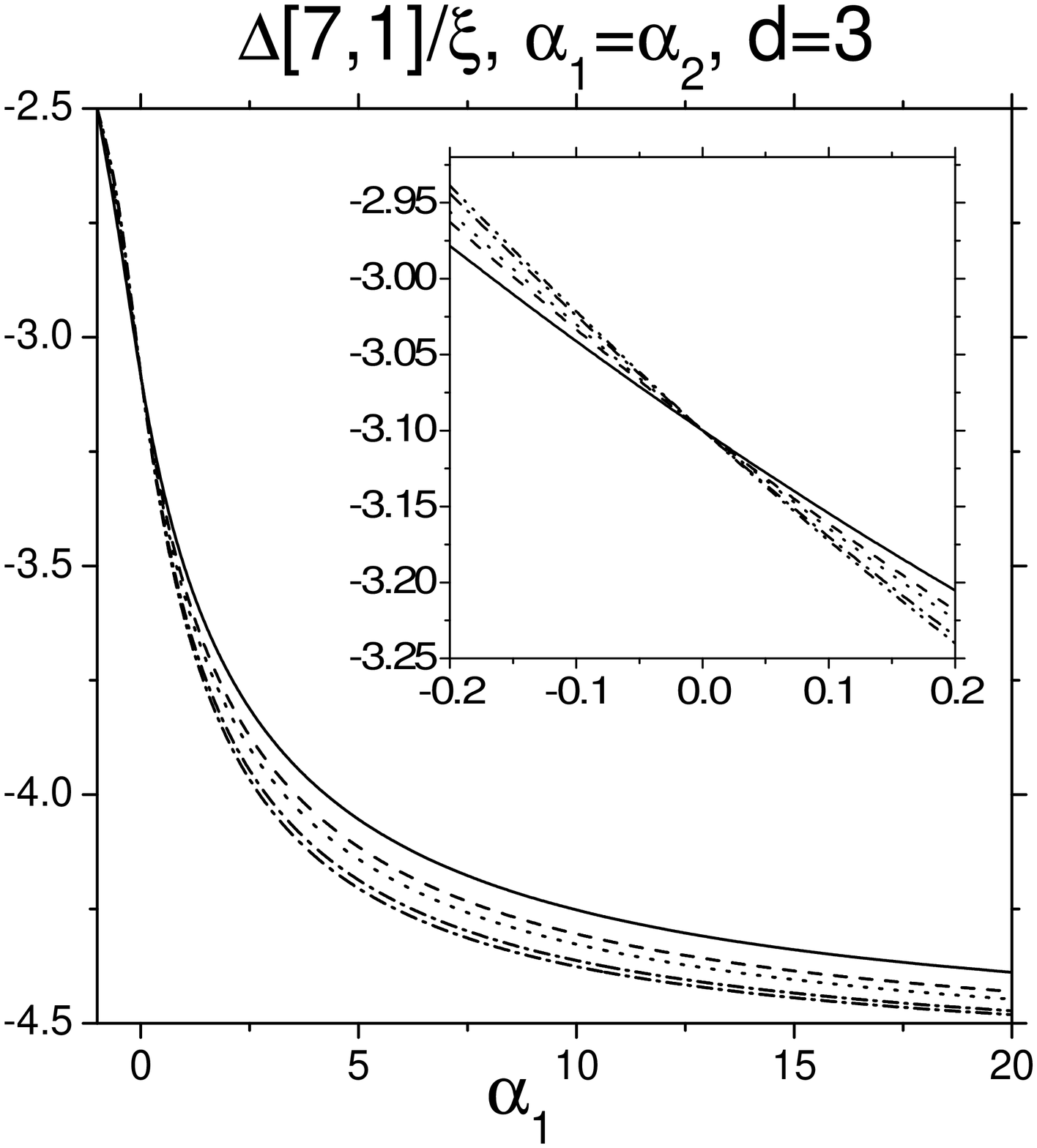}% Here is how to import EPS art
\vspace{-1.2cm} \caption{\label{fig34} Dependence of the critical
dimension $\Delta[7,1]/\xi$ on anisotropy parameter
$\alpha_1=\alpha_2$ for different fixed point values of the
parameter $u$ (for notation see the caption in Fig.\,\ref{fig5}).}
\end{figure}

\begin{figure}[t]
 \vspace{-0.8cm}
\includegraphics[width=70mm]{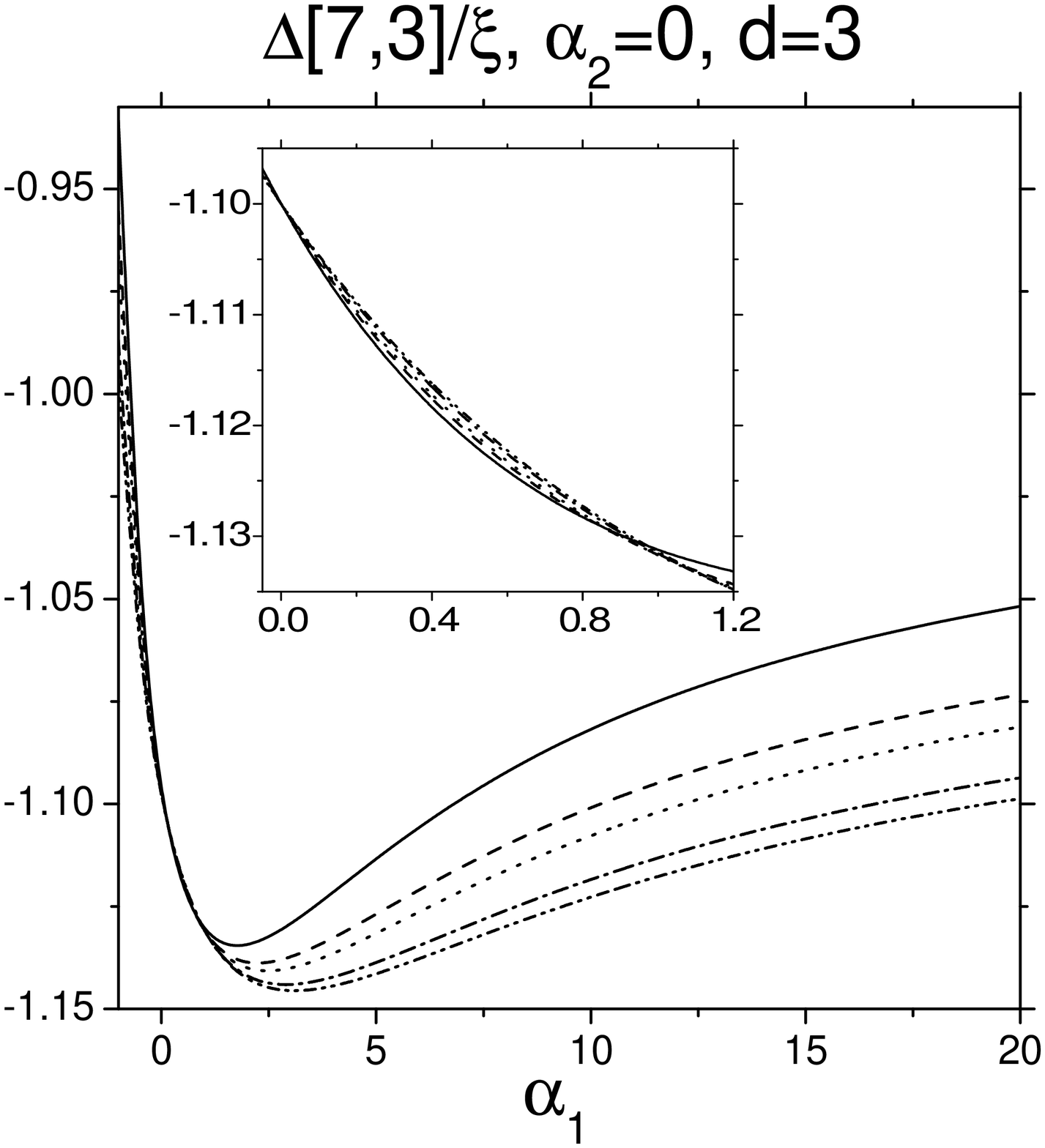}% Here is how to import EPS art
\vspace{-1.2cm} \caption{\label{fig35} Dependence of the critical
dimension $\Delta[7,3]/\xi$ on anisotropy parameter $\alpha_1$
($\alpha_2=0$) for different fixed point values of the parameter $u$
(for notation see the caption in Fig.\,\ref{fig5}).}
\end{figure}

The situation is different when presence of small scale anisotropy
is supposed. In this case, the matrix of critical dimensions is not
diagonal and the eigenvalues depend on anisotropy parameters, as
well as on the parameter $u$. It leads to the sufficient difference
between anomalous dimensions of the models with different time
correlations of the velocity field. On the other hand, the fact that
the matrix (\ref{Dnp}) is triangular in the isotropic case (it is
also triangular in the case with large-scale anisotropy) is also
important here because it allows us to assign uniquely the concrete
critical dimension to the corresponding composite operator even in
the case with small-scale anisotropy and study their hierarchical
structure as functions of $p$ (see Ref.\,\cite{AdAnHnNo00} for
details). As was shown in Ref.\,\cite{AdAnHnNo00} within the
Kraichnan model, as for anomalous scaling, the leading role is
played by the operators with the most negative critical dimensions:
for the structure functions (\ref{struc}) with even $N$ it is the
operator with $p=0$ and for the structure functions (\ref{struc})
with odd $N$ it is the operator with $p=1$. As we shall see, the
same situation also holds in the general case with the finite time
correlations.

\subsection{Anomalous scaling of the structure functions in one-loop approximation}

The combination of the RG representation (\ref{frscaling}) with the
OPE (\ref{ope}) leads to the final asymptotic expression for the
structure functions (\ref{struc}) within the inertial range, namely,
\begin{eqnarray}
S_N({\bf r})&\simeq&  r^{N(1-\xi/2)} \label{struc10}
\\ && \times \sum_{N^{\prime}\leq N} \sum_p
\{C_{N^{\prime,p}}\,(r/L)^{\Delta[N^{\prime},p]}+\dots\}\,,\nonumber
\end{eqnarray}
where $\xi$ is defined in Eq.\,(\ref{xi}), $p$  obtains all possible
values for given $N^{\prime}$, $C_{N^{\prime,p}}$  are numerical
coefficients which are functions of the parameters of the model, and
dots means contributions by the operators others than $F[N,p]$ (see,
e.g., \cite{Vasiliev,AdAnHnNo00} for details).

As was already mentioned in Introduction, our aim is twofold. First
of all, we shall find the dependence of the critical dimensions on
the parameter $u$, thus we shall answer the question whether the
system with finite time correlations of the velocity field with
presence of small-scale anisotropy is more anomalous, i.e., whether
the corresponding critical dimensions are less than those of the
Kraichnan rapid change model which was investigated in
Ref.\,\cite{AdAnHnNo00}. This question is interesting because the
model with finite correlation time of velocity field can be consider
as further step on the way to the model with velocity field driven
by the stochastic Navier-Stokes equation. Thus, the answer on the
aforementioned question in the framework of the present model can
also give preliminary answer, as for possible tendencies, on the
similar question in the framework of the scalar advection by the
Navier-Stokes velocity field. The second aim is to investigate
whether the system with finite correlation time of velocity field
together with the presence of small-scale anisotropy can lead to the
more complicated structure of critical dimensions than it was shown
in Ref.\,\cite{AdAnHnNo00}. There are two possibilities. First, it
is possible that the pairs of complex conjugate eigenvalues of the
matrix of critical dimensions can exist. In this case, the
oscillation behavior of the corresponding scaling function appears.
Therefore, the scaling functions in Eq.\,(\ref{struc10}) would
contain terms of the following form
\begin{equation}
(r/L)^{\Delta_{R}}\left\{ c_1 \cos\left[\Delta_I (r/L) \right] + c_2
\sin\left[\Delta_I (r/L) \right]\right\},
\end{equation}
where $\Delta_R$ and $\Delta_I$ are real and imaginary part of
$\Delta$, and $c_{1,2}$ are constants. Another, in general, possible
structure of the matrix (\ref{Dnp}) is related to the situation when
the matrix of critical dimensions cannot be diagonalized and has
only the Jordan form. Then a logarithmic correction would be
involved to the powerlike behavior of the form
\begin{equation}
(r/L)^{\Delta}\left[ c_1 \ln(r/L) + c_2\right],
\end{equation}
where $\Delta$ is the eigenvalue related to the Jordan cell.

\begin{figure}[t]
 \vspace{-0.8cm}
\includegraphics[width=70mm]{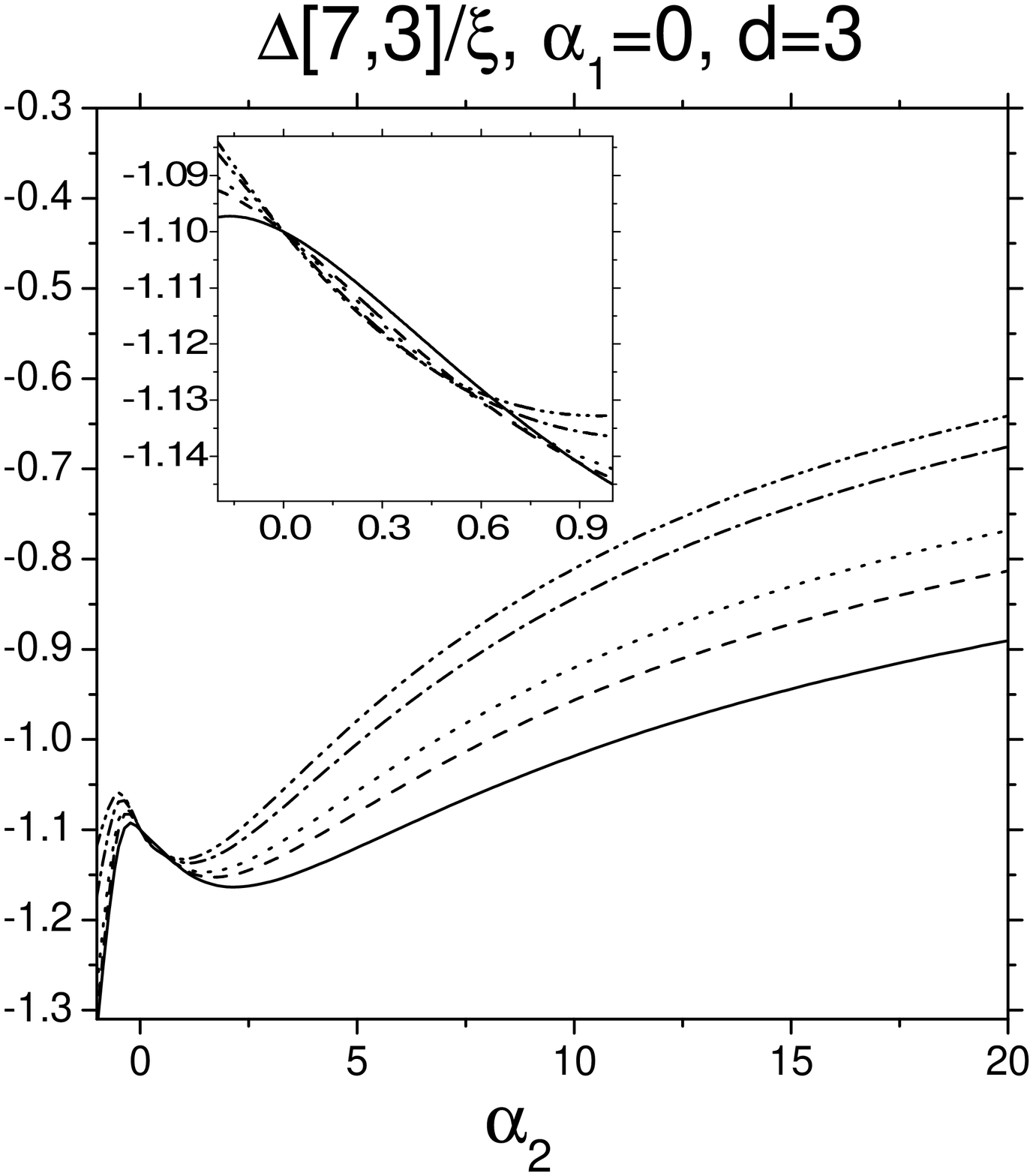}% Here is how to import EPS art
\vspace{-1.2cm} \caption{\label{fig36} Dependence of the critical
dimension $\Delta[7,3]/\xi$ on anisotropy parameter $\alpha_2$
($\alpha_1=0$) for different fixed point values of the parameter $u$
(for notation see the caption in Fig.\,\ref{fig5}).}
\end{figure}

\begin{figure}[t]
 \vspace{-0.8cm}
\includegraphics[width=70mm]{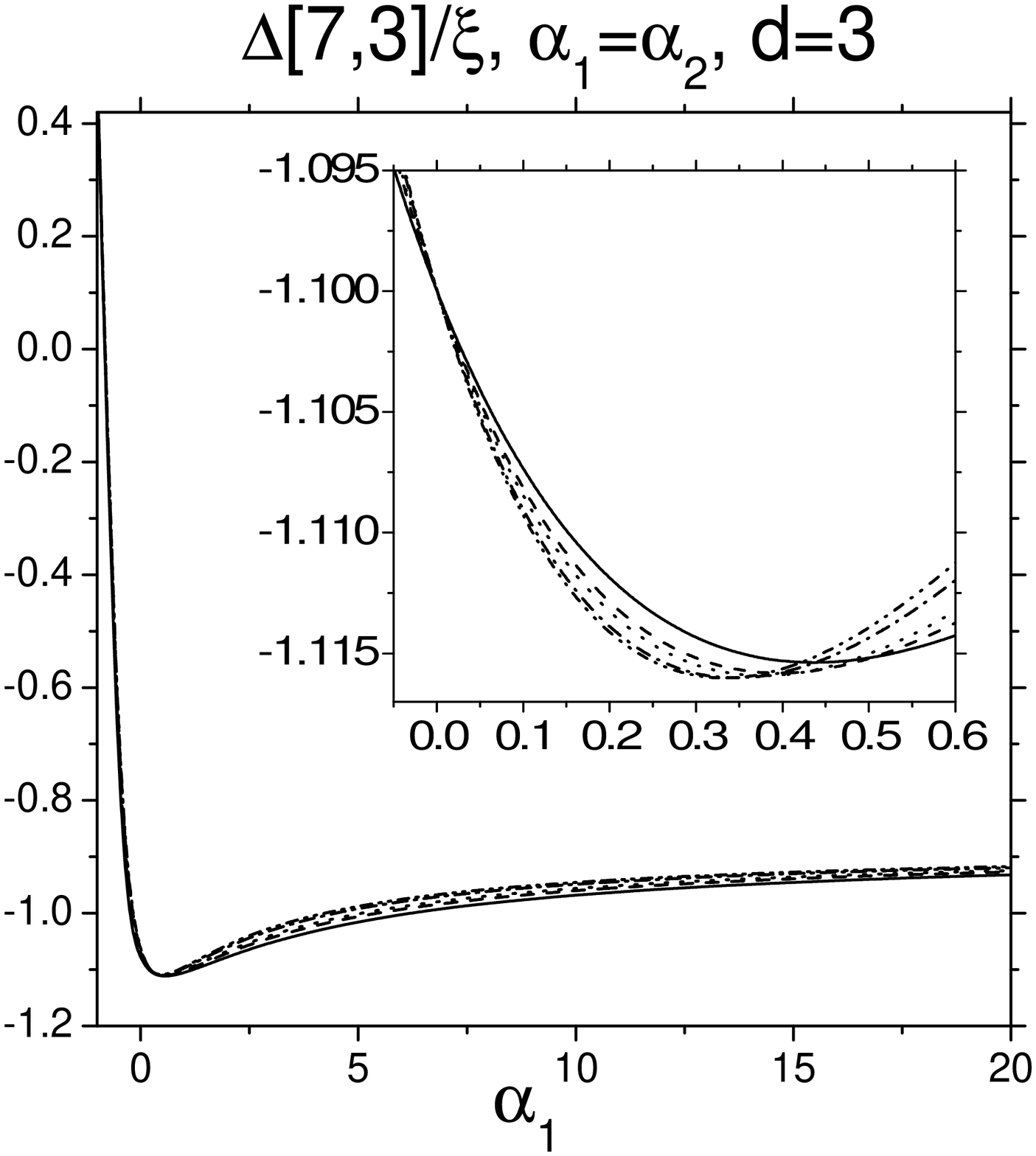}% Here is how to import EPS art
\vspace{-1.2cm} \caption{\label{fig37} Dependence of the critical
dimension $\Delta[7,3]/\xi$ on anisotropy parameter
$\alpha_1=\alpha_2$ for different fixed point values of the
parameter $u$ (for notation see the caption in Fig.\,\ref{fig5}).}
\end{figure}

In Figs.\,\ref{fig5}-\ref{fig37} behaviour of the eigenvalues of the
matrix of critical dimensions $\Delta[N,p]$ for various values of
$N=2,3,4,5,6,7$ ($p=0,2$ for even values of $N$ and $p=1,3$ for odd
values of $N$) is shown as function of the anisotropy parameters
$\alpha_1$ and $\alpha_2$ in three-dimensional case and for
different fixed point values of the parameter $u$. The dependence of
the critical dimension $\Delta[2,0]$ is not shown explicitly because
it is identically equal to zero for all fixed point values of the
parameter $u$. It can be shown either by direct calculation or by
using the Schwinger equation (see, e.g., Ref.\,\cite{AdAnHnNo00}).
At first sight one can conclude that there are different behaviors
of critical dimensions as functions of anisotropy parameters
$\alpha_1$ and $\alpha_2$ and of the parameter $u^*$ for odd and
even structure functions. Let us discuss it in detail.

First of all we shall concentrate on the even structure functions
($N=2,4,6$) and we shall discuss the behavior of the most important
critical dimensions with $p=0$ which define the anomalous scaling of
the corresponding structure functions. As was already mentioned in
the case $N=2$ the corresponding critical dimensions $\Delta[2,0]$
are identically equal to zero. On the other hand, one can see
identical qualitative behavior of the critical dimensions
$\Delta[4,0]$ and $\Delta[6,0]$ as functions of anisotropy
parameters as is shown in Figs.\,\ref{fig14}-\ref{fig16} and in
Figs.\,\ref{fig26}-\ref{fig28}. In the case when the anisotropy
parameter $\alpha_2$ is vanished the corresponding critical
dimensions (as functions of parameter $\alpha_1$) are the most
negative in the frozen limit of the model ($u^*=0$) as is shown in
Figs.\,\ref{fig14} and \ref{fig26}. On the other hand, in the case
when the anisotropy parameter $\alpha_1$ is vanished (see
Figs.\,\ref{fig15} and \ref{fig27}), as well as in the case when
$\alpha_1=\alpha_2$ (see Figs.\,\ref{fig16} and \ref{fig28}) the
situation is opposite, namely, the most negative critical dimensions
as functions of the corresponding anisotropy parameters are those
that corresponds to the rapid-change model limit ($u^*\rightarrow
\infty$). This is some kind of nonuniversality of the behavior of
the critical dimensions in the plane of anisotropy parameters
$\alpha_1-\alpha_2$. Thus we still have some kind of hierarchical
behavior in respect to $u^*$ but the hierarchy depends also on the
values of anisotropy parameters. It means physically that the answer
on the question which model is "more anomalous" can depend on the
form of the small scale anisotropy. Besides, it is evident that
there must exist a system of curves in the plane $\alpha_1-\alpha_2$
on which the pairs of models with different fixed point values of
the parameter $u$ have the same anomalous dimensions. We shall not
show them explicitly here because we suppose that their form will
strongly depend on the higher loop calculations which are ignored
here (we work in one-loop approximation) but we can assume that the
qualitative picture will be the same. Of course, all of the curves
must cross in the point $\alpha_1=\alpha_2=0$ (as is evident from
corresponding figures for the same value of $N$) as a result of the
fact that in the isotropic case the critical dimensions for
different values of $u^*$ are the same and they are given explicitly
in Eq.\,(\ref{deltais}).

Let us now briefly discuss the critical dimensions $\Delta[N,2]$ for
even values of $N$ with $p=2$. Of course, they are not so important
as critical dimensions $\Delta[N,0]$ but are interesting from the
point of view of their nontrivial behavior as functions of $u^*$ as
is shown in Figs.\,\ref{fig5}-\ref{fig7},\ref{fig17}-\ref{fig19},
and \ref{fig29}-\ref{fig31}. Again one can see different behavior of
the critical dimensions in different directions in the plane given
by the anisotropy parameters $\alpha_1$ and $\alpha_2$ but the most
interesting feature is the fact that the corresponding couples of
curves cross themselves in two points except for $N=2$ (one of the
two points is $\alpha_1=\alpha_2=0$).

As for the structure functions of odd order the situation is
slightly different. Again we start with the most negative critical
dimensions for which $p=1$. They are shown in
Figs.\,\ref{fig8}-\ref{fig10}, \ref{fig20}-\ref{fig22}, and
\ref{fig32}-\ref{fig34} for $N=3$, $N=5$, and $N=7$, respectively.
One can see immediately that again in different directions in the
$\alpha_1-\alpha_2$ plane different models are the most anomalous
(frozen limit of the model or rapid-change model limit), i.e., they
have the most negative critical dimensions $\Delta[N,1]$. But,
besides, the situation is also different for positive and negative
values of the anisotropy parameters. For example, in the case when
the anisotropy parameter $\alpha_2=0$ the corresponding critical
dimensions as functions of parameter $\alpha_1$ are the most
negative in the frozen limit of the model ($u^*=0$) for $\alpha_1>0$
and they are the most negative in the rapid-change model limit of
the model ($u^*=\infty$) for $-1<\alpha_1<0$  as is shown in
Figs.\,\ref{fig8}, \ref{fig20}, and \ref{fig32}. On the other hand,
in the case when the anisotropy parameter $\alpha_1=0$ (see
Figs.\,\ref{fig9}, \ref{fig21}, and \ref{fig33}), as well as in the
case when $\alpha_1=\alpha_2$ (see Figs.\,\ref{fig10}, \ref{fig22},
and \ref{fig34}) the situation is opposite. Thus, one can conclude
that the answer on the question which model is more anomalous can
depend on the form of the small scale anisotropy, i.e, on the
parameters of anisotropy.

In the end, let us briefly discuss behavior of the critical
dimensions $\Delta[N,3]$ for odd values of $N$ as shown in
Figs.\,\ref{fig11}-\ref{fig13}, \ref{fig23}-\ref{fig25}, and
\ref{fig35}-\ref{fig37} for $N=3$, $N=5$, and $N=7$, respectively.
As in the case with even values of $N$ one can see different
behavior of the critical dimensions in different directions in the
plane given by the anisotropy parameters $\alpha_1$ and $\alpha_2$
but the existence of two intersections between couples of curves
appears only for $N=7$, i.e., it is not present in the cases with
$N=3$ and $N=5$.

In present paper we have shown only the smallest critical dimensions
for concrete value of $N$, namely, $p=0,2$ for even value of $N$ and
$p=1,3$ for odd value of $N$. But corresponding analysis can be also
done for others critical dimensions which correspond to higher
possible values of $p$ ($p\leq N$). Detail analysis shows that no
exotic situation appears in their behavior as well. Thus, we can
also answer the second question whether the finite correlation time
of velocity field together with small scale anisotropy can lead to
the more complicated structure of critical dimensions (oscillations
or logarithmic corrections). Our answer is no, i.e, the matrices of
critical dimensions have real eigenvalues at least up to $N=7$.

\section{Conclusion\label{sec6}}

Using the field theoretic RG technique and operator product
expansion we have investigated the influence of uniaxial small-scale
anisotropy on a passive scalar advected  by a Gaussian solenoidal
velocity field with finite correlation time in one-loop
approximation. First of all we have found and classified all
possible scaling regimes of the model which are directly related to
the corresponding IR stable fixed points of the RG equations. The
{}``phase diagram'' of the scaling regimes in the plane
$\varepsilon-\eta$ is shown (see Fig.\,\ref{fig3}) and it is found
that the small-scale anisotropy has no influence on the stability of
the scaling regimes (on one-loop level), i.e., we have the same five
scaling regimes with the same regions of stability as in the
isotropic case of the model \cite{Antonov99}. Two of the scaling
regimes are related to {}``frozen limit'' of the model, another two
to the {}``rapid-change'' model and the last one corresponds to
general case with finite time correlations of the velocity field.

Further, we have studied the influence of small-scale anisotropy on
the anomalous scaling of the single-time structure functions of a
passive scalar using the OPE. The corresponding leading composite
operators with the smallest (the most negative) critical dimensions
are studied in detail and the critical dimensions are found as
functions of the anisotropy parameters and the fixed point value of
the parameter $u$ which represents the ratio of turnover time of
scalar field and velocity correlation time. We have shown that the
corresponding anomalous dimensions, which are the same (universal)
for all particular models with concrete value of $u$ in the
isotropic case, are different (nonuniversal) in the case with the
presence of small-scale anisotropy and they are continuous functions
of the anisotropy parameters, as well as the parameter $u$. It is
shown that there is different behavior of the anomalous dimensions
in the case of even order single-time structure functions than in
the case of odd order ones, as well as there is different behavior
of the anomalous dimensions in the different directions in the plane
of the anisotropy parameters  (see discussion in the end of the
previous section for details). Thus, the answer on the question
which special case of the general model (rapid-change limit or
frozen limit) is more anomalous in the presence of anisotropy is not
unique. Therefore, we are also not able to make definite conclusion
what one can expect in the case of more realistic model of a passive
scalar advection, namely, in the model of a passive scalar advected
by the Navier-Stokes velocity field.

It was also shown that even in the case with finite time
correlations of the velocity field the critical dimensions of the
corresponding composite operators have simple structure, i.e., the
matrices of the critical dimensions have real eigenvalues. It means
that no exotic situations, namely, oscillations or logarithmic
corrections to the critical dimensions, are present.

\begin{acknowledgments}
The work was supported in part by VEGA grant 6193 of Slovak Academy
of Sciences, and by Science and Technology Assistance Agency under
contract No. APVT-51-027904.
\end{acknowledgments}

\appendix
\section{}

The explicit form of the coefficients $A_{ij}$  with $i=1,...,4$ and
$j=0,...,3$ from Eq.\,(\ref{qecka})) is
\begin{equation}
A_{10} = \frac{p(p-1)[(d^2-5)(1+\alpha_1)+4
\alpha_2]}{d^2-1},\nonumber
\end{equation}
\begin{eqnarray}
A_{11} &=& -\frac{p(p-1)}{d (d+1)} \nonumber \\
& \times&  [d^2+d-4-\alpha_2 (d-7)+ \alpha_1 (2 d^2+d-9)], \nonumber
\end{eqnarray}
\begin{equation}
A_{12} = \frac{p(p-1)}{d (d+2)} [d+1 +(d-1)(\alpha_1(d+3)-2
\alpha_2)], \nonumber
\end{equation}
\begin{equation}
A_{13} = p(p-1)(\alpha_2-\alpha_1)\frac{(d+1)(d+3)}{d(d+2)(d+4)},
\nonumber
\end{equation}
\begin{eqnarray}
A_{20} &=& \frac{1}{d^2-1}\big\{48 n(n-1) (\alpha_2-\alpha_1-1)
\nonumber \\ && - p(p-1)[16 \alpha_2+(1+\alpha_1)(7+d^2)]\nonumber \\
&& + 2 n [4 \alpha_2(d+2) + (1+\alpha_1)(d^2-4d+9) \nonumber \\
&& \hspace{0.8cm}+ 2 p(4\alpha_2+(1+\alpha_1)(d^2-13))] \big\},
\nonumber
\end{eqnarray}
\begin{eqnarray}
A_{21} &=& \frac{1}{d(d+1)}\nonumber \\ &\times& \big\{ 4
n(n-1)[d+13-24\alpha_2+(d+25)\alpha_1] \nonumber \\
&& \hspace{0.2cm} + p(p-1)[2 d^2 +d+7+\alpha_2(7-d) \nonumber \\ &&
\hspace{1.9cm} +\alpha_1(3 d^2+d+14)] \nonumber \\
&& \hspace{0.2cm} - 2n[2p(d^2+3d-10)-3d-7 \nonumber \\
&&  \hspace{0.9cm} + \alpha_2(2p(15-d)+7d+15) \nonumber \\ && +
\alpha_1(2p(2d^2 +3 d-23)+ d^2-7d-16 )] \big\}, \nonumber
\end{eqnarray}
\begin{eqnarray}
A_{22} &=& -\frac{1}{d(d+2)} \big\{2
n(\alpha_1-\alpha_2)(7+3d)\nonumber \\ &&+
n(n-1)[12-60 \alpha_2+4 \alpha_1 (16+d)]   \nonumber \\
&&\hspace{-1.0cm} + np[16\alpha_2(d-2)-4\alpha_1(d-1)(d+7)-12(d+1)]
\nonumber \\ && + p(p-1)[d(d+1) - \alpha_2(d^2+2d+9)\nonumber
\\ && \hspace{1.6cm}+\alpha_1(3d^2+2d+7)] \big\},\nonumber
\end{eqnarray}
\begin{eqnarray}
A_{23} &=& \frac{(d+3)(\alpha_1-\alpha_2)}{d(d+2)(d+4)}\big\{
12n(n-1) \nonumber \\ && + p(d+1)[d(p-1)-12n] \big\},\nonumber
\end{eqnarray}
\begin{eqnarray}
A_{30} &=& \frac{2n}{d^2-1}\big\{  2
(n-1)[(d^2-1)(1+\alpha_1)-24\alpha_2]\nonumber \\ && \hspace{-1cm}-
(2p+1)(d^2-1)(1+\alpha_1)-8\alpha_2(d+2+4p)\big\},\nonumber
\end{eqnarray}
\begin{eqnarray}
A_{31} &=& \frac{2n}{d(d+1)}\big\{ 2(n-1)[(d+1) (d+6)\nonumber
\\ && \hspace{0.5cm}-\alpha_2 (d+25)+\alpha_1 (d+1) (2 d+5)] \nonumber \\ &&
+ (d+1)(d+2p(2 d +1)) \nonumber \\ &&+\alpha_2 (15+7 d+2p(15 - d ))
\nonumber \\ && +\alpha_1 (d+1) (d (2+6 p)-1) \big\}, \nonumber
\end{eqnarray}
\begin{eqnarray}
A_{32} &=& \frac{2n}{d(d+2)} \big\{ 2(n-1) [\alpha_1 (18+d (d+13))
\nonumber \\ && - 6 (\alpha_2-1) (d+2)] -(d+1) [d(\alpha_1-\alpha_2)
\nonumber
\\ && + 2 (d+2+ 3 \alpha_1 (d+1) -\alpha_2
(d+3)) p] \big\}, \nonumber
\end{eqnarray}
\begin{equation}
A_{33} = \frac{4n (\alpha_1-\alpha_2) (d+3) ((d+1) p - 6
(n-1))}{d(d+4)}, \nonumber
\end{equation}
\begin{equation}
A_{40} = -4 n(n-1) (1+\alpha_1), \nonumber
\end{equation}
\begin{equation}
A_{41} = \frac{4n(n-1) [2d+4-\alpha_2+3 \alpha_1 (d+1)] }{d},
\nonumber
\end{equation}
\begin{equation}
A_{42} = -\frac{4n(n-1) [3 \alpha_1 (d+2)-(\alpha_2-1) (d+4)]}{d},
\nonumber
\end{equation}
\begin{equation}
A_{43} = \frac{4n(n-1) (\alpha_1-\alpha_2) (d+3)}{d}, \nonumber
\end{equation}

\end{document}